\documentclass[aps, prd,11pt,tightenlines, secnumarabic, superscriptaddress,showpacs, preprintnumbers, nofootinbib]{revtex4-1}
\pdfoutput=1
\usepackage{xspace}
\usepackage{pifont}
\usepackage{soul}
\usepackage{todonotes}
\usepackage{multirow,array}
\usepackage[english]{babel}
\usepackage{amsmath,amssymb,amsfonts,bm,bbm,slashed}
\usepackage{graphicx}
\usepackage[sort&compress]{natbib}
\usepackage{xcolor}
\usepackage[normalem]{ulem}
\usepackage{hyperref}
\usepackage{cleveref}
\definecolor{red}{rgb}{1.0, 0, 0}
\usepackage{enumerate}
\usepackage{epsfig, subfigure}
\usepackage{epstopdf}
\usepackage{setspace}
\usepackage{booktabs, tabularx}
\usepackage{units}
\usepackage{tikz}
\usepackage{ctable}
\usepackage{comment}
\excludecomment{toexclude}
\includecomment{toinclude}

\usetikzlibrary{arrows,shapes}
\usetikzlibrary{trees}
\usetikzlibrary{matrix,arrows}                          
\usetikzlibrary{positioning}                            
\usetikzlibrary{calc,through}                           
\usetikzlibrary{decorations.pathreplacing}  

\usetikzlibrary{decorations.pathmorphing}       
\usetikzlibrary{decorations.markings}
\tikzset{
  vector/.style={decorate, decoration={snake,amplitude=2.0pt}, draw=blue},
      provector/.style={decorate, decoration={snake,amplitude=2.0pt}, draw=blue},
      antivector/.style={decorate, decoration={snake,amplitude=-2.0pt}, draw=black},
  fermion/.style={draw=black, postaction={decorate},
      decoration={markings,mark=at position .55 with {\arrow[draw=black]{>}}}},
  fermionbar/.style={draw=black, postaction={decorate},
      decoration={markings,mark=at position .55 with {\arrow[draw=black]{<}}}},
  fermionnoarrow/.style={draw=red},
  gluon/.style={decorate, draw=purple,
      decoration={coil,amplitude=4pt, segment length=5pt}},
  scalar/.style={dashed,draw=red, postaction={decorate},
      decoration={markings,mark=at position .55 with {\arrow[draw=red]{<}}}},
  scalarbar/.style={dashed,draw=red, postaction={decorate},
      decoration={markings,mark=at position .55 with {\arrow[draw=red]{<}}}},
  scalarnoarrow/.style={dashed,draw=red},
  electron/.style={draw=red, postaction={decorate},
      decoration={markings,mark=at position .55 with {\arrow[draw=red]{>}}}},
      bigvector/.style={decorate, decoration={snake,amplitude=4pt}, draw=blue},
}

\hypersetup{
    pdfnewwindow=true,   
    colorlinks=true,     
    linkcolor=blue,      
    citecolor=blue,      
    filecolor=blue,      
    urlcolor=blue        
}

\makeatletter

\newcommand{\Rmnum}[1]{\expandafter\@slowromancap\romannumeral #1@}
\makeatother

\definecolor{darkgreen}{rgb}{0,0.5,0}
\definecolor{darkblue}{rgb}{0,0,0.5}
\definecolor{newred}{rgb}{0.5,0.1,0}
\definecolor{gold}{rgb}{0.7,0.7,0}

\newcolumntype{N}{>{\centering\arraybackslash}m{4cm}}
\newcolumntype{G}{>{\bfseries\centering\arraybackslash}m{3cm+6\tabcolsep}}
\newcolumntype{M}[1]{>{\centering\arraybackslash}m{#1}}

\allowdisplaybreaks

\bibpunct{[}{]}{,}{n}{}{,}

\newcommand{\GeV}{\mathrm{GeV}}

\definecolor{seagreen}{rgb}{0.180392,0.545098,0.341176}

\newcolumntype{C}[1]{>{\centering\arraybackslash}p{#1}}
\newcolumntype{L}[1]{>{\let\arraybackslash}p{#1}}

\begin{document}

\title{Exposing Dark Sector with Future Z-Factories}

\author{Jia Liu}
\email{liuj1@uchicago.edu}
\affiliation{\mbox{Enrico Fermi Institute, The University of Chicago, 
5640 S Ellis Ave, Chicago, IL 60637, USA}}
\affiliation{PRISMA Cluster of Excellence \& Mainz Institute for Theoretical Physics, 
 Johannes Gutenberg University, 55099 Mainz,
  Germany}

\author{Lian-Tao Wang}
\email{liantaow@uchicago.edu}
\affiliation{\mbox{Enrico Fermi Institute, The University of Chicago, 
5640 S Ellis Ave, Chicago, IL 60637, USA}}
\affiliation{\mbox{Department of Physics, The University of Chicago, 
5640 S Ellis Ave, Chicago, IL 60637, USA}}
\affiliation{\mbox{Kavli Institute for Cosmological Physics, 
The University of Chicago, 5640 S Ellis Ave, Chicago, IL 60637, USA}}  

\author{Xiao-Ping Wang}
\email{xia.wang@anl.gov}
\affiliation{\mbox{High Energy Physics Division, Argonne National Laboratory, Argonne, IL 60439}}
\affiliation{PRISMA Cluster of Excellence \& Mainz Institute for
  Theoretical Physics, Johannes Gutenberg University, 55099 Mainz,
  Germany}
  
\author{Wei Xue}
\email{weixue@mit.edu}
\affiliation{\mbox{Theoretical Physics Department, CERN, CH-1211 Geneva 23, Switzerland}}
\affiliation{Center for Theoretical Physics, Massachusetts Institute of Technology,
             Cambridge, MA 02139, USA}

\date{\today}
\preprint{\parbox{6cm}{\flushright CERN-TH-2017-278 ~ EFI-17-28  ~ MITP/17-102 ~ MIT-CTP/4972 }}

\begin{abstract}

We investigate the prospects of searching dark sector models via exotic Z-boson decay at 
future $e^+ e^-$ colliders with Giga Z and Tera Z options. Four general 
categories of dark sector models: Higgs  portal dark matter, vector portal dark matter, 
inelastic dark matter and axion-like particles, are considered. 
Focusing on channels motivated by 
the dark sector models,
we carry out a model independent study of 
the sensitivities of Z-factories in probing exotic decays.
The limits on branching ratios of the exotic Z decay  
are typically $\mathcal{O} (10^{-6}   - 10^{-8.5}) $ for the Giga Z and 
$\mathcal{O} (10^{-7.5} - 10^{-11})$ for the Tera Z, and they are compared with the projection for 
the high luminosity LHC. 
We demonstrate that future Z-factories can provide its unique and leading
sensitivity, and highlight the complementarity with other experiments,  including the indirect 
and direct dark matter search limits, and the existing collider limits. Future Z factories will play 
a leading role to uncover the hidden sector of the universe in the future.

\end{abstract}

\maketitle

\tableofcontents


\newpage
\section{Introduction}

Searching for dark sector particles, including dark matter~(DM) itself and other associated states, 
is a central goal of many experimental programs around the world. In the mass range between 
$\mathrm{MeV}$ and $\mathrm{TeV}$, collider search remains a crucial method to look for these hidden particles. 
Since the dark sector particles typically only have  weak couplings with the Standard Model, colliders with higher 
luminosity are natural places to lead this quest. Recently, there have been a couple of proposals for future 
Z-factories based on circular $e^+ e^-$ colliders, including FCC-ee and CEPC 
\cite{Gomez-Ceballos:2013zzn, dEnterria:2016fpc, dEnterria:2016sca, CEPC-SPPCStudyGroup:2015csa}, which 
are considering both Giga-Z and Tera-Z options. Giga-Z (Tera-Z) means running the electron collider at Z pole energy
and accumulate $10^9$ ($10^{12}$) Z's respectively.
Given the measured cross-section of hadronic Z is $30.5 ~ \text{nb}$ \cite{ALEPH:2005ab}, the integrated luminosity for Giga Z ($10^9$ Z) and Tera Z 
($10^{12}$ Z in the plan of FCC-ee) are $22.9 ~\text{fb}^{-1}$ and $22.9 ~\text{ab}^{-1}$, respectively.
In this paper, we give projections on the sensitivities of 
Z-factory searches to a set of Z rare decay channels inspired by the dark sector models.

A coupling between Z and dark sector states, dubbed as a ``portal",  is quite generic in dark sector models. We can classify the portals based on the type of operators through which they are implemented, as following (For recent reviews, see \cite{Essig:2013lka,Alexander:2016aln,Battaglieri:2017aum})
\begin{list}{\textbullet}{\leftmargin=1em}
   \item Marginal operators: Higgs portal~\cite{Silveira:1985rk, McDonald:1993ex, Burgess:2000yq,Patt:2006fw, Kim:2006af, Barger:2007im, Kim:2008pp,Cline:2013gha} and vector portal DM models~\cite{Okun:1982xi,  Holdom:1985ag, Pospelov:2007mp, ArkaniHamed:2008qp, ArkaniHamed:2008qn, Pospelov:2008zw},   in which the dark sector interacts with Z boson via SM Higgs mixing or gauge boson mixing. The signal is exotic Z decay into SM final states with missing energy;
   \item Dim-5 operators: Axion-like particle (ALP)~\cite{Peccei:1977hh, Weinberg:1977ma, Wilczek:1977pj, Frere:1983ag,Nelson:1993nf,Bagger:1994hh, Conlon:2006tq, Svrcek:2006yi, Nomura:2008ru,
   	Arvanitaki:2009fg, Jaeckel:2010ni,Acharya:2010zx, Ringwald:2012hr}, 
with anomalous coupling to Z boson and photon. The signal is exotic Z decay into ALP and photon;
   \item Higher dimensional operators: Magnetic inelastic DM and Rayleigh DM models~\cite{Sigurdson:2004zp, Masso:2009mu, Chang:2010en, Weiner:2012cb, Weiner:2012gm},  in which the dark sector interacts with Z via magnetic dipole or Rayleigh operator.
   The signal is exotic Z decay into photon and missing energy.
\end{list}
In addition to using exotic decay measurements to probe these models, we also compare the reach with direct and indirect dark matter detection experiments, current limits from collider searches, and 
estimated sensitivities of high luminosity run of the LHC (HL-LHC). Our results demonstrate that the Z-factory 
measurement will provide the leading sensitivities in many cases. 
We also include thermal relic abundance, with the understanding that it should serve as an interesting benchmark point, rather than a strict limit.

There have been previous work on constraining dark sector related new physics from Z properties at future $e^+e^-$ colliders, including
dark photon~\cite{Hook:2010tw, Curtin:2014cca}, sterile neutrino model~\cite{Blondel:2014bra, Abada:2014cca}, Z invisible 
width \text{e.g.}~\cite{Carena:2003aj}, rare SM Z decays~\cite{Ke:2009sk, Jin:2010wg, Huang:2014cxa, Xu:2014ova, Durieux:2015hsi}, light CP-odd Higgs bosons and supersymmetric models~\cite{Cao:2010na, Wang:2011qz, Domingo:2011uf, Wang:2011zzt, Ghosh:2014rha}, . Recently, this topic has been addressed for some specific models~\cite{Flacke:2016szy, Blinov:2017dtk, Gao:2017tgx, Yu:2014ula, Fabbrichesi:2017zsc}.
LEP has also searched for exotic Z decays into  light Higgs~\cite{Acciarri:1996um, Guo:2013dc},  
two light Higgs in MSSM~\cite{Schael:2006cr}, photon and missing energy
\cite{Acciarri:1997im, Akers:1994vh, Acciarri:1999kp} and three photons
\cite{Acciarri:1994gb}.  There are also direct searches for DM particles at LEP-II via mono photon final states~\cite{Fox:2011fx}.

In \cref{sec:IDandDD}, we briefly outline the DM indirect, direct 
searches and the DM relic abundance, in order to compare with the Z decay searches in \cref{sec:models}.
Section \ref{sec:models} focuses on  well-defined and representative dark sector models 
to illustrate the power of exotic Z decay search at Z-factory. Certainly, we can not cover 
all the dark sector models for exotic Z decay. Therefore, we list the possible 
topologies for exotic Z decay according to the final states and number of
resonances in \cref{sec:ZdecayChannels}. For each decay topology, we 
comment on the origin of possible UV models, provide the appropriate cuts
for each topology and present the sensitivity on exotic Z decay BR.
In \cref{sec:conclusion}, we conclude.

\section{DM relic abundance, indirect and direct detection}
\label{sec:IDandDD}

In this section, we briefly describe the inputs from DM direct detection, indirect detection and
relic abundances employed in this study. 

DM direct detection experiments look for DM collision with nuclei in the detector,
which leaves visible energy in terms of phonon, electron and photon signals.
We are interested in the kind of collision which provides spin-independent
cross-section with nuclei, where Xenon type experiments 
such as XENON1T~\cite{Aprile:2017iyp}, LUX~\cite{Akerib:2016vxi},
PANDAX-II~\cite{Tan:2016zwf}, provide the best sensitivity for large DM mass.
For small DM mass, \textit{e.g.} $< 5 $ GeV, CRESST-II~\cite{Angloher:2015ewa} 
and CDMSlite~\cite{Agnese:2015nto} provide better sensitivity; because
their nucleus are lighter than Xenon, thus they can obtain more energy transfered 
from light DM collisions.

DM indirect detection experiments search for DM annihilation products like
photons, electrons, positrons and anti-protons from astrophysical sources. 
We consider the gamma-ray line searches by Fermi-LAT \cite{Ackermann:2013uma}, 
continuous gamma-ray limits from dwarf galaxies \cite{Ackermann:2015zua}, 
and $e^\pm$ flux measurements from AMS-02 \cite{Aguilar:2014mma}. 
And we consider constraints from cosmic microwave background~(CMB),
where DM annihilation products heat and ionize the plasma during the recombination epoch \cite{Ade:2015xua}.
When the DM annihilation cross-section is proportional to the DM velocity square $v^2$, dubbed as 
p-wave cross-section, 
the constraints from the indirect detection are negligible. 
If DM annihilate into two photon, the leading constraints from indirect detection normally are the
gamma-ray line and CMB searches.
 
The relic abundance $\Omega h^2 = 0.12$ from Planck collaboration~\cite{Ade:2015xua}  is used in this paper
as a benchmark point. We assume a standard thermal freeze out. Therefore, the relic abundance only depends 
on the thermal average of the DM annihilation cross-section $\sigma v$.\footnote{As a caveat, this 
choice relies on the assumption of the standard thermal freeze-out.
Some non-thermal process or other interesting model building of hidden sector for GeV DM can give us different
dark matter relic density predication, which is not the focus of this paper.}

\section{Hidden sector models and exotic Z decays}
\label{sec:models}

In this section, we discuss several classes of well motivated dark sector models, such as Higgs portal DM, 
vector portal DM, inelastic DM and ALPs. These models can be probed by the exotic 
Z decays in future $e^+e^-$ Z-factories. It is a demonstration of the capability of $e^+e^-$ collider as 
a new physics search machine and a novel intensity frontier experiment.
  
For each model, we point out how it could be probed by the exotic Z decays. The existing limits from cosmology,
astrophysics and collider are presented and compared with the reach of the Z-factories. 
If the model contains a dark matter candidate, we will derive the DM relic density by assuming thermal production. 
The limits from exotic Z decay are obtained from the general analysis presented in detail in \cref{sec:ZtoMET2l}.

\subsection{Higgs Portal Fermionic DM}

\label{sec:ModelSingletS}

Higgs portal is a particular simple possibility to extend the Standard Model and link it with hidden sectors.
After discovering Higgs, searching for another fundamental scalar will help us improve our understanding of
Electroweak Symmetry breaking. Interestingly, the other scalar is potentially related to some enigma in cosmology, 
such as Baryogenesis and DM. Here we will study the discovery potential of this scalar and its hidden sector by using the exotic Z decays 
from future $e^+e^-$ collider.

\subsubsection{Model}
We start with a fermionic DM, $\chi$, interacting with a singlet real scalar $S$. $S$ couples to SM via Higgs 
portal, and DM $\chi$ is stable due to the $U(1)_\chi$ symmetry
~\cite{ Kim:2008pp, Baek:2011aa, Baek:2012uj, Fairbairn:2013uta, Esch:2013rta, Baek:2014jga, Bagherian:2014iia, Freitas:2015hsa}.

The general Lagrangian of the simplified model is written down as follows~\cite{Kim:2008pp},
\begin{align}
\mathcal{L} & = \frac{1}{2} \partial_\mu S \partial^\mu S - \frac{\mu_S^2}{2} S^2 
-\frac{\lambda_3}{6} S^3- \frac{\lambda_4}{24} S^4
-\lambda_{1} \left(H^\dagger H \right) S-\lambda_{2} \left(H^\dagger H \right) S^2 \nonumber \\
& + \bar{\chi} \left(i \slashed{\partial} - m_\chi^0  \right) \chi - y_{\chi} S \bar{\chi} \chi
+ \left| D_\mu H \right| ^2 - \mu_H^2  \left(H^\dagger H \right) - \lambda_H \left(H^\dagger H \right)^2 \,.
\end{align}
We assume $\mu_H^2 <0$ and $\mu_S^2 <0$, which trigger spontaneous symmetry breaking of the SM and hidden sector. 
The tree-level vacuum stability condition requires $\lambda_H >0$, $\lambda_4 >0$; and if $\lambda_2 <0$, 
$|\lambda_2| > \sqrt{\lambda_H \lambda_4/24}$ should be satisfied.
In the broken phase, the Higgs and the singlet scalar obtain their vacuum expectation values (vevs)
$v_H$ and $v_S$, respectively,
\begin{align}
H = \frac{1}{\sqrt{2}} \left(v_H + h \right), \quad S = v_S + s \,.
\end{align}
Accordingly, the DM mass $m_\chi^0$ is shifted to $m_\chi = m_\chi^0 + y_\chi v_S$, which is treated as 
a free parameter here. Adding the extrema condition that $\partial_s V = 0 $ and $\partial_h V =0$, where
$V$ is the scalar potential, we will have the mass matrix of $s$ and $h$, 
\begin{align}
M^2_{11} &  = 2 \lambda_H v_H^2 \nonumber \,, \\
M^2_{12} & = M^2_{21} = \left(\lambda_1 + 2 \lambda_2 v_S  \right) v_H \nonumber \,, \\
M^2_{22} & = -\frac{\lambda_1 v_H^2}{2 v_S} + \frac{\lambda_3 v_S}{2} 
+ \frac{\lambda_4 v_S^2}{3}  \,.
\end{align}
The scalar mass eigenstates $\tilde{h}$ and $\tilde{s}$ are obtained via the following
rotation,
\begin{align}
\left(
\begin{array}{c} \tilde{h} \\ \tilde{s} \end{array}
\right) = \left(
\begin{array}{cc} 
\cos \alpha & -\sin \alpha \\ 
\sin\alpha & \cos\alpha 
\end{array}
\right) \left( 
\begin{array}{c} h \\ s \end{array}
\right) \,,
\end{align}
where 
\begin{align}
\tan (2\alpha) = \frac{2 M^2_{12} }{M^2_{22} - M^2_{11} } \,.
\label{eq:tan2alpha}
\end{align}
The mass of $\tilde{h}$ and $\tilde{s}$ are
\begin{align}
m_{\tilde{h}, \tilde{s}}^2   = \frac{1}{2}
\left( M^2_{11} + M^2_{22} \pm \sqrt{\left(M^2_{11} - M^2_{22} \right)^2 + 4 (M^2_{12})^2} \right) \,.
\end{align}

Let us pause here to count the relevant free parameters for the scalars.
There are nine parameters including $\mu_S$, $\mu_H$, $\lambda_{1,2,3,4}$, $\lambda_H$ and two vevs $v_H$ 
and $v_S$.
The extrema conditions eliminate two of them: $\mu_S$ and $\mu_H$.
By changing to mass eigenstate basis, the five physical observable are 
$m_{\tilde{h}}$, $m_{\tilde{s}}$, $v_H$, $v_S$, and mixing angle $\sin \alpha$, which are 
determined by seven parameters. 
Without losing generality, we set the coefficients $\lambda_1$ and $\lambda_3$ appearing in odd terms of $S$ to be 0,
which can be achieved by adding some additional quantum number or $Z_2$-symmetry for $S$. 
Having observed that the Higgs mass $m_{\tilde{h}} = 125$ GeV and $v_H =  246$ GeV, this leads to three final 
free parameters $m_{\tilde{s}}$, $v_S$ and $\sin\alpha$. 

The decay rates and branching  ratios relevant to the scalar searches are presented below.
In the case that $m_{\tilde{h}} > 2 m_{\tilde{s} }$, the SM Higgs decays to two $\tilde s$ with decay width
\begin{align}
\Gamma(\tilde{h} \to \tilde{s}\tilde{s}) = \frac{\sin^2\alpha \cos^2\alpha}{32 \pi} 
\sqrt{1- \frac{4 m_{\tilde{s}}^2}{m_{\tilde{h}}^2}}
\left( 1+ 2 \frac{ m_{\tilde{s}}^2}{m_{\tilde{h}}^2} \right)^2
\frac{m_{\tilde{h}}^3 \left(\cos\alpha v_H - \sin\alpha v_S \right)^2}{v_H^2 v_S^2} \,.
\end{align}
The singlet scalar $\tilde{s} $ can decay to pair of DM if kinematically allowed. This is the missing
energy signal in the collider. The decay width is 
\begin{align}
\Gamma(\tilde{s} \to \bar{\chi} \chi) = \frac{y_\chi^2 \cos^2\alpha}{8 \pi} 
m_{\tilde{s}} \left(1 - \frac{4 m_\chi^2}{m_{\tilde{s}}^2} \right)^{3/2} \,.
\end{align}

The SM Higgs $\tilde{h}$ can also decay to DM pair, with a similar decay width of $\tilde {s}$ by changing
$\cos^2\alpha$ to $\sin^2 \alpha$ and $m_{\tilde{s}}$ to $m_{\tilde{h}}$,
\begin{align}
\Gamma(\tilde{h} \to \bar{\chi} \chi) = \frac{y_\chi^2 \sin^2\alpha}{8 \pi} 
m_{\tilde{h}} \left(1 - \frac{4 m_\chi^2}{m_{\tilde{h}}^2} \right)^{3/2} \,.
\end{align}
In this model, the invisible decay branching ratio
for $\tilde{s}$ and $\tilde{h}$ are
\begin{align}
BR(\tilde{s} \to \text{inv}) & = \frac{ \Gamma (\tilde{s} \to \bar{\chi} \chi)}
{\Gamma (\tilde{s} \to \bar{\chi} \chi) + \sin^2\alpha \Gamma_{\tilde{h}, tot}^{\text{SM}} (m_{\tilde{s}})  } \,, \\
BR(\tilde{h} \to \text{inv}) & = \frac{ \Gamma (\tilde{h} \to \bar{\chi} \chi) + 
\Gamma (\tilde{h} \to \tilde{s}\tilde{s}) BR^2(\tilde{s} \to \text{inv})}
{\Gamma (\tilde{h} \to \bar{\chi} \chi) + \Gamma (\tilde{h} \to \tilde{s}\tilde{s}) 
+ \cos^2\alpha \Gamma_{\tilde{h}, tot}^{\text{SM}}   }  \,. 
\label{eq:HportalHinvis}
\end{align}

The mass of the singlet scalar relevant for the study of exotic Z decays is $m_{\tilde{s}} \lesssim m_Z$. 
If $m_\chi < \frac{1}{2} m_{\tilde{s}}$, the singlet decays to DM leading to missing energy signals.

\subsubsection{DM relic abundance, indirect and direct searches, and collider constraints}

\noindent $\bullet$ \textit{Relic abundance and indirect detection:}

In this model, the s-channel annihilation $\bar{\chi} \chi \to \bar{f} f$ is the dominant process for 
the thermal DM freeze-out. This process is p-wave suppressed, because the mediator is CP even, while
the initial state is CP odd \cite{Kumar:2013iva}. 
The analytic expression for the cross-section can be written as 
\begin{align}
\sigma v(\bar{\chi}\chi \to \bar{f}f ) = \frac{N_C}{8 \pi} \sin^2\alpha \cos^2\alpha y_\chi^2 
y_f^2 \frac{\left(1 - 4 \frac{m_f^2}{s} \right)^{3/2} (s - 4 m_{\chi}^2)(m_{\tilde{h}}^2 - m_{\tilde{s}}^2)^2}
{\left( (s- m_{\tilde{h}}^2)^2 + m_{\tilde{h}}^2 \Gamma_{\tilde{h}}^2  \right) 
\left( (s- m_{\tilde{s}}^2)^2 + m_{\tilde{s}}^2 \Gamma_{\tilde{s}}^2  \right)} \,,
\end{align}
where $y_f \equiv m_f/v$ and $s$ is the center of mass energy square. 
From this expression, it is clear that the annihilation cross-section is p-wave from the term 
$s - 4 m_\chi^2 \propto v_{rel}^2$.
As a result, DM indirect detection can not put strong limits on this model, since the velocity dispersion of 
the galaxies is relatively slow. However, the temperature during DM freeze-out is relatively high, and 
$v_{rel} \simeq  1/ 3 $. Therefore,  the p-wave suppression is not dramatic during this period.

\begin{figure*}[h]
\centering
 \includegraphics[width=0.48\textwidth]{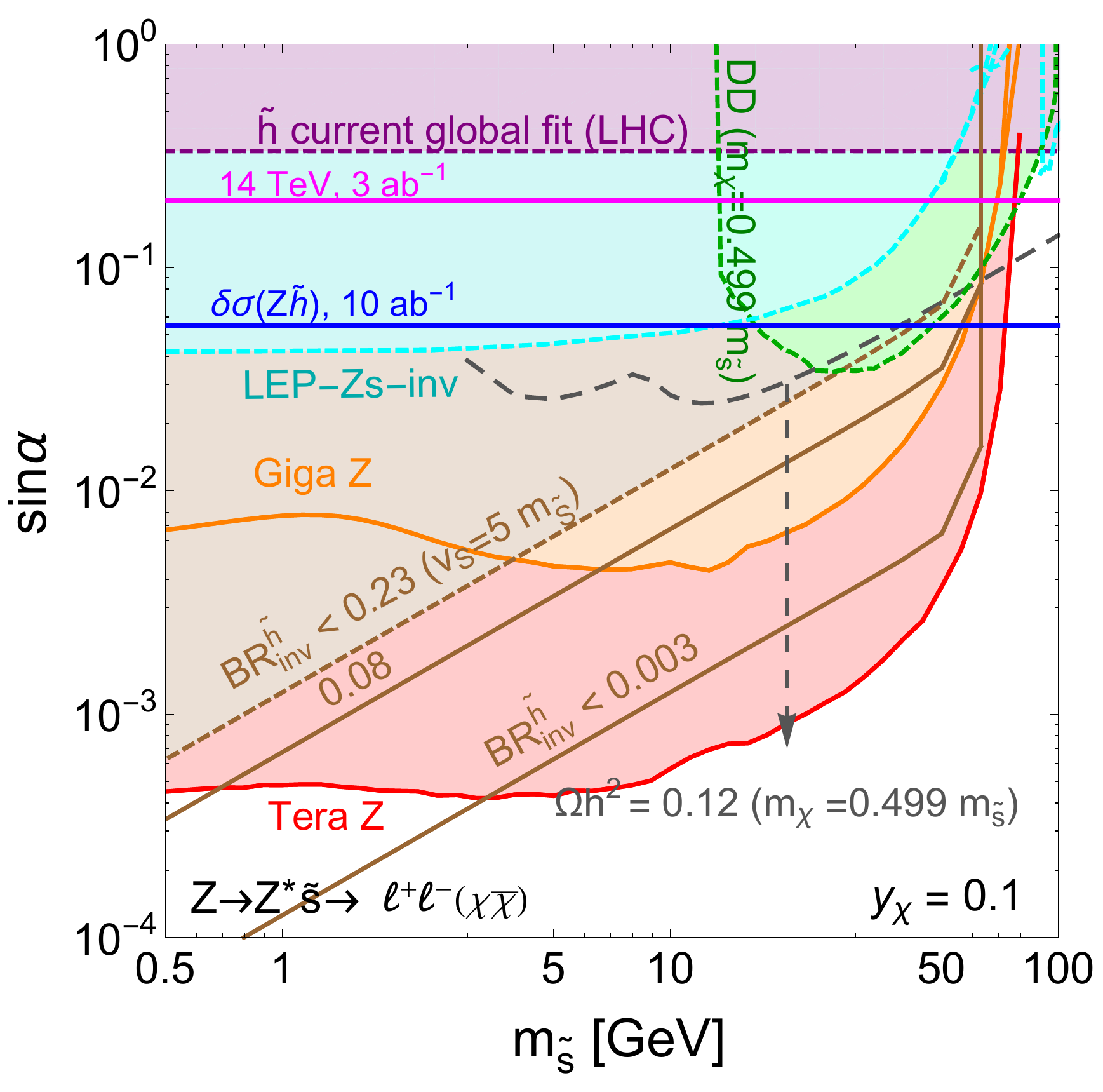} 
  \includegraphics[width=0.48\textwidth]{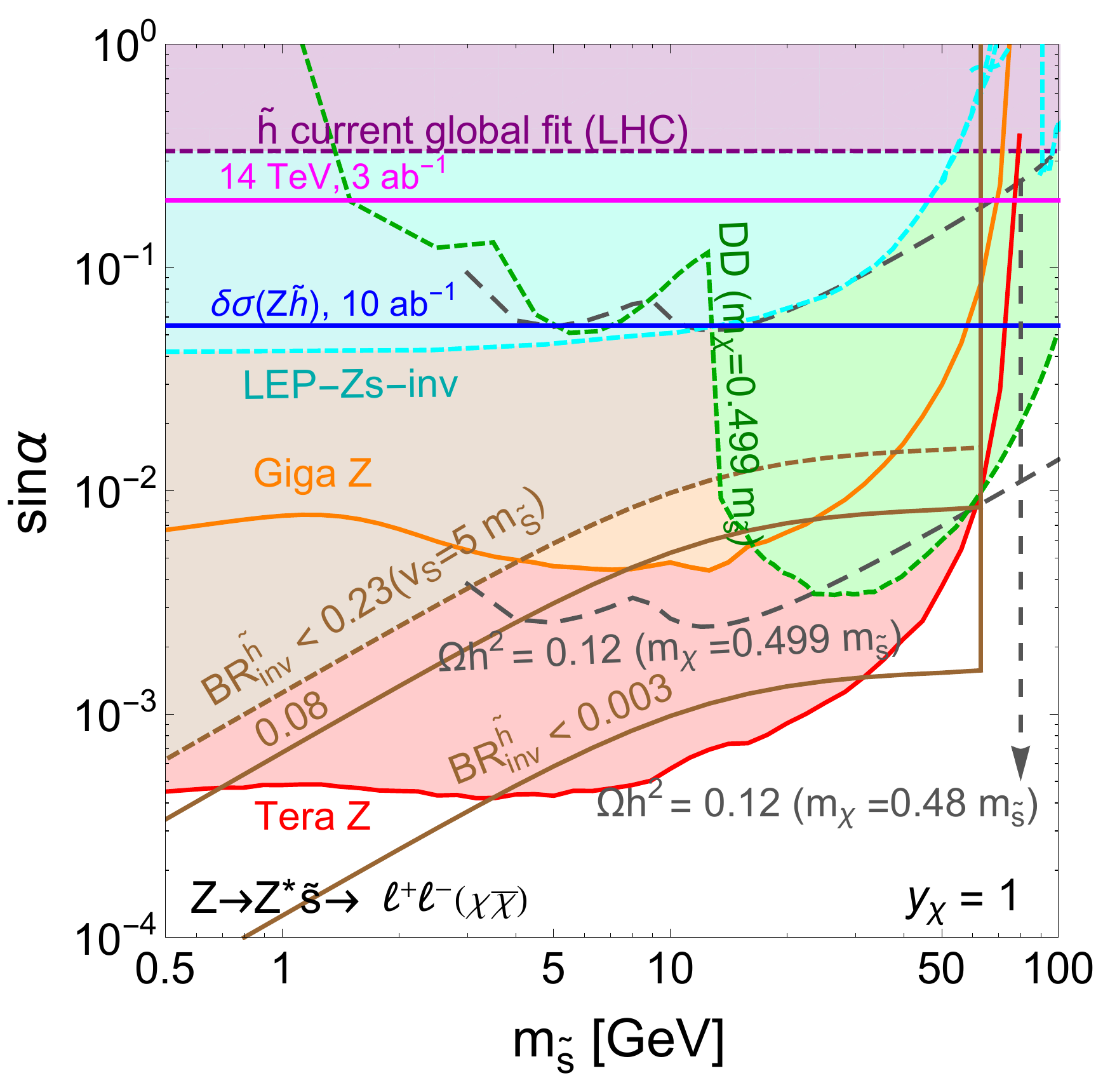} 
 \caption{The $95\%$ C.L. sensitivity for $\sin\alpha$ from exotic Z decay 
$Z  \to \tilde{s} Z^* \to (\bar{\chi}\chi) + \ell^+ \ell^-$ at Giga (Tera) Z-factory, 
with $y_\chi = 0.1 (1)$ in the left (right) panels. We also compare with limits from 
DM direct detection, relic abundance, invisible Higgs BR from the LHC~\cite{Aad:2015pla, Khachatryan:2016whc}
($\text{BR}^{\tilde{h}}_{\text{inv}}<0.23$), the high luminosity ($3 ~ \text{ab}^{-1}$) LHC projection
($\text{BR}^{\tilde{h}}_{\text{inv}} \lesssim 0.08 - 0.16$)~\cite{ATL-PHYS-PUB-2013-014, CMS:2013xfa} and future $e^+e^-$ collider ($\text{BR}^{\tilde{h}}_{\text{inv}} \lesssim 0.003$) \cite{CEPC-SPPCStudyGroup:2015csa, Fujii:2017vwa} 
, current and future Higgs global fit from ($h$ current global fit) \cite{Khachatryan:2016vau, Dawson:2013bba} 
with purple and magenta lines, low mass Higgs searches in invisible channels (LEP-$Zs$-inv)
\cite{Acciarri:1996um, Acciarri:1997tr, Abreu:1999vu, Searches:2001ab}, and precision 
measurement of $\sigma(Zh)$  ($\delta \sigma(Zh)$) \cite{Gomez-Ceballos:2013zzn, 
CEPC-SPPCStudyGroup:2015csa, Ruan:2014xxa}. The dashed (solid) lines are for existing 
constraints (future prospects). }
  \label{fig:SingletSinAlpha}
\end{figure*}

In the DM relic abundance calculation, we consider the fermions in the final states if the annihilations are 
kinetically allowed. In the mass range of $ m_b <  m_\chi <  m_Z/2$, the final states of quarks~(b and c) and
the $\tau$ lepton are included. The computation of the thermal relic density is restricted to $m_\chi > 1.5~\GeV$. For the smaller 
DM mass, QCD non-perturbative effects and some hadronic channels should be considered. 
To avoid other  limits, we choose $m_\chi$ close to $m_{\tilde{s}}/2$
in \cref{fig:SingletSinAlpha}.
For non-resonance case, relic abundance does not lead to competitive limits.

\noindent $\bullet$ \textit{Direct detection:}

The DM $\chi$ scattering with nuclei is mediated by t-channel scalar $\tilde{s}$
and $\tilde{h}$, which give the possibility to detect DM via spin-independent direct detection. 
The spin independent scattering cross-section with nucleon is \cite{Kainulainen:2015sva},
\begin{align}
\sigma_{SI} = \frac{\mu_n^2  f_n^2 m_n^2}{\pi v_H^2}  
g_\chi^2 \sin^2\alpha \cos^2\alpha 
\left(\frac{1}{m_{\tilde{h}}^2 - m^2_{\tilde{s}}} \right)^2 \,,
\end{align}
where $\mu_n$ is the reduced mass between DM and nucleon, $f_n \approx 0.3$ is the
Higgs-nucleon coupling, and $m_n$ is the nucleon mass. We compare $\sigma_{SI}$
with the limits from XENON1T~\cite{Aprile:2017iyp}, LUX~\cite{Akerib:2016vxi},
PANDAX-II~\cite{Tan:2016zwf}, and CRESST-II~\cite{Angloher:2015ewa} as
well as CDMSlite~\cite{Agnese:2015nto} for low mass DM, and show the constraints
in \cref{fig:SingletSinAlpha}. The limits drop around $m_{\chi} \sim 10$ GeV,
because below this mass Xenon scintillators looses its sensitivity and CDMSlite becomes the dominant one.

\noindent $\bullet$ \textit{Existing collider constraints:}

The current LHC limits from the Run I combination of ATLAS and CMS data constrains 
BR$(h \to \text{ inv}) \leq 0.23$ at $95\%$ C.L.~\cite{Aad:2015pla, Khachatryan:2016whc}. 
Following the $\tilde{h}$ invisible decay branching ratio in \cref{eq:HportalHinvis},
the limits on mixing angle $\sin\alpha$ are given in \cref{fig:SingletSinAlpha},
labeled as ``$\text{BR}^{\tilde{h}}_{\text{inv}}<0.23$". We also add the HL-LHC 
($3 ~\text{ab}^{-1}$) and future $e^+e^-$ collider projections on invisible Higgs search, which leads to $95\%$ C.L. limits 
$\text{BR}_{\text{inv}}^{\tilde h} \lesssim 0.08 - 0.16$~\cite{ATL-PHYS-PUB-2013-014, CMS:2013xfa} 
and $\text{BR}^{\tilde{h}}_{\text{inv}} \lesssim 0.003$~\cite{CEPC-SPPCStudyGroup:2015csa, Fujii:2017vwa}. 
Moreover, the global fit to Higgs data at  the LHC 7 TeV and 8 TeV runs can 
constrain the single scaling factor to Higgs interactions,  and this gives 
$\sin\alpha < 0.33 $~\cite{Khachatryan:2016vau} which is also added in
\cref{fig:SingletSinAlpha}, labeled as ``$\tilde{h}$ current global fit (LHC)". 
The HL-LHC can extend the reach to $\sin\alpha < 0.28 ~ (0.20) $ using
$300~ \text{fb}^{-1}$ $(3~ \text{ab}^{-1})$ luminosity~\cite{Dawson:2013bba}.

At LEP-II, a low mass Higgs has been searched in 
$e^+ e^- \to Z \to  Z^* h$ channel, where $Z$ decays visibly and $h$ decays invisibly,
with integrated luminosity of $\sim 114 ~\text{pb}^{-1}$~\cite{Acciarri:1996um}. 
The Higgs bremsstrahlung process $Z h$ is also used at higher $\sqrt{s}$ to set
limit on heavier Higgs up to $114.4$ GeV~\cite{Acciarri:1997tr, Abreu:1999vu, Searches:2001ab}. 
The searches can put constraint on $\sin\alpha$ for the similar process $Z \tilde{s}$,
which we give in \cref{fig:SingletSinAlpha} and labeled
as ``LEP-$Zs$-inv". For the on-shell production of $Z \tilde{s}$ at FCC-ee, the 
sensitivity on $\sin\alpha$ has been estimated to be $\sim 0.03$ for 
$m_{\tilde{s}} < 100$~GeV \cite{Liu:2017lpo}. The precision measurement
of the Higgs bremsstrahlung cross-section $\sigma(Zh)$ can reach the
accuracy of $\mathcal{O}(0.3\%-0.7\%)$ expected from 
$5 - 10 $~ab$^{-1}$ ~\cite{Gomez-Ceballos:2013zzn, 
CEPC-SPPCStudyGroup:2015csa, Ruan:2014xxa}, which can probe the scalar 
mixing  down to $0.055 -0.084$ \cite{Liu:2017lpo}, labeled
as ``$\delta \sigma(Zh)$".

\begin{figure*}[h]
\centering
 \includegraphics[width=0.4\textwidth]{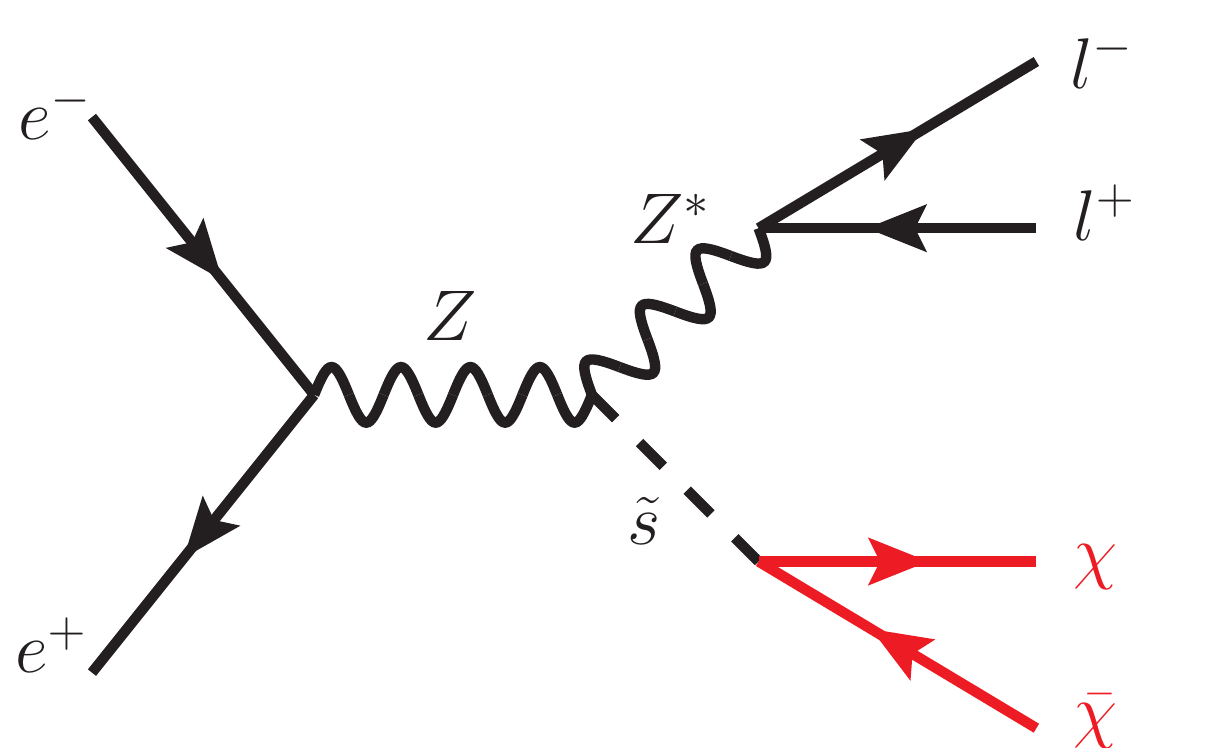}
 \caption{The Feynman diagram for exotic Z decay 
 $Z  \to \tilde{s} Z^* \to (\bar{\chi}\chi) + \ell^+ \ell^-$. Note the
 Z is produced on shell and followed by a three-body decay $\tilde{s} \ell^+ \ell^-$,
 and the parentheses for $\bar{\chi} \chi$ indicates they are from the decay of a resonance .}
  \label{fig:SingletSandDiagrams}
\end{figure*}

\subsubsection{Prospects from exotic Z decay}
\noindent $\bullet$ \textit{Exotic Z decay sensitivity:}

For the sensitivity at a Giga (Tera) Z-factory, we study the process $Z \to \tilde{s} Z^* \to 
(\bar{\chi} \chi) + \ell^+ \ell^-$, with Feynman diagram in \cref{fig:SingletSandDiagrams}, 
where $\tilde{s}$ decays to DM particles and  off-shell $Z^*$ goes to lepton pairs. 
We set constraints on $\sin\alpha$ using this process and plot them  in \cref{fig:SingletSinAlpha}. 
The previous LEP experiment~\cite{Acciarri:1996um} has searched the similar channel 
with $Z^*$ decay to both hadronic and leptonic channels. 
The details of the simulations and cuts are given in \cref{sec:ZtoMET2l}, where the limit on
the exotic decay BR has been calculated. After calculating the exotic decay
BR, one can translate the constraints of decay BR to physical variable $\sin\alpha$.
We have compare our analysis with LEP and found good agreement.
To be more specific, given ``LEP-Zs-inv" has also worked on Z pole with an 
integrated luminosity $114 ~\text{pb}^{-1}$, we normalize our result to the 
same luminosity and find the constraint is similar to the LEP.

In the SM, Higgs can decay to diphoton or $Z\gamma$ via top loop and W loop.
Due to the mixing between $\tilde{s}$ and $\tilde{h}$, the mono-photon process 
$Z\to \gamma \tilde{s} \to \gamma (\bar{\chi}\chi)$ is possible. We have checked
this process following the cuts in \cref{sec:monophoton} 
and found its constraint on $\sin\alpha$ is about
one order of magnitude weaker than $Z  \to \tilde{s} Z^* \to (\bar{\chi}\chi) + \ell^+ \ell^-$. 
The main reason is mono-photon decay is loop suppressed. Furthermore, 
mono-photon background is higher than $\ell^+\ell^- + \slashed{E}$ background.
Therefore, we do not put the constraint from mono-photon in \cref{fig:SingletSinAlpha}.

\noindent $\bullet$ \textit{Summary:}

From \cref{fig:SingletSinAlpha}, we see the relic abundance provides constraints on
$\sin \alpha$ only in the fine-tuned scenario with $2 m_\chi \sim m_{\tilde{s}}$. 
The indirect detection does not provide limits because it is
p-wave suppressed. The direct detection provides a useful constraint, which is not
sensitive to the resonant mass of $m_{\tilde{s}} \sim 2 m_\chi$. At the same time, it depends
on the size of the Yukawa coupling $y_\chi$. The existing and future Higgs global fit from the LHC
does not provide competitive limits in comparison with precision measurement of $\sigma(Zh)$,
while invisible decay BR of SM Higgs provides a pretty good limit down to 
$\sin\alpha \sim \mathcal{O}(10^{-2} - 10^{-3})$ via  the existing LHC data. 
At the HL-LHC ($3 ~\text{ab}^{-1}$), the reach of invisible BR is 
about $0.08 - 0.16$~\cite{ATL-PHYS-PUB-2013-014, CMS:2013xfa}, which provides
only a moderate improvement of the limit. 
The future sensitivity of $\text{BR}_{\text{inv}}^h$ is expected to reach 
$\sim 0.003$ at future $e^+e^-$ collider 
\cite{Gomez-Ceballos:2013zzn, CEPC-SPPCStudyGroup:2015csa, Fujii:2017vwa}, which can improve the limits by a
factor of $\sim 8.7$.

The proposed exotic Z decay $Z \to \tilde{s} Z^* \to (\text{inv}) + \ell^+ \ell^-$ can cover
$\sin \alpha$ down to $\sim 10^{-2}~(10^{-3})$ for Giga Z (Tera Z), 
and such constraints do not rely much on value of $y_\chi$ and $\chi$ mass. 
The constraints from exotic Z decay are superior than most of the existing and 
future searches, and only invisible SM Higgs decay search at the future Higgs-factories can provide
competitive limits.

\subsection{Vector portal DM} 
\label{sec:VportalDM}

Vector portal, as another simple extension of the SM physics, employs a massive $U(1)$ dark photon 
connecting the SM sector and the hidden sector~\cite{Okun:1982xi,  Holdom:1985ag, Pospelov:2007mp, 
ArkaniHamed:2008qp, ArkaniHamed:2008qn, Pospelov:2008zw}.
The searches for vector portal DM and the vector field itself attract world-wide effort 
(see review \cite{Essig:2013lka,Alexander:2016aln,Battaglieri:2017aum} and references therein). 
Various experiments, such as  fixed target,   
$e^+e^-$ and $pp$ colliders, are aiming to find such dark photon, especially utilizing its coupling to $\ell^+ \ell^-$. 
Aside from decaying to SM fermions, the invisible decays of the dark photon are directly 
related to DM, which can be searched by radiative return process, meson decay and missing energy events in scattering processes \cite{Essig:2013lka, Alexander:2016aln,Battaglieri:2017aum}. 

The dark photon $A'$, as a $U(1)$ gauge field in the hidden sector, can mix with SM 
hypercharge $U(1)_Y$ field $B_\mu$ through a renormalizable operator, 
\begin{align}
\mathcal{L} & = - \frac{1}{4} B_{\mu \nu} B^{\mu \nu} 
- \frac{1}{4} {A'}_{\mu \nu} {A'}^{\mu \nu} 
+ \frac{\epsilon}{2 c_W} B_{\mu \nu} {A'}^{\mu \nu}  + \frac{1}{2}m_{A'}^2 {A'}^\mu {A'}_\mu,
\label{eqn:DDP}
\end{align}
where $\epsilon$ is the kinetic mixing parameter and $c_W$ is the cosine of the weak angle.
The mass of the dark photon, $m_{A'}$ can be obtained from Higgs mechanism in the dark sector.
Interestingly, this underlying mechanism is related to our previous Higgs portal DM.   We ignore here (possibly interesting) dynamics of the dark Higgs~\footnote{The mass of $A'$ usually needs Higgs mechanism to break $U(1)_D$ and obtain a vev, therefore it requires a complex 
scalar $\phi$ charged under $U(1)_D$. It naturally provides exotic Z decay signature $Z \to A' \phi$ 
from $Z$-$A'$ mixing.} .
We can always rotate away the kinetic mixing terms and work in the mass
eigenstate basis. The rotation is non-unitary and is written down up 
to $\mathcal{O}(\epsilon^2)$~\cite{Cassel:2009pu},
\begin{align}
& \left( \begin{array}{c} 
Z_{\mu} \\ 
A_{\mu} \\
A'_\mu  
\end{array} \right) =  
 \left( \begin{array}{ccc}
1 
& 0 & \dfrac{  m_{A'}^2 t_W}{ - m_{A'}^2 + m_{Z}^2 } \epsilon \\
0 & 1 & \epsilon \\
\dfrac{  m_{Z }^2 t_W}{ m_{A'}^2 - m_{Z}^2} \epsilon  & 0 & 
1  
\end{array} \right) \left( \begin{array}{c}
\tilde{Z}_{\mu} \\
\tilde{A}_{\mu} \\
\tilde{A'}_{\mu} 
\end{array} \right)  \,,
\label{eqn:Amatrix}
\end{align}
where $t_W$ is the tangent of the weak angle. 
This formula does not apply to the region that $A'$ mass is pretty close to the mass of Z boson.
In the rest of the paper, we work on the mass eigenstates of these gauge fields; without ambiguities, 
$\tilde{A}$ and $\tilde{A'}$ are used to represent the mass eigenstates. After this rotation, the way that the currents
couples to gauge fields are changed, and
the interactions between vectors and currents up to $\mathcal{O}(\epsilon^2)$ are written as follows,
\begin{align}
\mathcal{L}_{\text{int}} 
&= \tilde{Z}_\mu \left( g J_Z^\mu - 
g_D \dfrac{m_{Z}^2 t_W}{m_{Z}^2 - m_{A'}^2} \epsilon J_D^\mu 
 \right) 
+ \tilde{A'}_\mu \left( g_D J_D^\mu + 
g \dfrac{m_{A'}^2 t_W}{m_{Z}^2 - m_{A'}^2} \epsilon J_Z^\mu 
+ e \epsilon J_{\text{em}}^\mu  \right)
+ \tilde{A}_\mu e J_{\text{em}}^\mu \,,
\label{eqn:currents}
\end{align}
The massless photon $\tilde{A}$ couples to the electromagnetic current $J_{\rm{em}}$.
The dark photon couples to dark $U(1)$ currents $J_D$; after the field rotating, 
a $\epsilon$ suppressed coupling to $J_{\rm{em}}$ and $J_Z$ arises. The $\tilde{Z}$ boson couples to $J_Z$, and
has the coupling to the dark currents with $\epsilon$ suppression.

\subsubsection{scalar vector-portal DM}
\label{sec:VportalSdm}

\noindent $\bullet$ \textit{Model:}

In this model, we introduce a complex scalar as DM, charged under the $U(1)_D$, and this scalar DM interacts with 
the SM particles via the dark photon $A'$. The relevant interactions can be written as follows,  
\begin{align}
\mathcal{L}_S & = (\partial_\mu S + i g_D {A'}_\mu S)^* 
(\partial^\mu S + i g_D {A'}^\mu S) - m_S^2 S^* S  \,.
\label{eq:lAS}
\end{align}
For $m_S^2 > 0$ and considering $Z_2$ symmetry, $\langle S \rangle = 0$ and $S$ is stable.
From eq.~(\ref{eqn:Amatrix},\ref{eqn:currents}), it is clear that there is coupling between $Z$, $A'$ and $S$,
\begin{align}
\mathcal{L}_S \supset g_D^2 S^*S \left( 
\tilde{A'}_\mu +   \epsilon
\frac{  m_Z^2 t_W }{(m_{A'}^2- m_{Z}^2 )} \tilde{Z}_\mu
\right)^2 \,,
\end{align}
which can provide interesting signal for the exotic Z decay, $\tilde{Z} \to \tilde{A'} S^* S$ from 
the leading $\epsilon$ terms in the Lagrangian. To have this signal, we must have this process kinematically allowed, 
$m_{\tilde{A}'} + 2 m_{S} < m_Z$.  
We will focus on the region that $m_S > \frac{1}{2} m_{\tilde{A}'}$, such that the $\tilde{A}'$ decay dominantly
to SM particles, rather than invisible DM pair.\footnote{This assumption can be relaxed, and the constraints
should be rescaled accordingly to the branching ratio of $\tilde{A}'$ to SM particles.}

The spontaneous symmetry breaking through dark Higgs $\phi$ is a simple mechanism to give mass to $S$ and $A'$. 
The difference from $S$ is that 
there is no exact $Z_2$ symmetry to make $\phi$ stable, but the Lagrangian is similar to \cref{eq:lAS}, 
\begin{align}
\mathcal{L}_\phi  = (\partial_\mu \phi + i g_D {A'}_\mu \phi)^* 
(\partial^\mu \phi + i g_D {A'}^\mu \phi)  
+ \lambda_2 S^* S \Phi^* \Phi  -\mu_\phi^2 |\phi|^2 - \frac{\lambda_4}{4} |\phi|^4
\label{eq:lAphi}
\end{align}
After symmetry breaking $\langle \phi \rangle \neq 0$, $A'$ and $S$ get their mass. 
When $\phi$ is much heavier than $S$ and $A'$, it can be integrated out, and \cref{eq:lAS} is enough to describe the 
process related to DM and various searches. When $\phi$ mass is smaller or comparable to the mass of $S$ and $A'$,  
$\phi$ need to be considered. In this case, $\phi$ can be produced at collider and decay back to 
$2 S$ or $2 \tilde{A}'$. 

\noindent $\bullet$ \textit{DM relic abundance and indirect detection:}

\begin{figure*}[h]
\centering
 \includegraphics[width=0.48\textwidth]{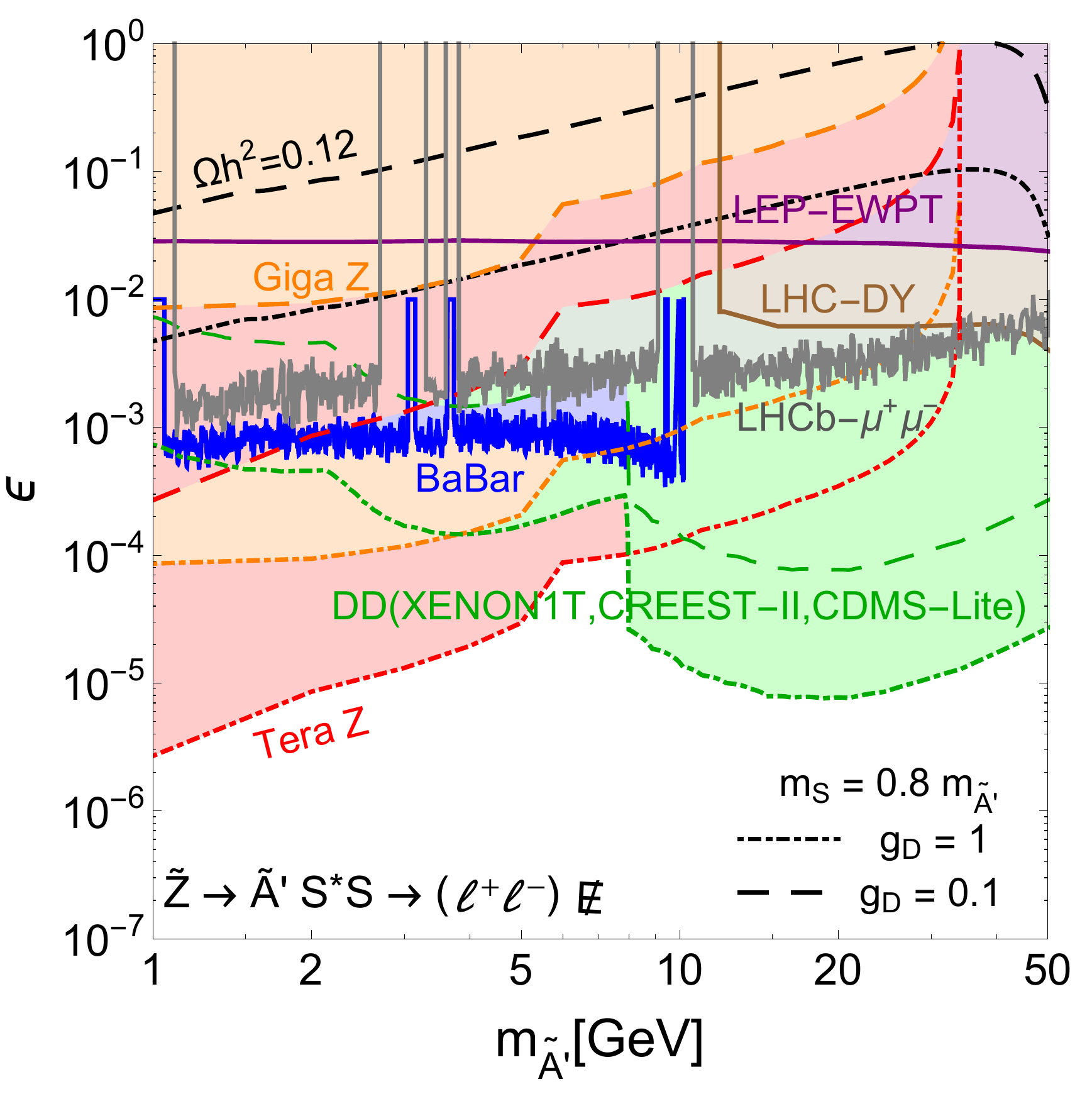} 
 \includegraphics[width=0.48\textwidth]{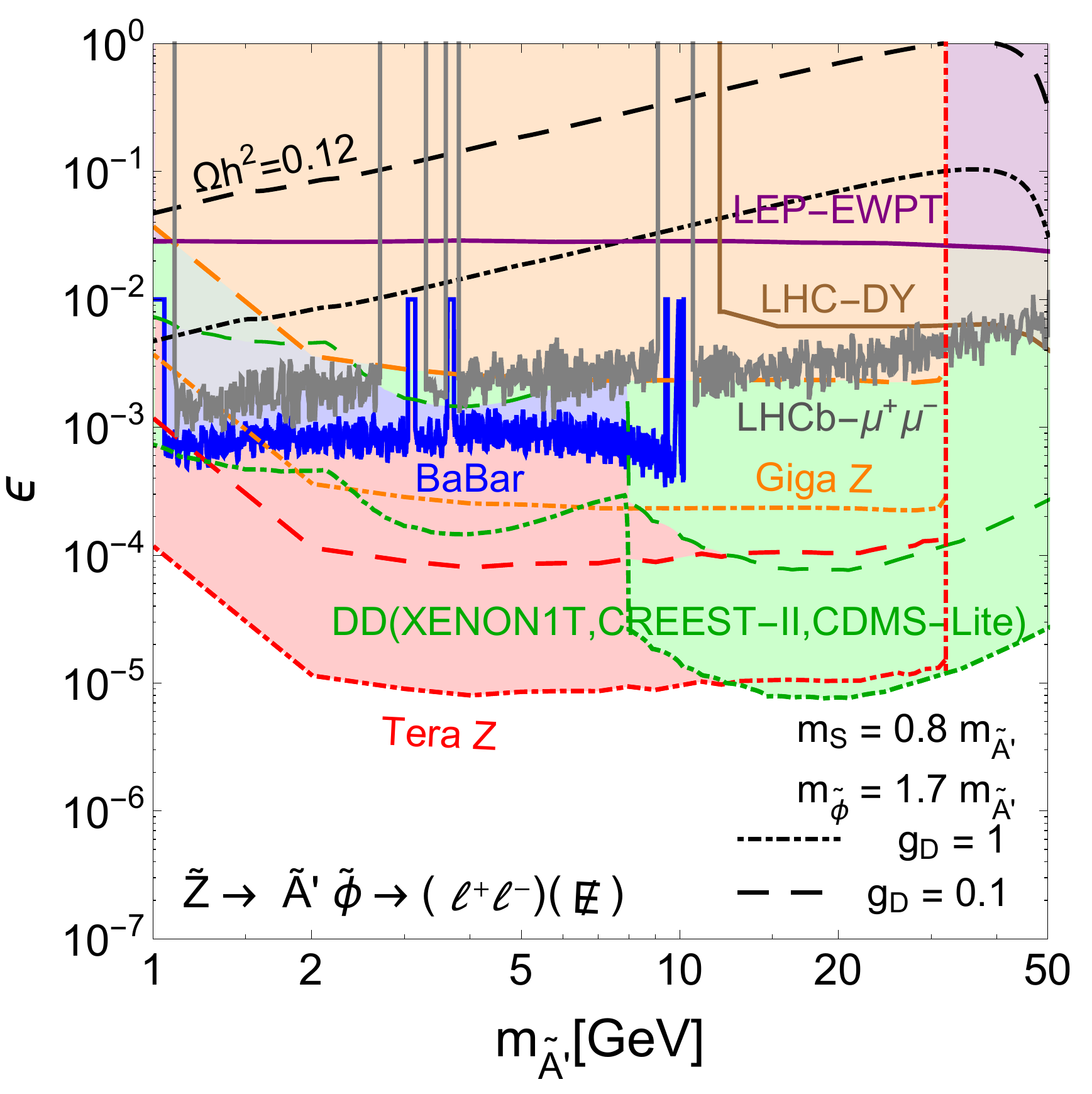} 
 \caption{The $95\%$ C.L. sensitivity for $\epsilon$ as a function of
  $m_{\tilde{A}'}$ from exotic Z decay $\tilde{Z}   \to (\ell^+ \ell^-)\slashed{E}$.
  The 3-body decay channel
  $\tilde{Z} \to \tilde{A}' S^* S\to (\ell^+ \ell^-)\slashed{E} $ is shown in the left panel, while 
  the 2-body cascade decay channel $\tilde{Z} \to \tilde{A}' \tilde{\phi}
  \to (\ell^+ \ell^-)(\slashed{E}) $ is shown in the right panel. 
 We take $g_D = 0.1$ and $1$ , $m_S = 0.8 m_{\tilde{K}}$. The constraints from
 exotic Z decay are labeled as Giga (Tera) Z, and also we show an illustrative
 line for LEP luminosity $114 ~\text{pb}^{-1}$. We also show limits from relic abundance,
 direct detection and existing collider searches for comparison. }
  \label{fig:scalarDMVportal}
\end{figure*}

If $m_S > m_{\tilde{A}'}$, the dominant process controlling the freeze-out is $S S^* \to \tilde{A}' \tilde{A}'$. 
The thermal cross-section is 
\begin{equation}   
  \sigma v ( S S^* \to \tilde{A}' \tilde{A}' )  =  \frac{g_D^4}{16 \pi m_S^2} 
  \frac{(8-8 y^2 + 3 y^4)  \sqrt{1-y^2}}{(2-y^2)^2} ,
\end{equation}   
where $y \equiv m_{\tilde{A}^\prime}/m_S$. By taking $s \to 4 m_S^2$, the leading term tell us that this process
is s-wave.
This thermal cross-section is not related to $\epsilon$, since the $\tilde{A}'$ are produced on-shell.
On the other hand, in the regime that $m_S < m_{\tilde{A}'}$, the dominant process is 
$S^* S \to  \tilde{A}' / Z \to  \bar{f} f$ via the off-shell Z and $\tilde{A}'$, and 
the thermal cross-sections for $ \sigma v ( S S^* \to f \bar{f} ) $ are given in
\cref{sec:appendix}. Since the thermal cross-section is proportional to $\epsilon^2$,
the relic abundance will rely on the size of the kinetic mixing. This can set the target for the search of exotic Z
decay. Without loss of generality, we will restrict to $m_S = 0.8 m_{\tilde{A}'}$ in the parameter space, 
to compare various limits from the complementary experiments, shown in \cref{fig:scalarDMVportal}. 

$S^* S \to  \tilde{A}' / Z \to  \bar{f} f$ is p-wave suppressed, which can be understood from the 
CP-symmetry of initial state \cite{Kumar:2013iva}. As we discussed before, the p-wave annihilation have the 
suppressed signal of the indirect detection. Therefore,  the corresponding limit is negligible.

\noindent $\bullet$ \textit{Direct detection:}

The scattering of $S$ off nuclei is mediated by t-channel $\tilde{A}'$ and $\tilde{Z}$. 
Interestingly, the contribution from $\tilde{Z}$ exchange has been canceled by the one from $\tilde{A}'$ 
coupling to $J_{Z}$ current~\cite{Liu:2017lpo}, hence only $\tilde{A}'$ coupling to $J_{\text{em}}$ current should be 
considered, which can be seen directly from \cref{eqn:currents}. Therefore,
the spin-independent scattering cross-section for $S$ and the nucleon has a simple expression and is given below,
\begin{align}
\sigma_n^{\text{SI}} \simeq \frac{e^2 g_D^2 
\epsilon^2\mu_{Sn}^2}{2\pi m_{\tilde{A}'}^4},
\end{align}
where $\mu_{Sn}=m_S m_n/(m_S+m_n)$ is the reduced mass of dark matter $S$ and 
nucleon $n$, and $e$ is the electron charge. 
We add the direct detection constraints as green shade area 
in \cref{fig:scalarDMVportal}.

\noindent $\bullet$ \textit{Existing collider limits:}

Focusing on the region of $m_{\tilde{A'}} < 2 m_S$, the decay mode of 
the dark photon, $\tilde{A}' \to \ell^+ \ell^-$, is the key channel to look for in the experiments:
beam-dump, fixed target, collider, and rare meson decay. 
In \cref{fig:scalarDMVportal}, we present the 
constraints from the experiments having the leading limits currently. There are also limits from LEP via 
electroweak precision observables \cite{Hook:2010tw}. 
For constraints from the LHC, the inclusive Drell-Yan process
$pp \to \tilde{A}' \to \ell^+ \ell^-$ can be used to constrain $\epsilon$
with the LHC 8 TeV data~\cite{Aad:2014cka, Khachatryan:2014fba}, 
which provides a stronger bound than the electroweak precision 
bounds \cite{Cline:2014dwa, Hoenig:2014dsa,Curtin:2014cca}.
For low mass $m_{\tilde{A}'} \sim \mathcal{O}(\text{GeV})$, the
limits from B-factory is the leading one from measuring visible decay products of the dark photon,
such as BaBar 2014 \cite{Lees:2014xha} having the limits of $\epsilon \lesssim 10^{-3}$.
Recently, the LHCb~\cite{Aaij:2017rft} performed dark photon search using the inclusive di-muon data. 
This will give the leading constraints in the mass window of ($10 ~ \text{GeV}$, $ 50 ~\text{GeV}$).

\noindent $\bullet$ \textit{Exotic Z decay search:}

\begin{figure*}[h]
\centering
 \includegraphics[width=0.7\textwidth]{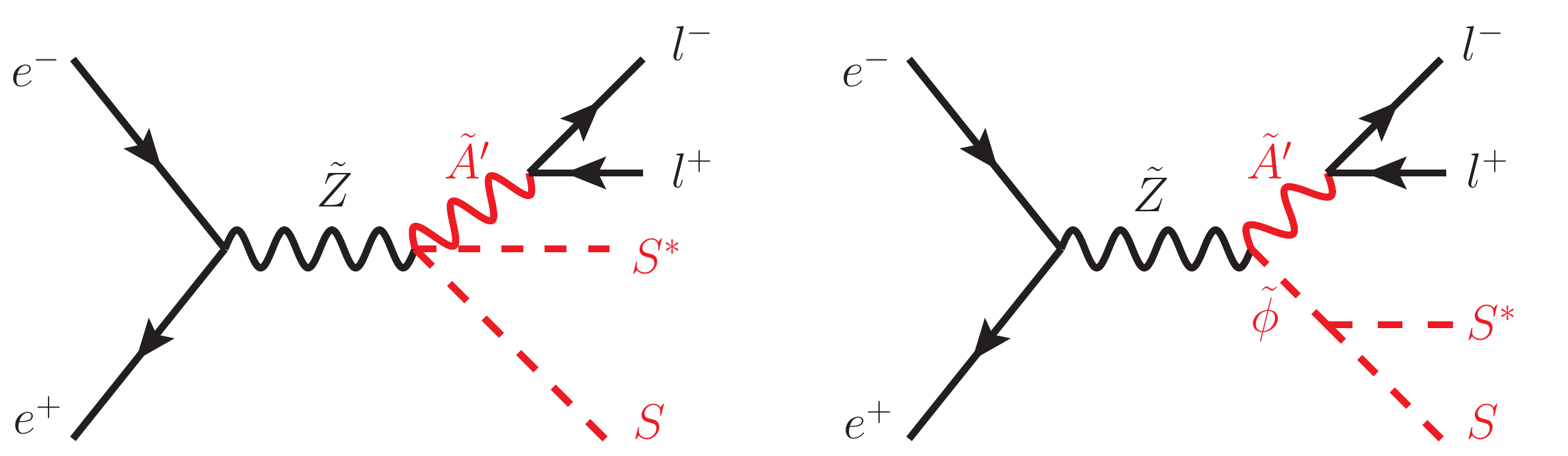}
 \caption{The Feynman diagrams for the 3-body
 decay process $\tilde{Z} \to \tilde{A}' S S^* \to (\ell^-\ell^+) \slashed{E}$ 
 from vector portal model with scalar DM and the Higgs bremsstrahlung process
 $\tilde{Z} \to \tilde{A}' \tilde{\phi} \to (\ell^-\ell^+) (\slashed{E}) $.}
  \label{fig:VportalScalarDM}
\end{figure*}

The first process we consider is the three-body decay
$\tilde{Z} \to \tilde{A'} S^* S \to (\ell^+ \ell^-) \slashed{E}$
shown in the left panel of \cref{fig:VportalScalarDM}. The limit on 
exotic Z decay branching ratio is given in \cref{sec:ZtoMET2l}.
Here we take the mass range of $\tilde{A'}$, $m_S < m_{\tilde{A'}} < 2 m_S$, such that $\tilde{A'}$ 
will not dominantly 
decay to invisible DMs, and DM relic density depends on the kinetic mixing $\epsilon$.  
To constrain kinetic mixing coupling $\epsilon$ at given $m_{\tilde{A}'}$, we fix coupling $g_D$, 
the mass ratio $m_S /m_{\tilde{A}'} $. 
The corresponding limit for $\epsilon$ as a function of
$m_{\tilde{A}'}$ is given in \cref{fig:scalarDMVportal}. The range of $m_{\tilde{A}'}$
starts from $1$ GeV. For smaller masses, other constraints like beam
dump experiments become quite strong. Moreover, the exotic Z search 
begins to lose its efficiency due to the small separation of lepton pair from $\tilde{A}'$
decay.

In addition to the three-body decay topology, we can also have the 2-body 
cascade decay $\tilde{Z} \to \tilde{A}' \tilde{\phi} \to (\ell^+ \ell^-)
(S^* S)$, shown in the right panel of \cref{fig:VportalScalarDM}. This channel has resonance 
in both lepton pair invariant mass and invisible mass. We still consider the regime
$m_S < m_{\tilde{A}'} < 2 m_S$. Therefore, $\tilde{A}'$ decays into lepton pairs. 
We assume $m_{\tilde{\phi}} < 2 m_{\tilde{A}'}$ and very small Higgs portal mixing,
so that $\tilde{\phi}$ decay to SM particles via Higgs mixing can be neglected. 
Therefore,  the decay branching ratio of $\tilde{\phi} \to S^* S$ is $\sim 100\%$. 
In the right panel of \cref{fig:scalarDMVportal}, we constrain $\epsilon$ as a
function of $m_{\tilde{A}'}$, with $m_S = 0.8 m_{\tilde{A}'}$ and $m_{\tilde{\phi}}
= 1.7 m_{\tilde{A'}}$. Given that $\tilde{\phi}$ has negligible coupling to SM sector,
the relic abundance, indirect detection and direct detection are similar
to the left panel of \cref{fig:scalarDMVportal}.

\noindent $\bullet$ \textit{Summary:}
As shown in \cref{fig:scalarDMVportal}, LEP electroweak precision test, 
LHC Drell-Yan, Babar radiative return and LHCb di-muon inclusive searches can provide the 
direct constraints on $\epsilon$.
For $m_{\tilde{A}'} < 10$ GeV,  Babar bounds
$\epsilon \lesssim 10^{-3}$, while LHC Drell-Yan and LHCb provide
complementary limits $\epsilon \gtrsim 5 \times 10^{-3}$ for 
$m_{\tilde{A}'} > 10$ GeV. LEP electroweak precision test
is the weakest constraint among the three.

The hint from the DM relic abundance and the constraints from direct detection and
exotic Z decay rely on coupling $g_D$. For a fixed $m_{\tilde{A}'}$,
DM annihilation cross-section and direct detection scattering
cross-section are proportional to $g_D^2$.
The coupling for the four point vertex $\tilde{Z}_\mu \tilde{A}'^\mu 
S^* S$ is proportional to $\epsilon g_D^2$, while the coupling
for three point vertex $\tilde{Z}_\mu \tilde{A}'^\mu \tilde{\phi}$
is proportional to $\epsilon g_D m_{\tilde{A}'}$. Therefore,
the 3-body decay width is proportional to $g_D^4$,
while the 2-body cascade decay width is proportional to $g_D^2$.
For $g_D =1$, we see Tera Z could provide the strongest bounds
at low $m_{\tilde{A}'}$, while direct detection provides
comparable limits to exotic Z decay at high $m_{\tilde{A}'}$.

In comparison with the 3-body cascade decay, one might expect better constraint 
from 2-body cascade decay
because there are resonances in both lepton pair and missing energy 
in this topology, while 3-body decay only has one resonance in
lepton pair. This intuition is indeed correct as the sensitivity on exotic
decay BR is better for 2-body cascade decay than 3-body in \cref{sec:ZtoMET2l},
but the difference is not significant.
The limits on $\epsilon$ in \cref{fig:scalarDMVportal} 
involve more parameters and couplings, which modify the dependence
of $m_{\tilde{A}'}$. 

For $m_{\tilde{A}'} \sim 1$ GeV, 3-body decay loses less efficiency from 
lepton separation requirement than 2-body cascade decay, since the energy of $\tilde{A}'$
in 3-body decay is generally softer than in 2-body cascade decay.

In summary, the exotic Z decay search in both topologies can provide 
good reach in $\epsilon$, which is complementary and competitive
to other constraints.

\subsubsection{(Inelastic) vector portal fermionic DM}
\label{sec:inelasticFdm}

For vector portal fermionic DM, we consider inelastic DM model here. The constrains
and future collider search of Z decays are similar to magnetic inelastic dark matter, 
which will be explored in \cref{sec:MIDMandRayDM}.

Starting from the fermionic DM charged under dark sector $U(1)_D$, we can write down its
Dirac mass term $m_D \bar{\chi} \chi$, and its Majorana mass is obtained through  Yukawa interaction with a scalar $\Phi$.
The Lagrangian is
\begin{align}
\mathcal{L}_{F} =  \bar{\chi} i\slashed{\partial} \chi
+ g_D \bar{\chi} A'_\mu \gamma^\mu \chi - m_D \bar{\chi} \chi
+ \left( \Phi^* \left(y_L \bar{\chi}^c P_L \chi + y_R \bar{\chi}^c P_R \chi \right) + h.c.\right)
\,.
\end{align}
The ratio of $U(1)_D$ charge of $\Phi$ and $\chi$ equals to 2. Once $\Phi$ gets vev, 
the DM $\chi$ gets Majorana mass along with its Dirac mass. 

As a result, the Dirac fermion splits itself into 
two Majorana fermions, which provide the DM $\chi_1$ and its excited state
$\chi_2$, dubbed as inelastic dark matter (IDM) \cite{TuckerSmith:2001hy, TuckerSmith:2004jv}. 

We work with Weyl spinor and analyze the interactions for $\chi_1$
and $\chi_2$. If we write $\chi = \left\lbrace \eta , \xi^\dagger 
\right\rbrace$, the mass term is given as \cite{TuckerSmith:2001hy,
Izaguirre:2015zva},
\begin{align}
- \mathcal{L}_F \supset \dfrac{1}{2} 
\left( \begin{array}{cc} \eta & \xi \end{array}  \right)
\left(
\begin{array}{cc} 
m_\eta & m_D \\ 
m_D & m_\xi 
\end{array}
\right) 
\left( 
\begin{array}{c} \eta \\ \xi \end{array}
\right) + h.c.  \,,
\end{align}
where $m_\eta = - \sqrt{2} y_L v_D $ and $m_\xi = - \sqrt{2} y_R^* v_D$.
The mass matrix can be diagonalized by a rotation, 
\begin{align}
\left( 
\begin{array}{c} \eta \\ \xi \end{array}
\right)  =  
\left(
\begin{array}{cc} 
\cos \beta & \sin \beta \\ 
-\sin \beta & \cos \beta 
\end{array}
\right)
\left( 
\begin{array}{c}  \chi_1 \\ \chi_2 \end{array}
\right) \,,
\end{align}
where $\tan 2\beta = 2 m_D/(m_\xi - m_\eta)$.
The mass of $\chi_1$ and $\chi_2$ are 
\begin{align}
m_{\chi_1,\chi_2} = 
\frac{1}{2} \left(m_\eta + m_\xi \mp 
\sqrt{\left(m_\eta - m_\xi \right)^2 + 4 m_D^2} \right) \,.
\end{align}
The vector current of the DM couples to $U(1)_D$ gauge field $A'$. We can write both of them in 
the mass basis as follows,   
\begin{align}
& \mathcal{L}_F \supset A'^\mu J_\mu  = \left( \tilde{A}'^\mu  + t_W \epsilon
\frac{  m_{Z}^2  }{(m_{A'}^2- m_{Z}^2 )} \tilde{Z}^\mu\right)
(\eta^\dagger \bar{\sigma}^\mu \eta - \xi^\dagger \bar{\sigma}^\mu \xi) 
\nonumber
 \\
& = \left( \tilde{A}'^\mu  + t_W \epsilon
\frac{  m_{Z}^2  }{(m_{A'}^2- m_{Z}^2 )} \tilde{Z}^\mu\right) 
\left(
\frac{1}{x} (\chi_1^\dagger \bar{\sigma}^\mu \chi_1 
- \chi_2^\dagger \bar{\sigma}^\mu \chi_2 ) 
- \frac{2 m_D }{(m_\xi - m_\eta)x} (\chi_1^\dagger \bar{\sigma}^\mu \chi_2 
+ \chi_1^\dagger \bar{\sigma}^\mu \chi_2 ) 
\right) \,,
\end{align}
where we have define $x \equiv \sqrt{1 + 4 m_D^2 / (m_\xi - m_\eta)^2}$.
For the scalar interaction with DM can be written as,
\begin{align}
 \mathcal{L}_F & \supset \frac{-1}{2}\left(1 + \frac{\phi}{v_D} \right)
\left( m_\eta \eta \eta + m_\xi \xi \xi   \right)  + h.c. 
   \nonumber \\
& = \frac{-1}{2}\left(1 + \frac{\cos \alpha \tilde{\phi} - \sin\alpha \tilde{h}}{v_D} \right)  
\times
\nonumber
\\
&\left( 
\frac{1}{2} \left(m_\xi + m_\eta + \frac{-m_\xi + m_\eta}{x} \right) \chi_1 \chi_1 
+ \frac{1}{2} \left(m_\xi + m_\eta + \frac{m_\xi - m_\eta}{x}\right) \chi_2 \chi_2 
- \frac{2 m_D}{x} \chi_1 \chi_2
\right)  
\,.
\end{align}

There are two interesting parameter regions for this model. In the first one,  the 
Majorana mass is much larger than its Dirac mass, $m_\eta, m_\xi \gg m_D$, 
such that the mixing angle $\beta$ is small and the mass of $\chi_1$ and
$\chi_2$ have small corrections to its Majorana masses, where $m_{\chi_1} \approx m_\eta + m_D^2/(m_\eta - m_\xi) $
and $m_{\chi_2} \approx m_\xi + m_D^2/(m_\xi - m_\eta) $. The interactions
with vector boson and scalar are mainly diagonal, while the off-diagonal interactions
for $\chi_1$ and $ \chi_2$ are suppressed.

In the second case, Dirac mass is dominant, $m_\eta, m_\xi \ll m_D$. Therefore,
the mixing angle $\beta$ is very close to its maximal value $\pi/4$. 
The mass of ${\chi}_1$ and $\chi_2$ are $m_{{\chi}_1} \approx
m_D - (m_\xi  + m_\eta)/2$ and $m_{\chi_2} \approx
m_D + (m_\xi  + m_\eta)/2$, with the mass splitting $\Delta = m_\xi + m_\eta$.
We have $x \approx 2 m_D / (m_\xi - m_\eta) $ and $|x| \gg 1$. It suggests that
the diagonal interactions with vector boson are suppressed while
the off-diagonal interaction to ${\chi}_1$ and $\chi_2$ is dominant.
For the special case of $m_\xi = m_\eta$, the diagonal interaction with vector
boson vanishes. Since the IDM relies on the off-diagonal interactions with vector boson, 
the DM scattering only happens when the final states are its excited ones
and provides very different phenomenology from ordinary elastic
scattering in direct detection \cite{TuckerSmith:2001hy}.
However, for scalar interactions, the diagonal interaction with the fermionic DM
is proportional to $m_\xi + m_\eta$, while the off-diagonal interaction is
proportional to $m_\xi - m_\eta$. The scalar mediation to diagonal terms
can potentially spoil the IDM setup when the Higgs portal coupling is large.

Coming back to the exotic Z decays, we see that vector portal IDM motivates the exotic decays of 
$ \tilde{Z} \to \chi_2 \chi_1$ and $\chi_2 \chi_2$, followed by the subsequent cascade decay $\chi_2 
\to \tilde{A}' \chi_1 \,, \tilde{\phi} \chi_1$ and $\tilde{A}' \,, \tilde{\phi} \to
\bar{f} f \,, \chi_1 \chi_1$. It shows that  IDM with vector portal 
can motivate the topologies of exotic Z decay in \cref{sec:ZdecayChannels}.

\subsection{Magnetic inelastic DM, Rayleigh DM}
\label{sec:MIDMandRayDM}
The coupling of DM to the Standard Model particles can be very weak.  One possible scenario is that the 
hidden sector interacts with the Standard Model via  high dimensional operators. 
The representative models, the Magnetic inelastic DM~(MIDM) and Rayleigh DM model~(RayDM) \cite{Sigurdson:2004zp, Masso:2009mu, 
Chang:2010en, Weiner:2012cb, Weiner:2012gm}, 
are introduced, and their relevance to the exotic Z decay is studied in this section.

\subsubsection{Model}
\label{sec:MIDMmodelDetail}
\begin{figure*}[h]
\centering
 \includegraphics[width=0.7\textwidth]{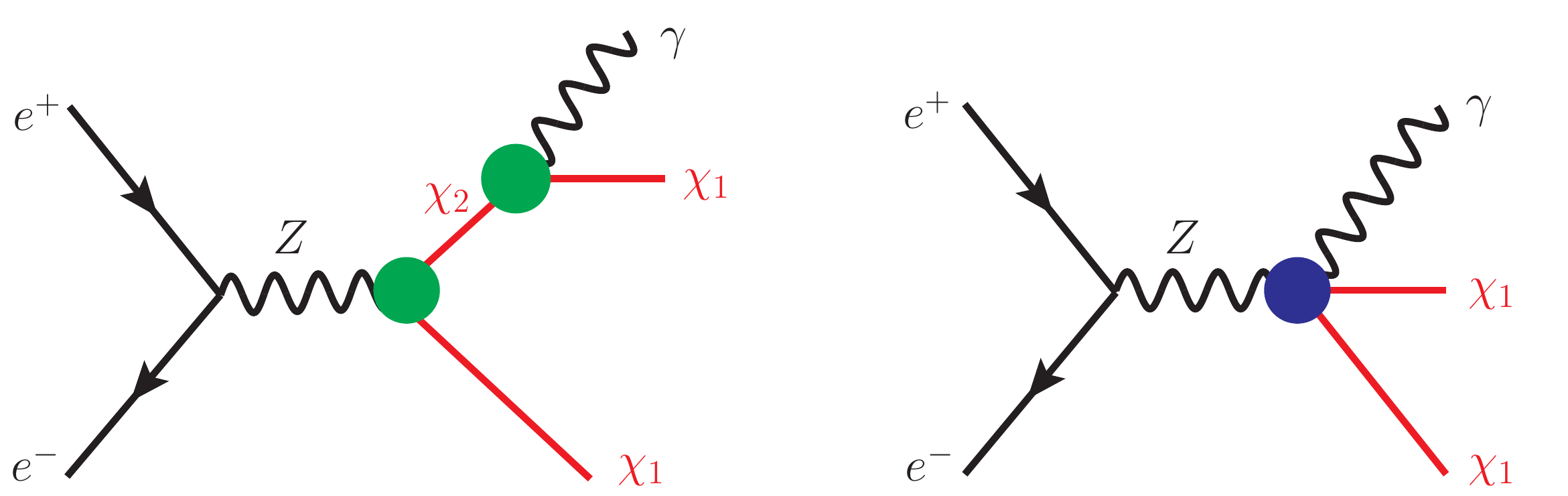}
 \caption{The Feynman diagrams for the 
 cascade decay process $Z \to \chi_2 \chi_1 \to \chi_1\chi_1 \gamma$ 
 from $O_{\text{MIDM}}$ and the three-body process 
 $Z  \to \chi_1\chi_1 \gamma$ from $O_{\text{RayDM}}$.}
  \label{fig:MIDMRayDMandDiagrams}
\end{figure*}

The two models, the MIDM and 
RayDM, can be derived from the same UV model \cite{Weiner:2012gm},
\begin{align}
\mathcal{L}  = \bar{\chi} (i \slashed{\partial} - m_\chi) \chi - \frac{1}{2} \delta m
\bar{\chi}^c \chi  + \bar{\psi} (i \slashed{D} - M_\psi )\psi + (D^\mu \phi)^\dagger (D_\mu \phi)
-M_\phi^2 \phi^\dagger \phi   + (\lambda \bar{\psi} \chi \phi + h.c.).
\label{eq:LagofMIDMRay}
\end{align}
$\chi$ is fermionic DM with a Dirac mass term $m_\chi$ and Majorana mass term $\delta m$. It interacts 
with scalar $\phi$ and another fermion $\psi$ via a Yukawa coupling.
The Dirac and Majorana mass terms can split DM $\chi$ into two Majorana fermion $\chi_1$ and $\chi_2$,
where we assume $m_{\chi_2}  > m_{\chi_1}$.
The fermion $\psi$ and scalar $\phi$ have the same charge under SM gauge 
group $\text{SU}(2)_L \times \text{U}(1)_Y$ \cite{Weiner:2012gm}.
The dark matter will couple to photon via $\psi$ and $\phi$ loop.
Integrating out $\psi$ and $\phi$ will generate two higher dimensional operators. 
The first operator is MIDM operator~\cite{Sigurdson:2004zp, Masso:2009mu, 
Chang:2010en}, and the second is RayDM operator~\cite{Weiner:2012cb}.
Both of them are given below, 
\begin{align}
O_{\text{MIDM}}= \frac{1}{\Lambda_{\text{MIDM}}} \bar{\chi}_2 \sigma^{\mu \nu} 
\chi_1 B_{\mu \nu}+h.c.,
\quad O_{\text{RayDM}}=\frac{1}{\Lambda_{\text{RayDM}}^3} \bar{\chi}_1 \chi_1 B^{\mu \nu}B_{\mu \nu} \,.
\label{eq:MIDM-Reyleigh}
\end{align}
Note there are also operators including $\gamma_5$ in the DM bilinear, which corresponds the
electric dipole operator. For RayDM, the corresponding one is 
\begin{align}
O_{\text{RayDM}}^{\gamma_5} = 
\frac{i}{\Lambda_{\text{RayDM}}^3} \bar{\chi}_1 \gamma_5 \chi_1 B^{\mu \nu} \tilde{B}_{\mu \nu} ,
\end{align}
where  $\tilde{B}_{\mu \nu} = \epsilon_{\mu\nu\alpha\beta} B^{\alpha\beta}$ and $\epsilon_{\mu\nu\alpha\beta}$
is the anti-symmetric Levi-Civita symbol.
The interaction scale $\Lambda$ has been calculated in \cite{Weiner:2012gm}
\begin{align}
\frac{1}{\Lambda_{\text{MIDM}}} \approx \frac{\lambda^2 g_Y}{64 \pi^2 M_\psi},
\quad
\frac{1}{\Lambda^3_{\text{RayDM}}} \approx \frac{\lambda^2 g_Y^2}{48 \pi^2 M_\psi^3} \,,
\label{eq:LambdatoMf}
\end{align}
where we have assumed that $\psi$ and $\phi$ are singlet under $SU(2)_L$ and
are charged under $U(1)_Y$. In \cref{eq:LambdatoMf}, we have assumed $\phi$ mass
is similar to $M_\psi$ and we take the form factor function to be $\mathcal{O}(1)$.
These two operators can lead to the cascade decay 
$Z \to \chi_2 \chi_1 \to  (\chi_1 \gamma) \chi_1$ 
and the three-body decay $Z \to \chi_1 \chi_1 \gamma$
at Z-factory, with Feynman diagrams given in \cref{fig:MIDMRayDMandDiagrams}. 
In the exotic Z decay study,
we will choose a significant mass splitting between $\chi_1$ and $\chi_2$,
to get a hard photon signal which can be detected at Z-factories.

With this setup, we see that decay topologies $Z \to \chi_2 \chi_1 \to  (\chi_1 \gamma) \chi_1$ 
and $Z \to \chi_1 \chi_1 \gamma$ in \cref{fig:MIDMRayDMandDiagrams} can be easily achieved. 
In the perspective of model building, the cascade decay channel 
$Z \to \chi_2 \chi_2 \to  (\chi_1 \gamma) (\chi_1 \gamma)$ would be more complicated. 
In particular, if $\chi_2$ is Majorana fermion, the dipole term
$\bar{\chi}_2 \sigma_{\mu \nu }\chi_2$ will vanish. 
One would add new species of Dirac fermion DM $\chi$, then the Yukawa term in
\cref{eq:LagofMIDMRay} becomes $\lambda_i \bar{\psi} \chi_i \phi $, where $i$
is the number of species~\cite{Primulando:2015lfa}. In this case, one can have
$\bar{\chi}_i \sigma_{\mu \nu }\chi_j$ in MIDM operator and $\bar{\chi_i} \chi_j$
in RayDM operator, which provide rich cascade decays for exotic Z decay.

\subsubsection{DM relic abundance, indirect and direct searches, and collider constraints}
\noindent $\bullet$ \textit{Relic abundance and Indirect detection:}

\begin{figure*}[h]
\centering
 \includegraphics[width=0.32\textwidth]{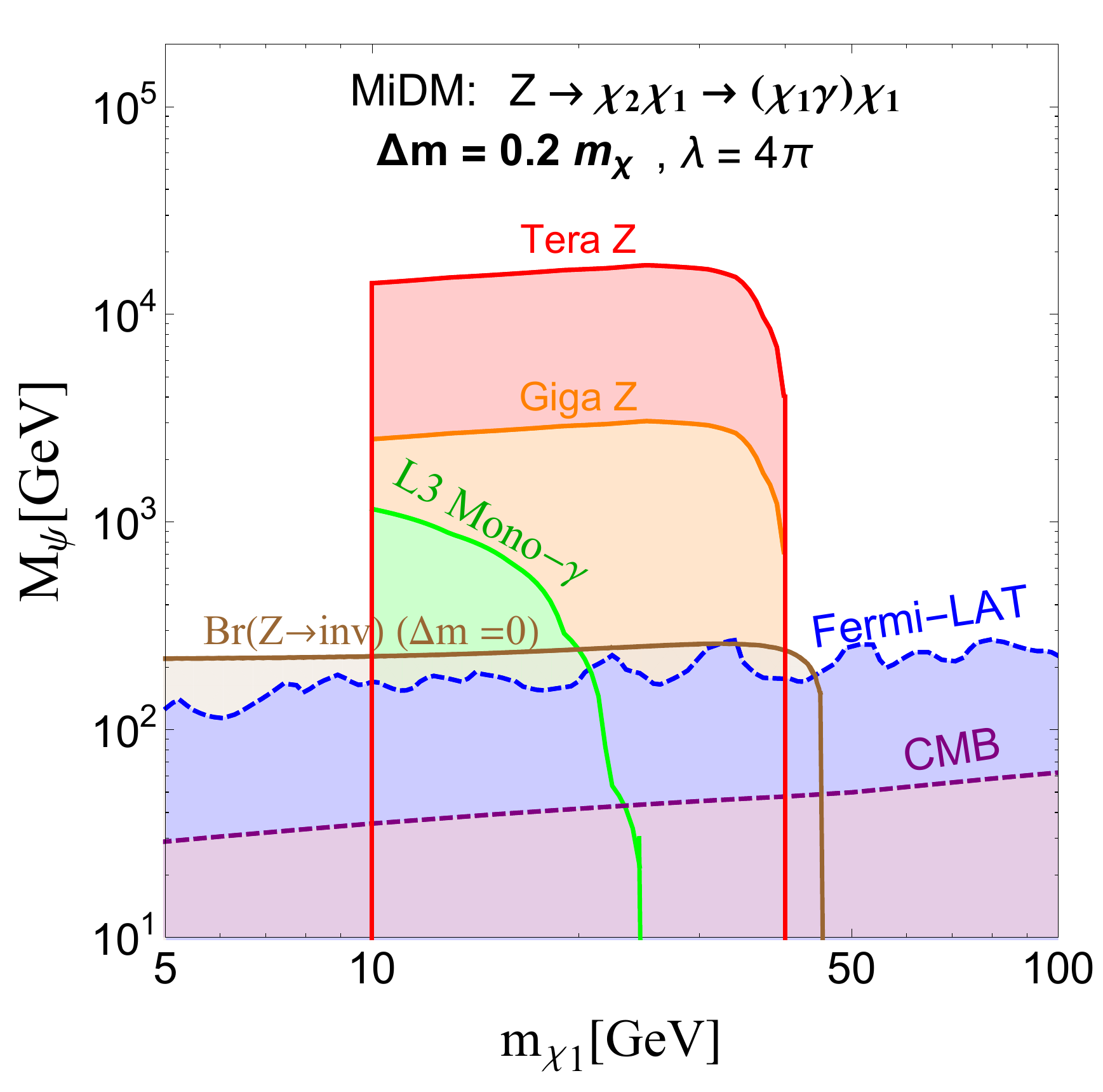} 
 \includegraphics[width=0.32\textwidth]{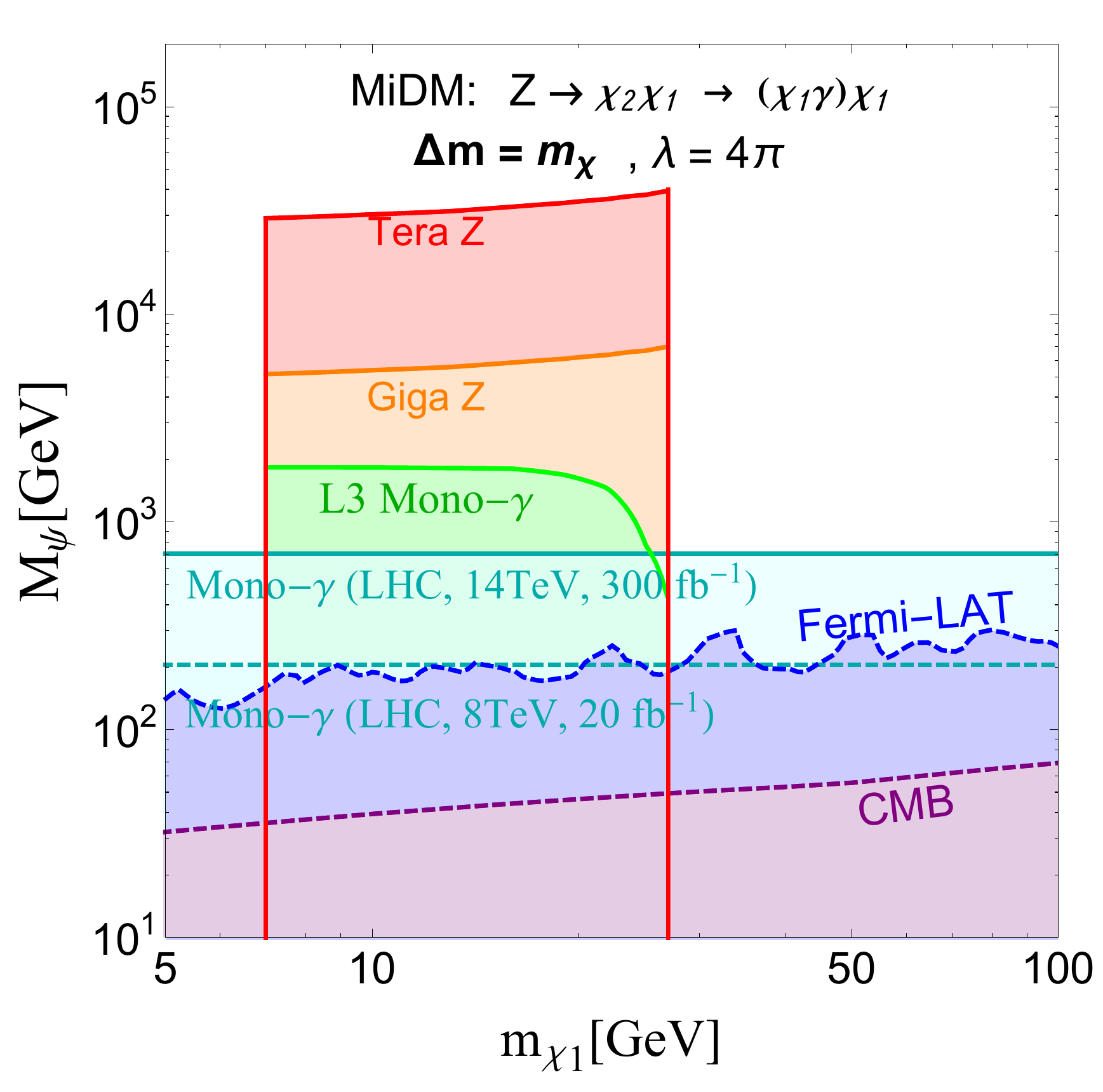} 
 \includegraphics[width=0.32\textwidth]{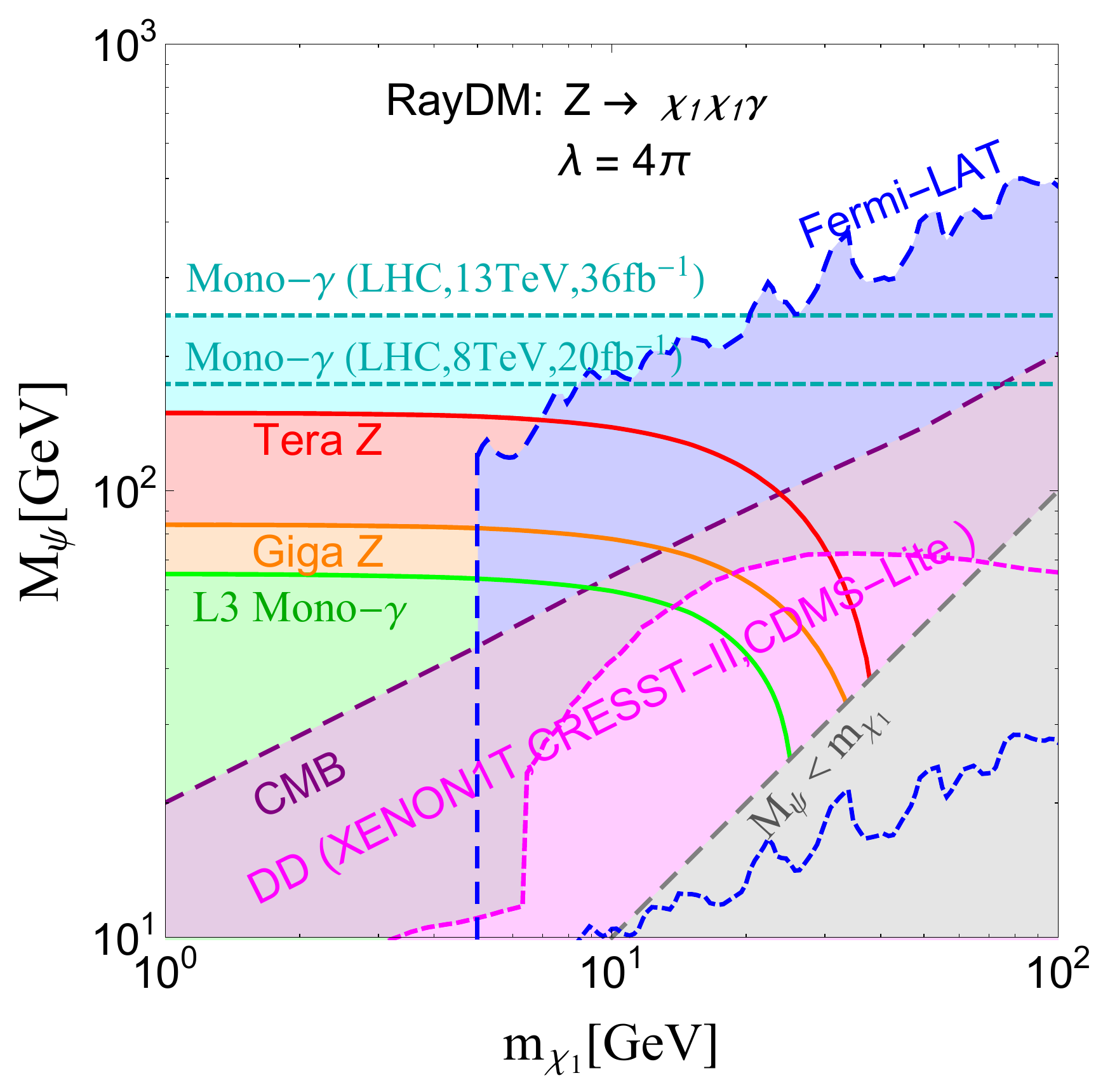} 
 \caption{The $95\%$ C.L. sensitivity for $M_{\psi}$ from exotic Z decay 
  $Z   \to \slashed{E}+ \gamma$, for MIDM operator in the left (middle) panels with
  different mass splitting and for Rayleigh operator in the right panel. 
  The constraints are labeled
  as Giga Z and Tera Z for future Z-factory with $\lambda = 4\pi$, and the LEP limit
  from \cite{Acciarri:1997im} is shown.
  We also compare the limits from DM direct detection, indirect detection searches,
  mono-photon, and mono-jet searches at the LHC. For RayDM, 
  the gamma-ray constraints from Fermi-LAT and CMB use long dashed line for $ O_{\text{RayDM}}^{\gamma_5}$
  and dashed line for $ O_{\text{RayDM}}$. For collider limits, the two operators are
  similar and for spin-independent direct detection limits, only $ O_{\text{RayDM}}$
  is constrained.   }
  \label{fig:MIDMandRayDM}
\end{figure*}

We focus on the case that there is a significant mass splitting between $\chi_2$ and $\chi_1$, 
which can give rise to interesting photon signal in exotic Z-decay. In this case, 
 the relevant annihilation initial state contains only
$\chi_1$. The annihilation rate is dominated by the Rayleigh operator into
$\gamma \gamma$, $\gamma Z$, $ZZ$ and $W^+ W^-$. For the mass range
$m_{\chi_1} < m_Z$, we find only the following annihilation cross-section
relevant \cite{Weiner:2012cb},
\begin{align}
\sigma v(\chi_1 \chi_1 \to \gamma \gamma)_{\text{MIDM}} & = \frac{ \cos^2\theta_w m_{\chi_1}^2}{\pi \Lambda_{\text{MIDM}}^4}
\frac{16 y^6 - 9 y^4 - 2 y^2 -2}{y^4 (y^2+2)^2} \,,  \label{eq:MIDM2gamma} \\
\sigma v(\chi_1 \chi_1 \to \gamma \gamma)_{\text{RayDM}} & = \frac{ \cos^2\theta_w}{\pi}
\frac{m_{\chi_1}^4}{\Lambda_{\text{RayDM}}^6} v_{\text{rel}}^2 , \label{eq:RayDM2gamma} \\
\sigma v(\chi_1 \chi_1 \to \gamma \gamma)_{\text{RayDM}}^{\gamma_5} & = \frac{16 \cos^2\theta_w}{\pi}
\frac{m_{\chi_1}^4}{\Lambda_{\text{RayDM}}^6} ,
\label{eq:RayDM2gammaG5}
\end{align}
where $y \equiv m_{\chi_2}/m_{\chi_1}$. The two annihilation cross-section for RayDM are for 
$ O_{\text{RayDM}}$ and $ O_{\text{RayDM}}^{\gamma_5}$ respectively, where
the former one is p-wave suppressed while the second one is s-wave.  
The annihilation process for the MIDM scenario is two loop suppressed. This can be seen in 
\cref{eq:MIDM2gamma}, \cref{eq:RayDM2gamma} and \cref{eq:RayDM2gammaG5}
 through the dependence on $\Lambda_{\text{MIDM}}$ and $ \Lambda_{\text{RayDM}}$, respectively.  
The annihilation into gamma-ray lines is 
constrained by Fermi-LAT search \cite{Ackermann:2013uma} (blue shaded region )
and also by CMB~\cite{Ade:2015xua} (purple shaded region), which we constrain $M_\psi$ as a 
function of $m_\chi$ in \cref{fig:MIDMandRayDM}. 
The long dashed lines are for $ O_{\text{RayDM}}^{\gamma_5}$, while the dashed lines are for 
$ O_{\text{RayDM}}$ which is very weak due to the p-wave suppression.

\noindent $\bullet$ \textit{Direct detection:}

In the case of large splitting, only Majorana $\chi_1$ is relevant for direct detection, 
because inelastic scattering into $\chi_2$ is kinetically forbidden.
Therefore, the scattering cross-section is dominated by the loop exchange of 
two photons from Rayleigh operator, and the spin-independent cross-section per 
nucleon is given below \cite{Weiner:2012cb},
\begin{align}
\sigma^{\text{SI}}_n \approx \frac{4 \alpha^2_{\text{EM}} Z^4}{\pi^2 A^4}
\frac{m_N^2 Q_0^2}{\Lambda_{\text{RayDM}}^6} \,,
\end{align}
where $m_N$ is the mass of nuclei $N$, $A$ is the nucleon number, $Z$ is the
proton number of nuclei and $Q_0$ is the nuclear coherence scale
$Q_0 = \sqrt{6} (0.3+0.89 A^{1/3})^{-1} ~\text{fm}^{-1}$. The current 
leading constraints on spin-independent cross-sections are 
XENON1T~\cite{Aprile:2017iyp}, LUX~\cite{Akerib:2016vxi},
PANDAX-II~\cite{Tan:2016zwf}, and CRESST-II~\cite{Angloher:2015ewa} as
well as CDMSlite~\cite{Agnese:2015nto}. The limits
from direct detection constraints are shown as magenta in \cref{fig:MIDMandRayDM}.
Only $ O_{\text{RayDM}}$ is shown as in dashed lines, because $ O_{\text{RayDM}}^{\gamma_5}$
produces spin-dependent cross-section and spin-independent cross-section is suppressed.
 
\noindent $\bullet$ \textit{Existing collider constraints:}

Besides DM indirect and direct detection, the MIDM and RayDM operators 
can also get constraints from mono-jet and mono-photon searches at LHC
and LEP.   

The Rayleigh operator $O_{\text{RayDM}}$ has been studied
in mono-photon, mono-jet and mono-V searches \cite{Crivellin:2015wva}, 
where V stands for
vector gauge boson W and Z. The authors found that the limits from mono-photon
provides the strongest bound and constrain $\Lambda_{\text{RayDM}} \gtrsim 510$
GeV at $95\%$ C.L, for $m_{\chi_1} \lesssim 100$ GeV from the LHC 
8 TeV at $20 ~\text{fb}^{-1}$~\cite{Khachatryan:2014rwa, Aad:2014tda}.
Very recently, ATLAS \cite{Aaboud:2017dor} has explored 13 TeV data to
search mono-photon signature with integrated luminosity $36 \text{fb}^{-1}$, and it
pushes the limit to $\Lambda_{\text{RayDM}} \gtrsim 725$ GeV.
These limits has been integrated in the right panel of \cref{fig:MIDMandRayDM},
and denoted as ``mono-$\gamma$". For $\quad O_{\text{RayDM}}^{\gamma_5}$,
the limits are similar as $O_{\text{RayDM}}$ and therefore we only show the results
for $O_{\text{RayDM}}$.

For the MIDM operator $O_{\text{MIDM}}$, \cite{Primulando:2015lfa} has studied
the limits from the mono-jet, mono-photon and di-photon searches at 8 TeV and
14 TeV LHC. For a significant splitting, they found mono-photon search 
\cite{Khachatryan:2014rwa} is the most stringent, similar to the RayDM operator case. 
For $m_{\chi_1} = 10$ GeV, it requires $\Lambda_{\text{MIDM}} \gtrsim 2400$
GeV, and the result is roughly unchanged for $m_{\chi_2} > 20$ GeV.
In the left and middle panels of \cref{fig:MIDMandRayDM}, we vary DM mass 
$m_{\chi_1}$ from $0$ to $40$ GeV. Since its mass is much smaller than 
the required photon $p_T$ and MET, we expect the constraint to be similar 
as $m_{\chi_1} = 10$ GeV. For mono-photon search at the LHC 14 TeV with 
$300 ~\text{fb}^{-1}$, the corresponding limit is estimated to be 
$\Lambda_{\text{MIDM}} \gtrsim 8200$ GeV \cite{Primulando:2015lfa},
and labeled as ``mono-$\gamma$" in \cref{fig:MIDMandRayDM}.

For the MIDM case, it is interesting to note that, when $m_{\chi_2} = m_{\chi_1}$, 
the exotic Z decay $Z \to \slashed{E } \gamma$ loses its sensitivity at Z -factory, 
and also for mono-photon search at the LHC. 
The mono-jet search will be better than the mono-photon search in this case.
Moreover, \cite{Primulando:2015lfa} pointed out that actually the invisible decay width 
measurement of Z can beat the mono-jet search at the LHC 14 TeV with 
$3 ~\text{ab}^{-1}$ integrated luminosity, 
which suggest $M_{\psi} \gtrsim 226 $ GeV
for $m_{\chi_{1,2}} = 10$ GeV.
We have plotted the invisible Z width constraint in panel (a) of \cref{fig:MIDMandRayDM}.

Given the high center of mass energy at the LHC, it can search for 
the EW charged particles $\psi$ and $\phi$ directly from Drell-Yan production and
their subsequent cascade decays \cite{Liu:2013gba}. 
The Drell-Yan search could be more restrictive than mono object searches, 
but this conclusion is very model dependent, see \cite{Liu:2013gba}.
For example, when $\psi$ and $\phi$ are $SU(2)_L$ singlet, or
they decay dominantly to tau lepton and (or) gauge bosons,
the sensitivity from Drell-Yan is very poor, even at the LHC 14 TeV 
with $300~ \text{fb}^{-1}$.

For mono-photon at LEP, the L3 collaboration has collected data with $137 ~\text{pb}^{-1}$
at the Z pole, which can limit the BR of exotic decay $Z \to \gamma \slashed{E}$
down to $1.1 \times 10^{-6}$ if photon energy is greater than $\sim 30$ GeV
\cite{Acciarri:1997im}. The OPAL collaboration has a similar study at Z pole but
with only $40.5 ~\text{pb}^{-1}$ \cite{Akers:1994vh}. There are also many off-Z
peak measurements on single photon final state. The one with $176 ~\text{pb}^{-1}$
data taken at $189$ GeV has been carried out by the L3 collaboration, which looks for
MIDM topology $Z \to \chi_2 \chi_1 \to  (\chi_1 \gamma) \chi_1$, and bounds
the cross-section of such topology to be smaller than $0.15 - 0.4$ pb with some
dependence on $m_{\chi_1}$ and $m_{\chi_2}$ \cite{Acciarri:1999kp}.
The leading constraint is from  L3 measurement at Z pole due to large resonant 
cross-section, 
and we label the constraints as ``L3 Mono-$\gamma$" in \cref{fig:MIDMandRayDM}. 
We see that this constraint is comparable to the future LHC reach in middle panel 
of \cref{fig:MIDMandRayDM}.

\subsubsection{Prospects from exotic Z decay}
\noindent $\bullet$ \textit{Exotic Z decay sensitivity:}

For exotic Z decay with final state $\slashed{E} \gamma$, we summarize the results of the 
cascade decay process $Z \to \chi_2 \chi_1 \to \chi_1\chi_1 \gamma$ 
from $O_{\text{MIDM}}$ and of the three-body process 
$Z  \to \chi_1\chi_1 \gamma$ from $O_{\text{RayDM}}$ in \cref{fig:MIDMRayDMandDiagrams}.
The limits on such exotic decay BR is given in \cref{sec:monophoton},
and we can calculate the limits for $\Lambda_{\text{MIDM}}$ and 
$\Lambda_{\text{RayDM}}$ accordingly, then convert them into
constraints for $M_\psi$ by \cref{eq:LambdatoMf}.
The limits are given in \cref{fig:MIDMandRayDM}, and labeled as
``Giga Z" and ``Tera Z".

The results of the MIDM operator are presented 
in the left and middle panel of \cref{fig:MIDMandRayDM}. 
We find exotic Z decay can reach $M_\psi \sim \mathcal{O}(10^{4})$ GeV, 
which is much better than mono-photon searches at the HL-LHC 
with $M_\psi \sim 10^3$ GeV. 
The production cross-sections for $\chi_2 \chi_1$ at Z-factory
and LHC both scales as $1/\Lambda_{\text{MIDM}}^2$. However,
the cross-section at Z-factory benefits from Z resonance
comparing with at the LHC, therefore have larger statistics. Moreover, 
the $\slashed{E}+\gamma$ searches at Z-factories have much cleaner 
environment than hadron collider. As a result, exotic Z decay can give 
the $M_\psi$ reach two orders better than mono-photon search at the LHC
or HL-LHC. The indirect detection of gamma lines at Fermi-LAT 
provides a similar constraint to the LHC 8 TeV. The direct detection
does not provide any constraint for MIDM operator, because the mass
splitting between $\chi_1$ and $\chi_2$ is too large.

In the right panel of \cref{fig:MIDMandRayDM}, for RayDM operator, 
we find the mono-photon search at the LHC can reach $M_\psi$ to a 
few hundreds of GeV, which is better than exotic Z decay with $M_\psi \gtrsim 100$ 
GeV. The reason is that the cross-section for $\chi_1 \chi_1 \gamma$ is proportional to 
$s^2/ \Lambda_{\text{RayDM}}^6$,. Since the Z-factory has a small 
center of mass energy square $s \sim m_Z^2$,  it has less sensitivity.  The constraint from direct detection
is very weak, because it is a two loop process. The gamma line constraint
from Fermi-LAT is comparable to other constraints and is strongest at
$m_{\chi_1}$ around 100 GeV.

\noindent $\bullet$ \textit{Summary:}

We find complementarity between exotic Z decay 
$Z \to \slashed{E} \gamma$ at Z-factory and mono-jet or mono-photon search at the LHC
with large mass splitting between $\chi_1$ and $\chi_2$. 
For very small mass splitting, the photon from cascade decay
$\chi_2 \to \chi_1 \gamma$ becomes very soft, thus the mono-photon
and mono-jet search via initial state radiation are better. However, invisible
Z width measurement can provide a better limit $M_{\psi} \gtrsim 226 $ GeV.
For MIDM operator, future Z-factory can provide the leading constraints, while
for RayDM operator, the HL-LHC can provide better constraints.

\subsection{Axion-like particle}
\label{AxionPortal}

Axion-like particle~(ALP) is a light pseudo-scalar 
which couples to gauge fields via anomalous terms and interacts with fermions with 
derivatives, $\partial_\mu a \bar{\psi} \gamma^\mu \psi$. 
Its presence is quite generic in UV theories, such as string theory~\cite{Svrcek:2006yi,Arvanitaki:2009fg,Acharya:2010zx}, 
and Supersymmetry~\cite{Frere:1983ag,Nelson:1993nf,Bagger:1994hh}.
It can be a portal connecting dark matter with the standard model sector~\cite{Nomura:2008ru}, 
and ultralight ALP is dark matter candidate by coherent oscillating in the 
universe~\cite{Preskill:1982cy,Abbott:1982af,Dine:1982ah}.
Recently the dynamics of ALP in the universe has also been proposed to solve the Higgs hierarchy 
problem~\cite{Graham:2015cka}. 
For our Z-factory study, we are focusing on the mass range of ALP from $ 0.1~\mathrm{GeV}$ to Z boson mass.
Although we focus on the case of ALP, our analysis and results in this section 
can be applied to scalar easily.

\begin{figure}[h]
	\includegraphics[width=0.4\textwidth]{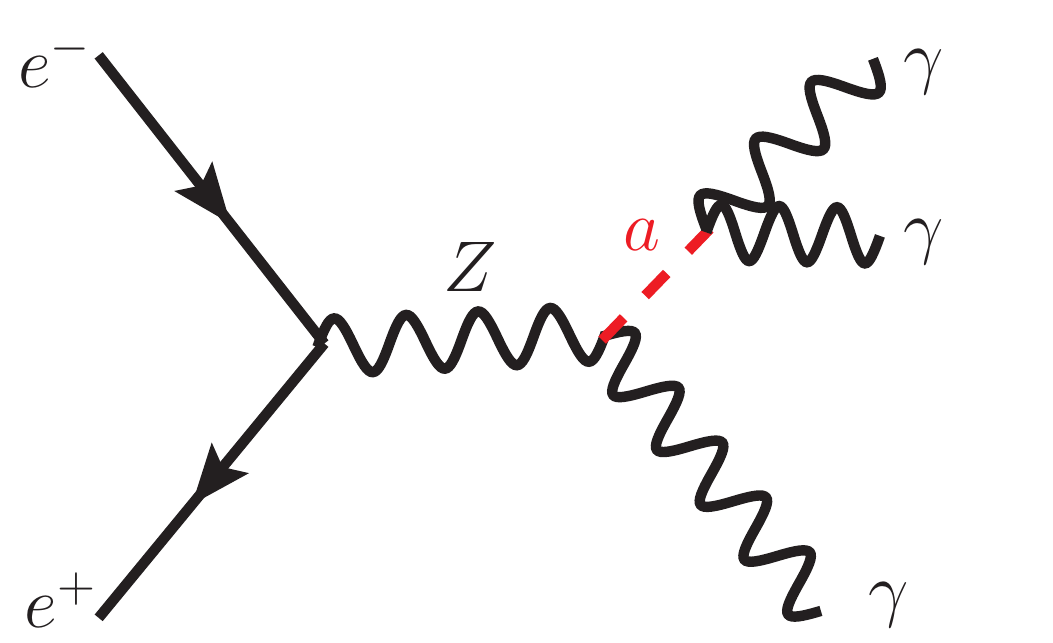} 
	\caption{The Feynman diagram for the exotic Z decay $Z \to a \gamma \to (\gamma \gamma) \gamma $. 
		The final state is $3\gamma$ and in case $m_a$ is too small to separate the two photons, the final state
		is $2\gamma$. }
	\label{fig:axionFeyn}
\end{figure}

\begin{figure}[h]
	\includegraphics[width=0.6\textwidth]{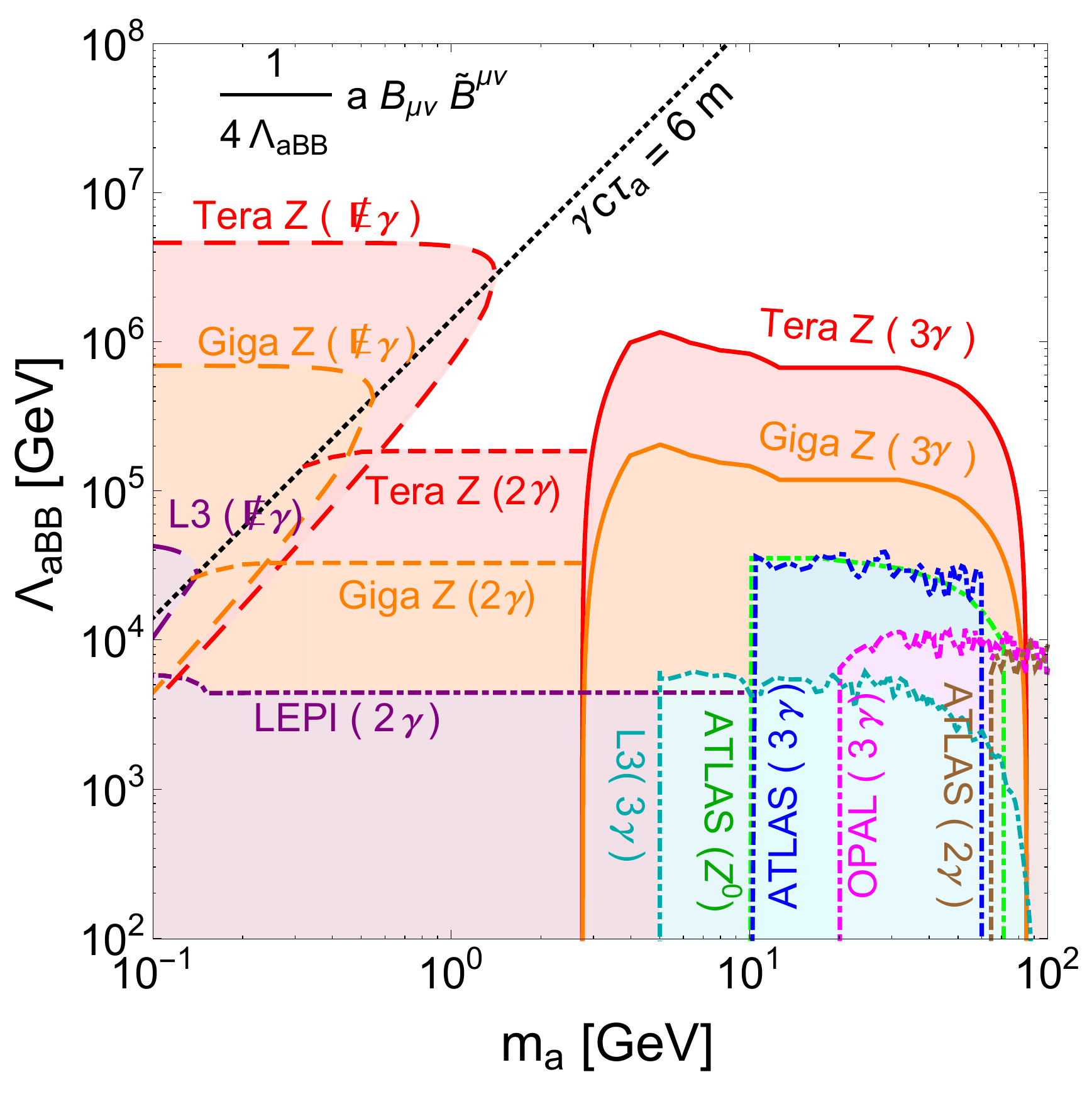} 
	\caption{The limit on $\Lambda_{a \text{BB}}$, ALP coupling to hypercharge field, from 
		future Z-factory. The limits from LEP~\Rmnum{1}~\cite{Jaeckel:2015jla} $\gamma\gamma$ search,
		LEP~\Rmnum{2}~(OPAL) $ 2\gamma $ and $ 3 \gamma$ searches~\cite{Abbiendi:2002je},
		, LEP (L3) $ 3 \gamma$ searche at Z pole \cite{Acciarri:1994gb},
		ATLAS $ 3 \gamma$ and $Z \to 3 \gamma$~\cite{Aad:2014ioa,Aad:2015bua} search are translated
		to limits on $\Lambda_{a \text{BB}}$ following \cite{Knapen:2016moh}. 
		There are three type of signals $Z \to 2\gamma, ~3\gamma$ and $\slashed{E} \gamma$, 
		depending on $m_a$. In $\slashed{E} \gamma$ final state
		where $a$ decay outside the detector, we have set the detector length to be 6 meter and 
		LEP limits on this final state from L3 collaboration 
		\cite{Acciarri:1997im} has been plotted. }
	\label{fig:axionconstraint}
\end{figure}
  
ALPs can have interactions with standard model particles fermions, gauge fields, Higgs obeying the 
(discrete-)shift symmetry. Here, we focus on the ALP coupling to the 
$U(1)_Y$ gauge field $B_\mu$ ~\footnote{The coupling to fermions are neglected here for simplicity.
The ALP coupling to fermion	is $c_{f} m_f /\Lambda$ where $c_f$ coefficient is model dependent. 
$a\rightarrow \gamma \gamma$ is the 
dominant decay channel for very light ALP, and the decays to fermions are suppressed by $m_f^2/m_a^2$
when ALP is significantly heavier than fermion. If the fermion coupling comes through the gauge field loops, 
this will get further suppression via the loop effects.
},

\begin{equation}
\mathcal{L}_{\text{ALP}}  = \frac{ 1 }{ 4 \Lambda_{aBB} } a B_{\mu\nu} \tilde{B}^{\mu\nu}  \ ,
\end{equation}
This interaction gives the decay rate of the ALP as 
\begin{equation} 
   \Gamma({a \to \gamma \gamma}) = \frac{1}{64 \pi} \frac{1}{ \Lambda_{a BB }^2} 
   \cos \theta_w^4 m_a^3  \ ,
\end{equation} 
and the rate of the Z decay,
\begin{equation} 
   \Gamma ( Z \to \gamma a )  = \frac{1}{96 \pi} \frac{1}{ \Lambda_{a BB }^2} \cos \theta_w^2 \sin \theta_w^2 m_Z^3 
   \left(1- \frac{m_a^2}{m_Z^2}  \right)^3\ .
\end{equation} 
 
Depending on the $a \to \gamma \gamma$ decay length, the analyses are performed in the two separate regimes:
one is ALP decaying inside the detector,
and the other is decaying outside the detector. 
For decay inside the detector, we focus on the prompt search, 
and leave the interesting case of displaced vertex to future work. For decay outside the detector, 
the signal is mono-photon $+ \slashed{E}$. The transverse radius of the detector radius is taken to be $6$ meters. 
The decay length of the ALP is computed according to the boost $\gamma_a$ of the ALP, 
$\text{D} \equiv \gamma_a c \tau_a$, where the 
$\gamma_a = E_a / m_a$ is the boost and $\tau_a = 1/\Gamma_a$ is the lifetime of $a$. 
Since the initial state is Z boson at rest and the final state is $a\gamma$, 
the energy $E_a$ is fixed by $m_a$. $\text{D} = 6 ~\rm{m}$ is plotted in \cref{fig:axionconstraint} 
as a dotted black line. Below it, the ALP has a decay length D smaller than
$6 ~\rm{m}$. However, it can still decay outside the detector with a probability of $1- e^{- \text{D} / (6~ \rm{m})}$. We
account for this probability to rescale the signal events in the detector, which leads to sensitivity below the line.
In the prompt decay region, for the high mass axion, the boost of axion is small, the dominant channel to search for
ALPs is $3 \gamma$. When the mass of the ALP is below $\mathcal{O}(1)$ GeV, the boost of axion makes
the two photons from axion decay close to enough, and cannot be resolved. The $2\gamma$ search channel 
is more relevant.

The current constraints for this operator are given by LEP and LHC photon searches. 
In \cref{fig:axionconstraint}, the LEP~\Rmnum{1}~\cite{Jaeckel:2015jla} uses inclusive 
di-photon search $ e^+ e^- \to 2 \gamma + X$ covering the small mass region.
In the higher mass region, the boost of the axion decreases and $3 \gamma$ channel is considered.
The LEP~\Rmnum{2}~(OPAL) have $ 2\gamma $ and $ 3 \gamma$ data~\cite{Abbiendi:2002je}, 
which are employed to put the bounds on the process, 
$e^+ e^- \to \gamma/Z^\star \to a \gamma \to 2 \gamma + \gamma $. 
The L3 collaboration has searched the process $ Z \to a \gamma \to (\gamma\gamma) \gamma$
at Z pole, with limit on BR of order $10^{-5}$ \cite{Acciarri:1994gb}. 
ATLAS $ 3 \gamma$ and $Z \to 3 \gamma$~\cite{Aad:2014ioa,Aad:2015bua} search can be 
translated to the ALP bound as derived in \cite{Knapen:2016moh}.

For $\slashed{E} + \gamma$ search, the strongest bound from LEP comes from L3 
collaboration with $137 ~\text{pb}^{-1}$ data at the Z pole \cite{Acciarri:1997im} 
as discussed in \cref{sec:MIDMandRayDM}. It can limit the BR of exotic decay 
$Z \to \gamma \slashed{E}$ down to $1.1 \times 10^{-6}$ if photon energy 
is greater than $\sim 30$ GeV. It directly excludes $\Lambda_{\text{aBB}} <
 4.3\times 10^4$ for $Z \to \slashed{E} + \gamma$ decay, and we label it as 
 ``L3 $(\slashed{E} \gamma)$" in \cref{fig:axionconstraint}.

In the Z-decay search, the ALP will give topologies $Z \to \slashed{E} + \gamma$ and 
$Z \to 3 \gamma , 2 \gamma$, depending on the life-time and boost of the ALP. Z-factory limits on the 
ALP are given in \cref{fig:axionconstraint}, which is about two order of magnitude better 
than the current constraints from LEP and LHC.

\section{Searching for Exotic Z Decays at Future Z-Factories}
\label{sec:ZdecayChannels}

\begin{table*}[htbp]
  \centering
  \begin{tabular}{|C{0.15\textwidth}|L{0.34\textwidth}|C{0.05\textwidth}|L{0.44\textwidth}|}
  \hline
      exotic decays                            & topologies  &  $n_{res}$  & models \\
    \hline
    \multirow{4}{*}{$Z\to \slashed{E}+ \gamma$}  & $Z\to\chi_1\chi_2, \chi_2\to \chi_1 \gamma $& 0
   & 1A:$\frac{1}{\Lambda_{1A}}\bar{\chi_2}\sigma^{\mu\nu}\chi_1 B_{\mu\nu}$ (MIDM)\\ \cline{2-4}    
   & $Z\to\chi \bar \chi \gamma$ & 0 & 1B: $\frac{1}{\Lambda_{1B}^3}\bar{\chi}\chi B_{\mu\nu}B^{\mu\nu}$ (RayDM) \\ \cline{2-4}  
   & $Z\ \to a \gamma \to (\slashed{E}) \gamma$ & 1 & 1C: $\frac{1}{4 \Lambda_{1C}} a B_{\mu\nu} \tilde{B}^{\mu\nu}$ (long-lived ALP)\\ \cline{2-4}
      & $Z\ \to A' \gamma \to (\bar{\chi}\chi) \gamma$ & 1 &1D: 
      $\epsilon^{\mu \nu \rho \sigma}      A'{}_\mu B_\nu \partial_\rho B_\sigma$ 
      (WZ terms)\\ \hline  
   \multirow{3}{*}{$Z\to \slashed{E}+ \gamma\gamma$} & $Z\to \phi_d A'$ ,$\phi_d\to (\gamma\gamma)$, $A'\to (\bar{\chi}\chi)$
   & 2 & 2A: Vector portal \\ \cline{2-4}
   & $Z\to \phi_H \phi_A$, $\phi_H\to (\gamma\gamma)$, $\phi_A\to (\bar{\chi}\chi)$ 
   & 2 & 2B: 2HDM extension \\  \cline{2-4}  
   & $Z\to \chi_2\chi_1$, $\chi_2\to\chi_1 \phi$, $\phi \to (\gamma\gamma)$ & 1 & 
   2C: Inelastic DM \\  \cline{2-4}  
   & $Z\to \chi_2\chi_2$, $\chi_2\to\gamma\chi_1$  & 0 & 
   2D: MIDM  \\  \hline
   \multirow{3}{*}{$Z\to \slashed{E}+ \ell^+ \ell^-$} & $Z\to \phi_d A'$, $A'\to (\ell^+\ell^-)$, $\phi_d\to (\bar{\chi}\chi)$ 
   & 2 & 3A:  Vector portal \\  \cline{2-4}  
    & $Z\to A' S S \to (\ell \ell) SS$ & 1 & 3B:  Vector portal\\  \cline{2-4} 
    & $Z\to \phi (Z^*/\gamma^*) \to \phi \ell^+\ell^-$ & 1 & 3C:  Long-lived ALP, 
    Higgs portal\\  \cline{2-4} 
    & $Z\to \chi_2 \chi_1 \to \chi_1 A' \chi_1\to  (\ell^+\ell^-) \slashed{E}$ 
    & 1 & 3D:  Vector portal and Inelastic DM \\  \cline{2-4} 
    & $Z\to \chi_2\chi_1$, $\chi_2\to \chi_1 \ell^+\ell^- $ & 0 & 
    3E:  MIDM, SUSY\\  \cline{2-4} 
    & $Z \to \bar \chi \chi \ell^+ \ell^- $ & 0 & 3F: RayDM, slepton,
    heavy lepton mixing \\  \hline
     \multirow{3}{*}{$Z\to \slashed{E}+ JJ$} 
     & $Z \to \phi_d A' \to (\bar{\chi} \chi) (jj)$ & 2 & 4A:  Vector portal \\  \cline{2-4} 
     & $Z \to \phi_d A' \to (bb)(\bar{\chi} \chi) $ & 2 & 4B:  Vector portal + Higgs portal\\  \cline{2-4} 
     & $Z \to \chi_2 \chi_1 \to bb \chi_1 + \chi_1 \to bb \slashed{E}$ & 0 & 4C:  MIDM \\  
     \hline
    \multirow{3}{*}{$Z\to (JJ)(JJ)$} & $Z\to \phi_d A' , \phi_d \to jj, A'\to jj$ & 2 & 
    5A:  Vector portal + Higgs portal \\ \cline{2-4} 
    & $Z\to \phi_d A' , \phi_d\to b \bar{b} , A'\to jj$ & 2 & 5B:  vector portal + Higgs portal \\  \cline{2-4}     
    & $Z\to \phi_d A'  , \phi_d \to  b\bar{b} , A'\to b\bar{b}$ & 2 & 5C:  vector portal + Higgs portal \\  \hline 
    $Z\to \gamma\gamma\gamma$  & $ Z\to \phi \gamma \to (\gamma\gamma) \gamma$ & 1 & 6A: ALP, Higgs portal \\  \hline  
  \end{tabular}
  \caption{ Classification of exotic Z decay channels by particles in final states and 
      number of resonances ($n_{res}$).
  The $\chi$ and $\chi_1$ are fermionic DM, $\chi_2$ is an excited state of DM,
  and $S$ denotes scalar DM. The final state $J$ represents either light flavor jet $j$ 
  or heavy flavor jet $b$. $A'$ is the dark photon, and the $\phi$ is
  intermediate scalars. The parentheses $()$ indicates a resonance in the final states.
  The details of these models are discussed in the text.}
  \label{tab:finalstate}
\end{table*}

In this section, we make projections for the sensitivity of  exotic Z decay searches at future Z-factories. 
Motived by the previous discussed dark sector models, we classify decay channels by final states, 
the number of intermediate resonances, and different topologies. In most of the cases, we clarify the 
connections between the potential models and each topology. 
As Z is neutral, the final states of its decay can be described as
\begin{align}
Z \to  \slashed{E} + n_{\gamma} \gamma + n_{\ell^+ \ell^-} {\ell^+ \ell^-}  
+ n_{\bar q q} \bar q q \,.
\end{align}
Since lepton and quark are charged, they will show up in pairs. 
The $n$ is referred to as the number of particle or pair of particles. In our analysis, we choose to 
consider the number of final state particles to be less than $5$.
The $\slashed{E}$ can be considered as two particles, since normally it is constituted of two DM particles. 
It also can be a neutral particle which does not interact with detector and decays outside of it. 
The final states can be further grouped according to whether they are the decay products of some intermediate 
resonance.
This resonance can be the mother particles for $(\gamma\gamma)$, ${\ell^+ \ell^-}$, $(\bar q q)$ and $\slashed{E}$.
The kinematic information of the resonance decay can help us improve the search strategies. 
The details of classification are given in \cref{tab:finalstate}. 
The first set of channels has the missing energy  in the final states.
Since electron collider has  full kinematic information of initial states, 
the missing 4-momentum  can be fully reconstructed. This is the major advantage of electron 
collider compared with hadron collider in searching for exotic Z decay with missing energy. 
The second set of channels does not include missing energy. They are pure jet final states $(jj)(jj)$, $(jj)(bb)$, 
$(bb)(bb)$ and three photon final state $\gamma \gamma \gamma$. They can come from dark sector particles 
decays, which do not involve dark matter. 
Due to the cleaner environment of electron collider, 
it is better than hadron colliders to measure pure hadronic final states. 
For $jjjj$ final state, since it has large SM background, we concentrate on the 
case where it has two resonances. When generating 
corresponding SM backgrounds, one additional photon is included to count the 
initial state radiation~(ISR). The on-shell intermediate particles 
should be neutral, since LEP searches have already put severe constraints on charged particles with mass 
smaller than $m_Z /2$.

In the following subsections, we will 
discuss the possible models and the sensitivity of each channel at future Z-factory. 
The \cref{sec:FCCeesetup} introduces the basic setup and performance for future Z-factories at FCC-ee
and CEPC, and explores the sensitivity of exotic Z BR at this future Z-factory for different topologies 
from \cref{sec:monophoton} to \cref{sec:ZTo3gamma}.
To compare the future Z-factory and HL-LHC, 
\cref{sec:LHCcompare} presents the reach on those exotic Z BR for the HL-LHC.
The summary of this comparison between the future Z-factory and HL-LHC is in \cref{fig:summary}.

\subsection{Performance of Future Z-factories}
\label{sec:FCCeesetup}

The exotic Z decay phenomenology at future Z-factories at studied in this section. 
A Z-pole run has been considered for 
both FCC-ee and CEPC \cite{FCC-ee, CEPC}. Given that the measured cross-section of hadronic Z is 
$30.5 ~ \text{nb}$ \cite{ALEPH:2005ab}, the integrated luminosity for Giga Z ($10^9$ Z) and Tera Z 
($10^{12}$ Z in the plan of FCC-ee) are $22.9 ~\text{fb}^{-1}$ and $22.9 ~\text{ab}^{-1}$, respectively.

We simulate the backgrounds and signals in the electron-positron colliders   
at the Z mass energy using  
MadGraph5$\textunderscore$aMC$@$NLO~\cite{Alwall:2014hca} and analyze them at parton 
level. Assuming that the detector performance is similar for different future electron colliders,
we follow the detector effects at CEPC~\cite{CEPC-SPPCStudyGroup:2015csa} and apply 
the following Gaussian smearing in our analysis:
\begin{align}
& \text{Photon energy resolution: } \frac{\delta E_\gamma}{E_\gamma} = 
\frac{0.16}{\sqrt{E_\gamma / \rm{GeV}}} \oplus 0.01 \,, \label{eq:photonRES}
\\
&\text{Lepton momentum resolution: } \Delta \frac{\rm{GeV}}{p_T^\ell} = 
2\times 10^{-5}\oplus \frac{10^{-3} \rm{GeV}}{p_T^\ell \sin \theta} \,,
\\
&\text{Jet energy resolution:  } \frac{\delta E_j}{E_j} = \frac{0.3}{\sqrt{E_j / 
\rm{GeV}}} \oplus 0.02  \,. \label{eq:jetRES}
\end{align}
We make conservative assumptions about the tagging efficiency: $80\%$ for about b-tagging efficiency, $9\%$ for c quark mis-tagging rate and $1\%$ for light flavor mis-tagging 
rate~\cite{CEPC-SPPCStudyGroup:2015csa}. 
We also require that all visible particles 
satisfy $| \eta | <2.3$ ($\cos \theta < 0.98$). In addition, 
the photon, lepton and jet energy should be larger than $10~\rm{GeV}$.
For events with missing energy, we require $\slashed{E} > 10~\rm{GeV}$ as well. 
Lastly, both the photons and electrons in final state are separated by 
$\theta_{ij} \gtrsim 10^\circ = 0.175$ radian.
The charged leptons normally have better resolution than photons, thus the separation
requirement that we choose here is conservative. For jets, we use a conservative separation requirement
$\theta_{ij} \gtrsim 0.4$ radian corresponding to $\Delta R \geq 0.4$ at LHC 
\footnote{For other separation condition, see \cite{Liu:2016zki}.}. 
The study for LEP3 (a 240 GeV circular $ee$ collider using LHC 
tunnel) with the CMS detector \cite{Azzi:2012yn} shows the jet angular resolution
can be 30 milli-radian for energies below 100 GeV. The separation requirement 
for jets at lepton collider could be optimized due to much less QCD backgrounds than LHC in principal.
We leave the optimization for lepton collider as the future study. 

To derive the exclusion limits, the confidential level for the sensitivity 
calculation adopts Poisson
probability 
\cite{Patrignani:2016xqp}. When background event number 
$B \gg 1$ , the significance is about $S/\sqrt{B}$ which is proportional
to $\sqrt{L}$, where $L$ is the integrated luminosity. Therefore, the sensitivity reach
of Giga Z and Tera Z differ by about $10^{1.5}$. When background event number $B \ll 1$, 
the Poisson distribution with zero background assumption leads to a constant limit for 
signal. In this case, the exclusion limit is linear to $L$, thus Giga Z and Tera Z
differ by about $10^3$. If $B \ll 1$ for Giga Z while $B > 1$ for Tera Z, the difference of the sensitivity
reach is in the range of $10^{1.5} - 10^{3}$.

\subsection{$ Z \to \slashed{E}+ \gamma$ }
\label{sec:monophoton}

\begin{figure}[ht]
  \begin{tabular}{c c c}
    \includegraphics[width=0.32\textwidth]{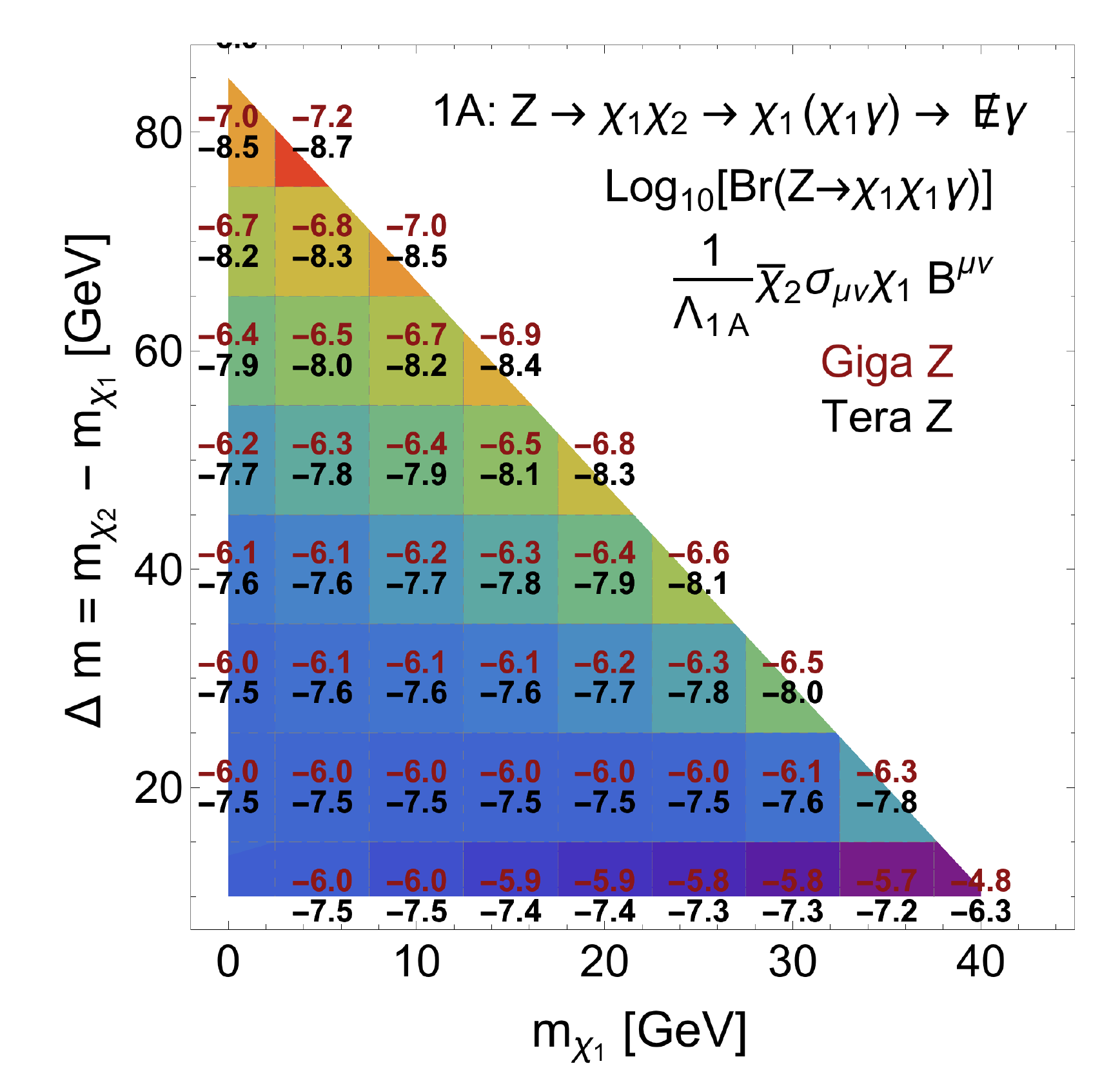}  &
    \includegraphics[width=0.32\textwidth]{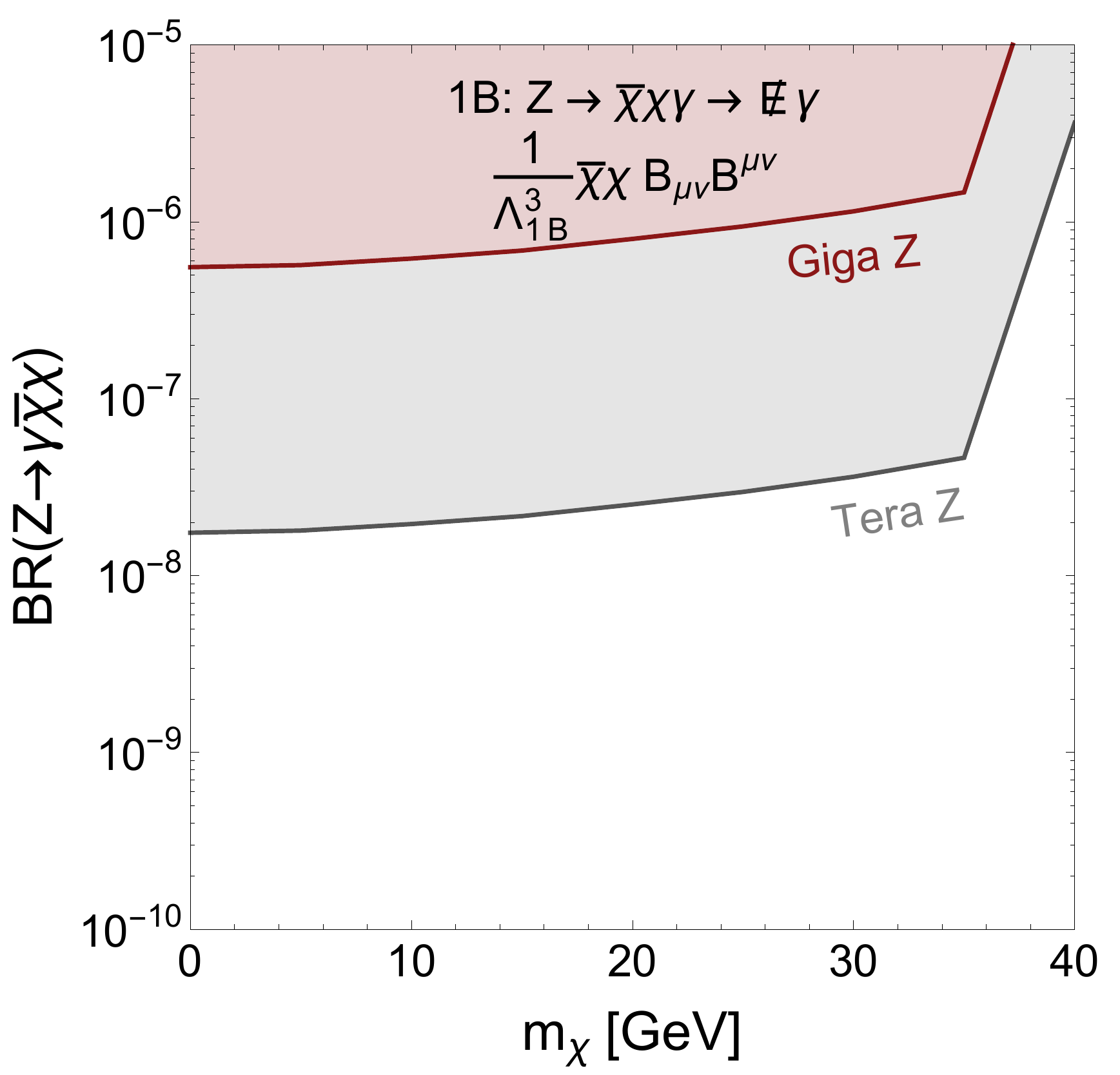} &
    \includegraphics[width=0.32\textwidth]{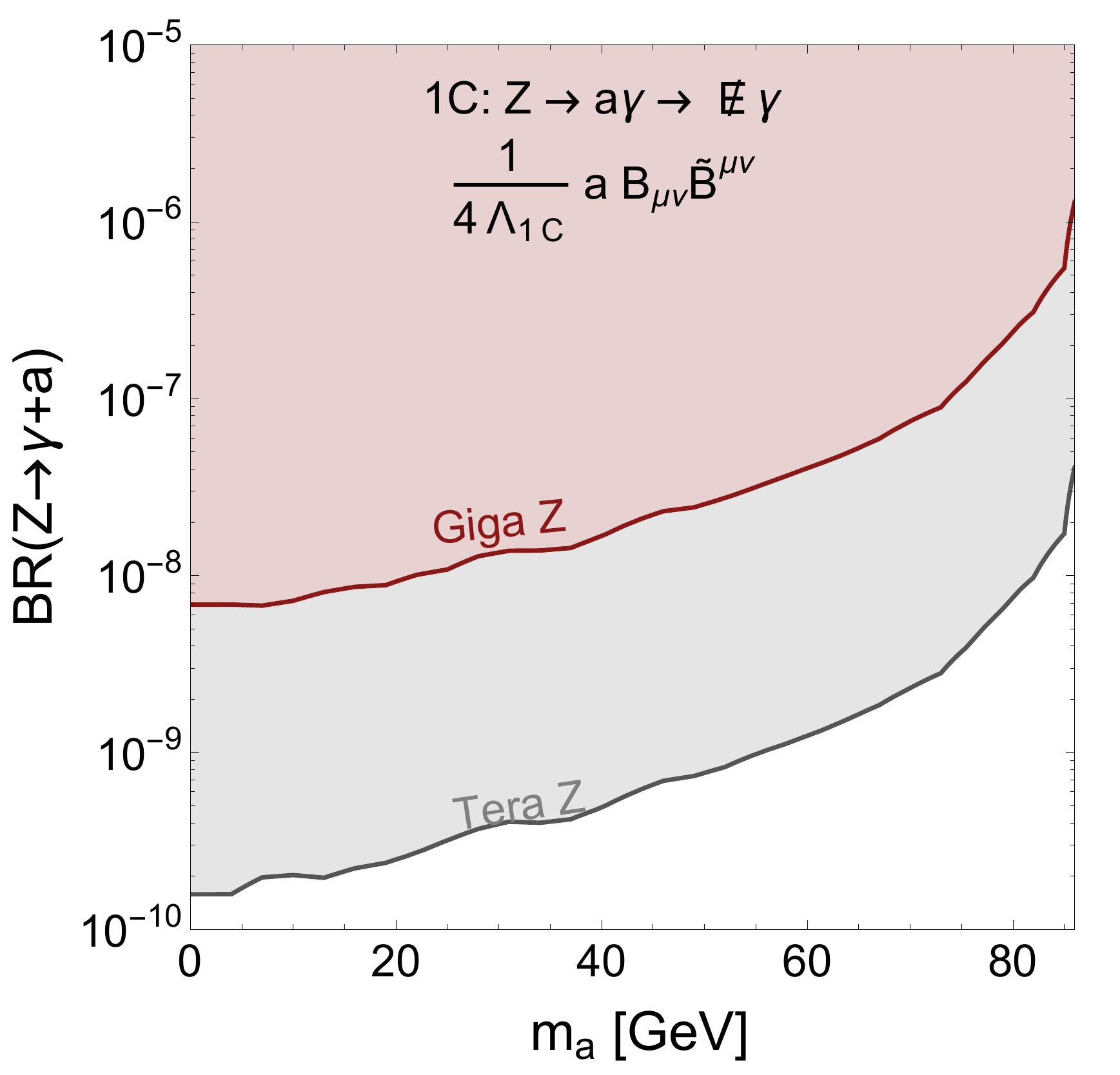}  \\
    (a) &  (b) &   (c)
  \end{tabular}
  \caption{ The $95\%$ C.L. exclusion on exotic Z decay BR for the final 
  state $Z \to \slashed{E} \gamma$.
  (a): the decay topology 1A, $Z \to \chi_2 + \chi_1 \to \chi_1 \gamma+ \chi_1 $
  from MIDM model. The numbers in each block are the sensitivity reach for the exotic Z 
  decay BR in $\text{log}_{10}$
  for Giga Z and Tera Z respectively, while the color mapping is coded for Tera Z.
  (b): the decay topology 1B, $Z\to \chi \chi \gamma$ from RayDM model.  
  (c): the decay topology 1C, $Z\to a \gamma \to \slashed{E} \gamma$ 
  from axion-like particle model.}
  \label{fig:ZtoMETgamma}
\end{figure}

In this section, we discuss the exotic Z decay with the final state $ \slashed{E}+ \gamma$. 
We consider the decay topologies  $Z \to \chi_2 \chi_1 \to \chi_1 \gamma + \chi_1 $ and $Z\to \chi \chi \gamma$, 
where $\chi$ and $\chi_{1}$ are fermionic DM. $\chi_2$ is an excited DM state which decays back to $\chi_1$. 
We also consider  2-body decay $Z \to a \gamma \to (\slashed{E}) \gamma$, where $a$ is a pseudo-scalar 
as missing energy signal if it is stable at collider scale or it decays to dark matter particles.
We denote these three topologies as 1A, 1B and 1C, respectively, shown in \cref{tab:finalstate}. 
The UV models for the 1A and 1B are the MIDM and RayDM model, while 1C is motivated by ALP.
The fourth topology, denoted as 1D, is $Z \to A' \gamma \to (\bar{\chi} \chi) \gamma$. It can come from
the Wess-Zumino term $\epsilon^{\mu \nu \rho \sigma}  A'_\mu B_\nu \partial_\rho B_\sigma$, 
when the dark photons couple to
anomalous currents \cite{Dror:2017ehi, Ismail:2017fgq, Ismail:2017fgq}. After integrating by parts, the longitudinal
part of $A'$ have similar interaction as the topology 1C; thus the limit on
exotic Z decay BR is similar as 1C.

The SM backgrounds for these final states are mainly $e^+ e^- \to \gamma \nu \bar \nu$. 
In our simulation, we include one more photon to account for the ISR effect. 
For $\gamma \nu_e \bar \nu_e$, this process is mediated by either off-shell 
$W^{\pm *}$ or off-shell $Z^*$, while $\gamma \nu_{\mu } \bar \nu_{\mu }$ 
and $\gamma \nu_{\tau} \bar \nu_{\tau}$ are mediated by off-shell $Z^*$. 
In these processes, most of the $\gamma$s come from ISR, or 
internal bremsstrahlung via the t-channel W boson. The background photons are generally quite \textit{soft} due to their origin as ISR.

The three models have the different kinematic distributions for the mono-photon. 
For the topology 1A, $Z \to \chi_2 + \chi_1 \to \chi_1\gamma+ \chi_1 $,
the photon energy spectrum has a box shape due to the cascade decay in this process.
The minimum and maximum of the photon energy are
\begin{align}
E_\gamma^{\rm{max},1A}=\frac{m_2^2-m_1^2}{4 m_2^2}\frac{s+m_2^2-m_1^2 
+\sqrt{s^2+(m_2^2-m_1^2)^2 -2s(m_2^2+m_1^2)} }{\sqrt{s}} \\
E_\gamma^{\rm{min},1A}=\frac{m_2^2-m_1^2}{4 m_2^2}\frac{s+m_2^2-m_1^2 
-\sqrt{s^2+(m_2^2-m_1^2)^2 -2s(m_2^2+m_1^2)} }{\sqrt{s}} \,.
\end{align}
The distribution of photon energy is flat between $\left[E_\gamma^{\rm{min},1A} , E_\gamma^{\rm{max},1A} \right]$,
and the edge of photon energy distribution can be used to determine the mass of DM.
Therefore, aside from the pre-selection cuts, we further impose the cuts below,
\begin{align}
\rm{1A}: \quad  E_\gamma^{\rm{min},1A} < E_\gamma < E_\gamma^{\rm{max},1A}, 
\quad m_{\rm{inv}} \geq 2 m_{\chi_1}\,,
\end{align}
where $m_{\rm{inv}}$ is the invariant mass of missing energy. The second cut 
comes from momentum conservation that the invariant mass of a set of particles is 
larger or equal to the sum of individual masses. According to the recoil mass relation,
$E_\gamma$ and $m_{\rm{inv}}$ are not independent with each other. If we 
apply the first cut, the second cut is automatically satisfied. 
Nevertheless, we list the second cut, sine this is not redundant in other cases.

For the topology 1B, $Z\to \chi \chi \gamma$, it has a broad distribution in photon
energy. 
The recoil mass $m_{\rm{inv}}$ is related to the photon energy $E_\gamma$ by
\begin{align}
E_\gamma & = \frac{s - m_{\rm{inv}}^2}{2 \sqrt{s}} \,.
\end{align}
In the mean time, the relation $m_{\rm{inv}} \geq 2 m_\chi$ gives the maximum 
allowed photon energy
\begin{align}
E_\gamma^{\rm{max},1B} = \frac{s - (2 m_{\chi})^2}{2 \sqrt{s}} \,.
\end{align}
Thus, in addition to the pre-selection cut, we imposes the following cuts to further suppress 
the SM background,
\begin{align}
\rm{1B}: \quad  \frac{1}{2} E_\gamma^{\rm{max},1B} < E_\gamma < E_\gamma^{\rm{max},1B},
 \quad  \ .
\end{align}
The lower bound of $E_\gamma$ is chosen to keep significant amount of signal event,
and to reject SM background as much as possible. 

For the topology 1C, $Z \to a \gamma \to (\slashed{E}) \gamma$, the 
photon energy spectrum is a delta function with $E^{\text{2C}}_\gamma= (s-m_{\phi_a}^2)/(2\sqrt{s})$.
Considering the photon energy $\sim 10 $ GeV, the energy resolution for this photon energy
is around $5\%$ according to \cref{eq:photonRES}. 
Therefore, we can choose a $2$ GeV window on the photon energy,
\begin{align}
\rm{1C}: \quad  E^{\text{1C}}_\gamma - 1~\rm{GeV} < E_\gamma < E^{\text{1C}}_\gamma + 1~\rm{GeV}\,.
\end{align}

After applying the pre-selection cuts and the specific cuts for the topologies 1A, 1B and 1C, 
we obtain the $95\%$ C.L. exclusion on the exotic Z decay BR in \cref{fig:ZtoMETgamma}.
In the panel (a) of \cref{fig:ZtoMETgamma} for the topology 1A, the numbers in each block are 
$\text{log}_{10}(\text{BR})$ for Tera Z (black) and Giga Z (dark red). It is clear
that the sensitivity on BR for Giga Z and Tera Z differ by a factor of $10^{1.5}$. 
The reason is SM background $\gamma \nu \bar{\nu}$ has event number much 
larger than 1 for both Tera Z and Giga Z, thus the sensitivity is scaled as $S/\sqrt{B}$. As a result,
the sensitivity scales with luminosity as $\sqrt{L}$, so the $\rm{BR}$ sensitivity gets a factor of 
$10^{1.5}$ increase from Giga Z to Tera Z. For Giga Z, the limit on BR falls in the range $10^{-6} 
- 10^{-7}$, while reaches $10^{-7} - 10^{-8}$ for Tera Z.  
In the panel (b) of \cref{fig:ZtoMETgamma} for the topology 1B, the luminosity scaling between
Giga Z and Tera Z is the same as in 1A. The limits on BR for Giga Z is close to $\sim 10^{-6}$, which
is a little bit weaker than 1A due to its 3-body decay topology. 
In the panel (c) of \cref{fig:ZtoMETgamma} for the topology 1C, the luminosity scaling between
Giga Z and Tera Z is similar to 1A and 1B. However, the sensitivity on BR for Giga Z is close 
to $\sim 10^{-8}$, which is about 2 orders better than 1A and 1B. The massive resonance
in $\slashed{E}$ implies that  the energy of photon is mono-chromatic, which greatly
reduces SM background.

\subsection{$Z\to \slashed{E}+ \gamma \gamma$ }

\begin{figure}[ht]
  \begin{tabular}{c c c}
    \includegraphics[width=0.33\textwidth]{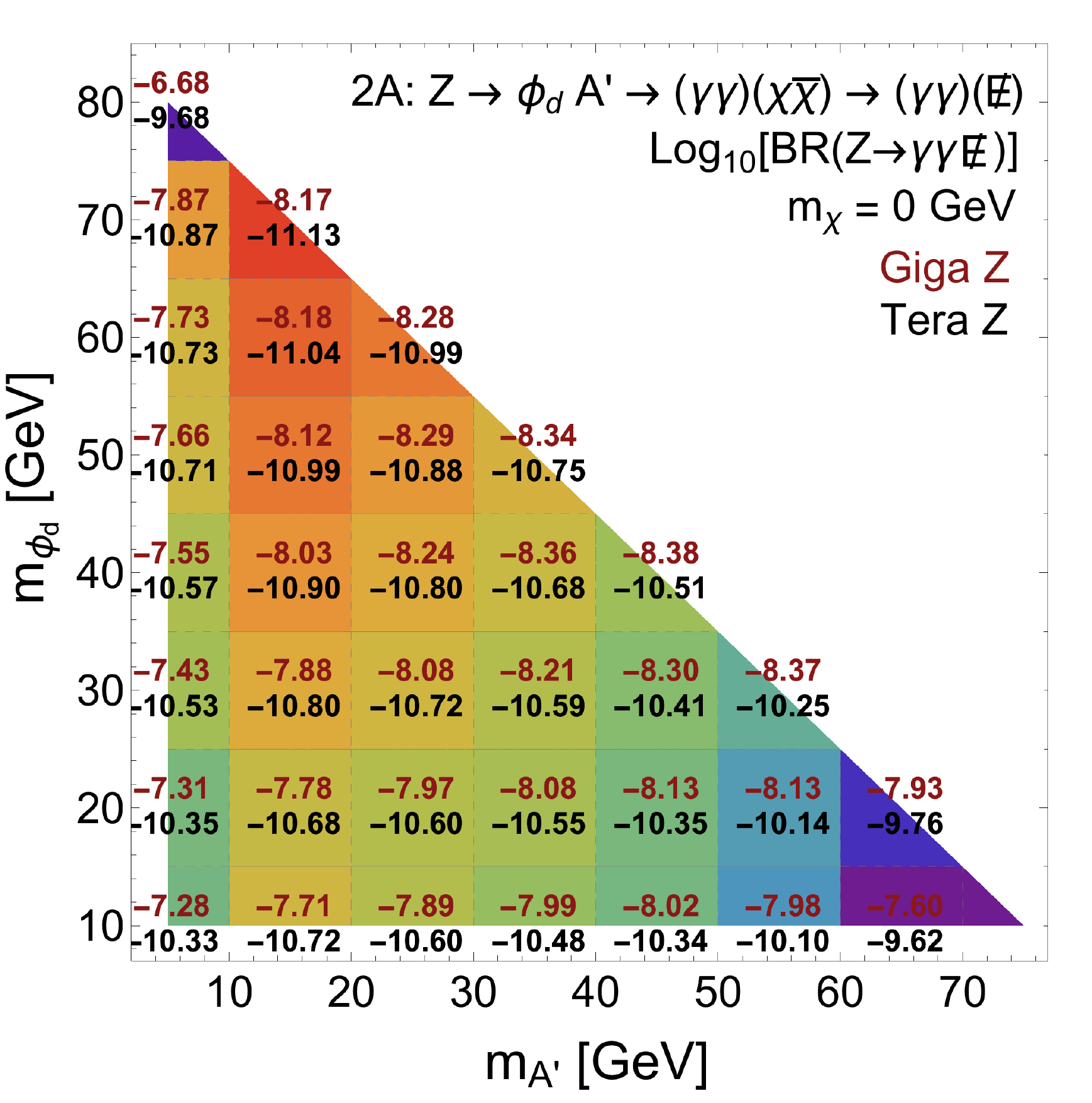} &
    \includegraphics[width=0.33\textwidth]{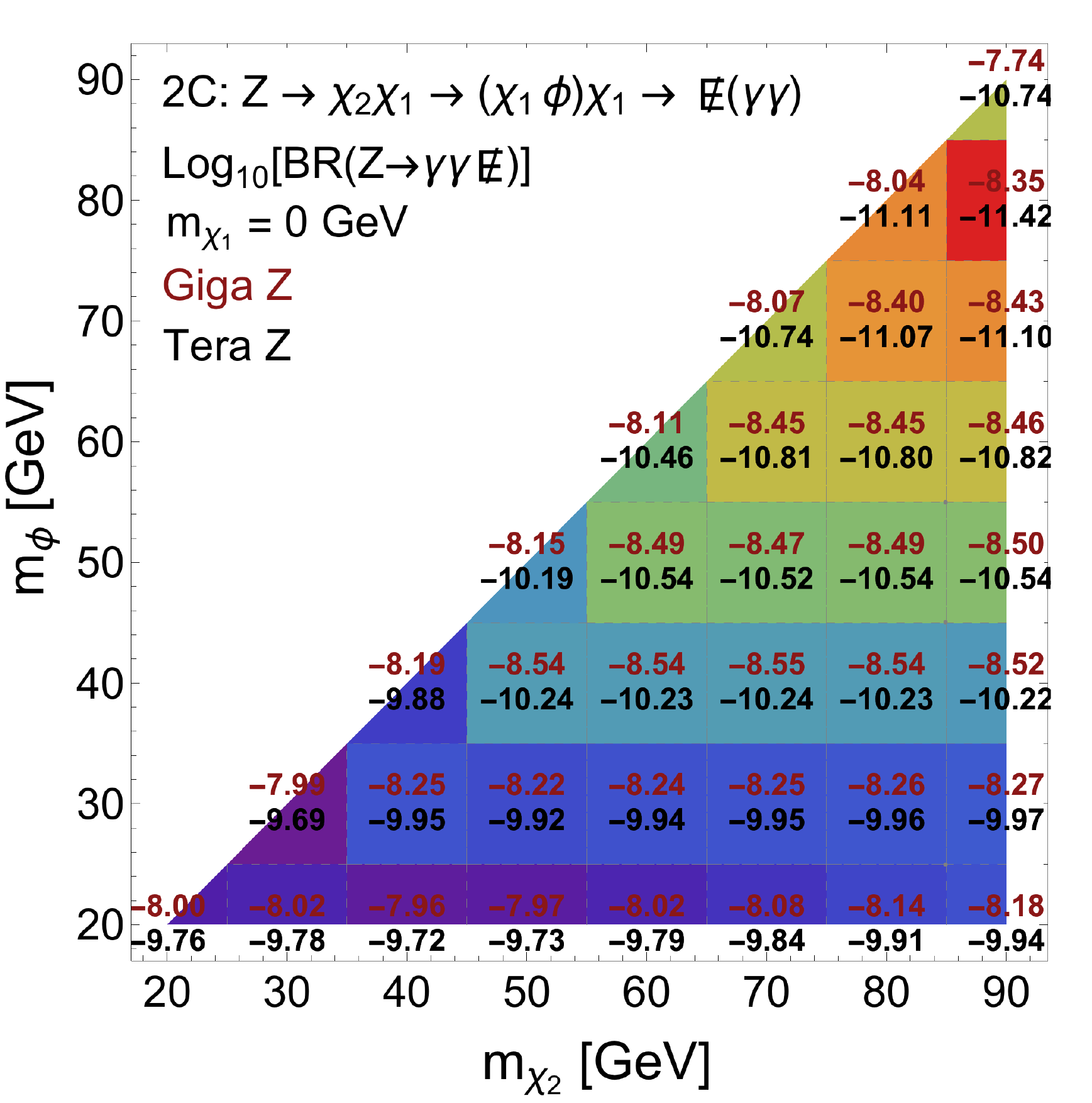} &
    \includegraphics[width=0.33\textwidth]{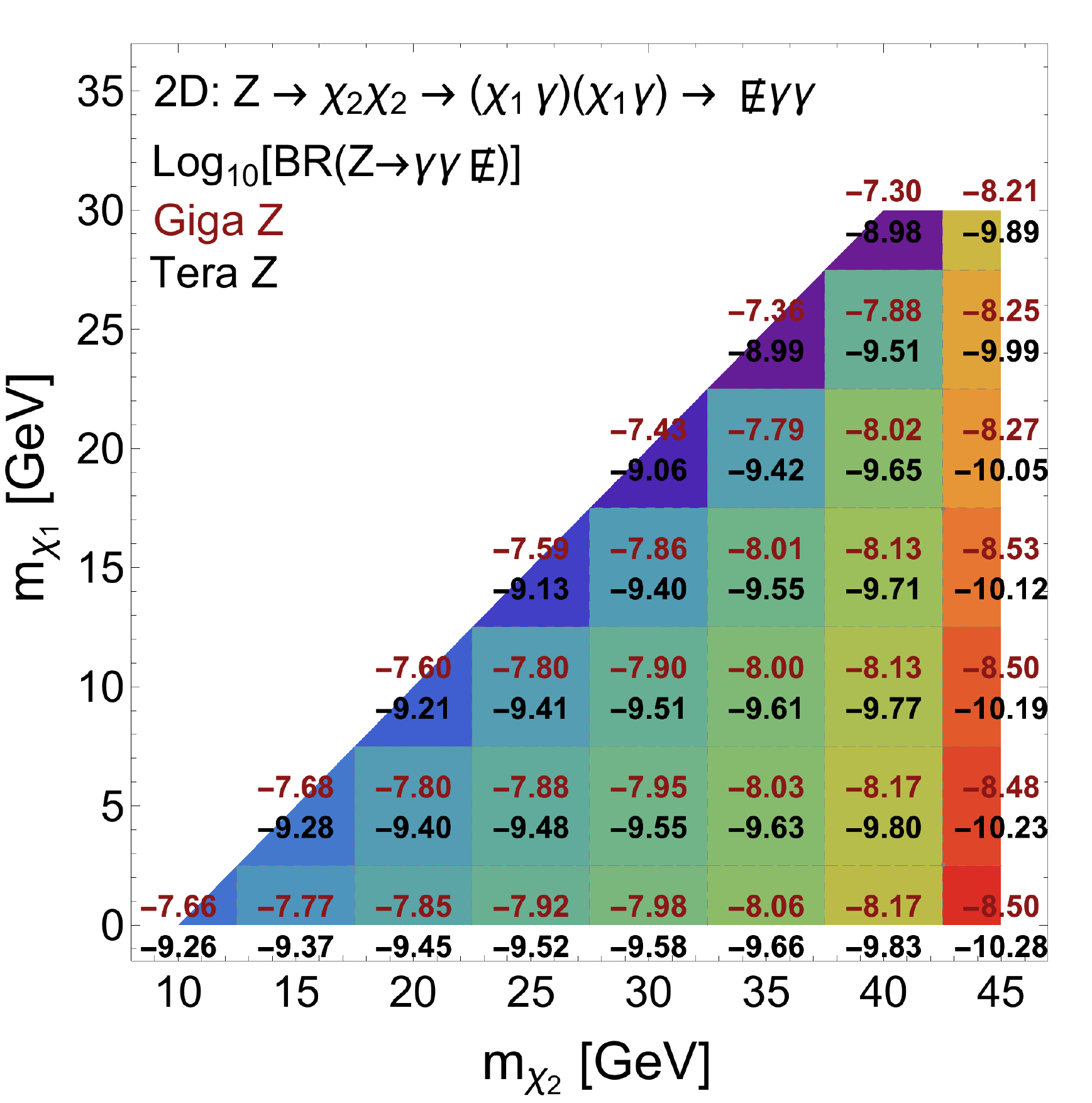} \\
      (a) &  (b) & (c)
  \end{tabular}
  \caption{ The $95\%$ C.L. exclusion on exotic Z decay BR for the final 
  state $Z \to \slashed{E} \gamma \gamma$. 
  The numbers in each block are the sensitivity reach for the exotic Z 
  decay BR in $\text{log}_{10}$
  for Giga Z and Tera Z respectively, and the color coding is based on Tera Z.
  (a): the decay topology 2A, $Z\to \phi_d A' \to (\gamma \gamma) (\bar \chi\chi)$
  from vector portal model. 
  (b): the decay topology 2C, $Z\to \chi_2 \chi_1 \to \phi \chi_1 \chi_1 \to (\gamma \gamma)
  \slashed{E} $ from IDM embedded in the vector portal model.  
  (c): the decay topology 2D, $Z\to \chi_2 \chi_2 \to \gamma \gamma \chi_1 \chi_1 $
  from MIDM model. 
} 
  \label{fig:ZtoMET2gamma}
\end{figure}

In this section, we focus on the exotic Z decay to the final states $ \slashed{E}+ 
\gamma\gamma$. The decay topologies can be classified by number of
resonances. SM background for this final state is coming from 
$e^+ e^- \to \gamma \gamma \nu \bar \nu$. The general feature
of the background is the same as $\gamma \nu \bar \nu$, where the photons dominantly 
come from ISR and tend to be \textit{soft}. 

For topologies with 2 resonances, the first one is the topology 2A, 
$Z\to \phi_d A' \to (\gamma \gamma) (\bar \chi\chi)$, where $A'$
is a vector boson which decays into a pair of DM and $\phi_d$ is a 
scalar which decays into a pair of photons. It can be motivated by the vector portal 
model in \cref{sec:VportalSdm}. 
The dark Higgs $\phi_d$ decays to diphoton via SM Higgs mixing, or by the loop of heavy vector-like charged particles. The dark photon
decays to fermionic DM which is charged under this $U(1)'$.

The second topology with 2 resonances is the topology 2B.
$Z\to \phi_A \phi_H \to (\bar \chi\chi) (\gamma \gamma)$, where $\phi_A$ and $\phi_H$ are CP odd and CP even scalar respectively.
The topology 2B can be well motivated by the two Higgs doublet model (2HDM).
The CP even scalar $\phi_H$ is the mixture of CP even scalars in the 2HDM, 
and can decay to diphoton via loop. For the CP odd scalar $\phi_A$, decaying to $\bar{\chi} \chi$, one needs to add a singlet CP odd scalar $\phi_a$
which couples to DM via $i \phi_a  \bar{\chi}\gamma^5 \chi$. The $\phi_a$
can further couples to scalars by $i \phi_a H_1^\dag H_2 + h.c.$~\cite{Ipek:2014gua},
where $H_{1,2}$ are the doublet Higgs in 2HDM.
After working out the mass eigenstate, $\phi_A$ is the mixture of
singlet CP odd scalar $\phi_a$ and doublet CP odd scalar in $H_{1,2}$. As
a result, it can have the decay topology $Z \to \phi_A \phi_H$ and $\phi_A$ 
can further decay to $\bar{\chi} \chi$.

Since the topology 2A and 2B has the same kinetic feature, the
sensitivity to them are similar. Due to the similarity,
we take the topology 2A as an example.
With the presence of  two resonances in $\gamma\gamma$ and $\bar \chi \chi$,
we propose to use the following cuts besides the fiducial selection,
\begin{align}
\rm{2A}: \quad   \left| m_{\gamma \gamma} - m_{\phi_d}  \right| < 2.5 ~\text{GeV}, 
\quad \left| m_{inv} - m_{A'}  \right| < 2.5 ~\text{GeV} \,.
\end{align}
Note our invariant mass window cut for diphoton $\gamma\gamma$ and 
missing energy $\bar \chi \chi$ are conservative. The resolution for
diphoton invariant mass is about 0.5 GeV at LEP \cite{Abbiendi:2002je}.
The invariant mass of missing energy is determined by the energy resolution
of the diphoton system, which should be smaller than $\lesssim 2$ GeV according
to \cref{eq:photonRES}.

For the topology with 1 resonance, we have the topology 2C,
$Z\to \chi_2 \chi_1$, with the subsequent decays of $\chi_2 \to \chi_1 \phi_d \to \chi_1 
(\gamma \gamma) $, where $\chi_{1,2}$ are the light and heavy DM, and $\phi_d$ 
is a scalar. This topology can be realized by either MIDM model in 
\cref{sec:MIDMandRayDM} or IDM embedded in vector model in \cref{sec:inelasticFdm}.
Since there is a resonance in $\gamma\gamma$,
one can propose the following cuts besides the pre-selection cuts,
\begin{align}
\rm{2C}: \quad   \left| m_{\gamma \gamma} - m_{\phi_d}  \right| < 2.5 ~\text{GeV}, 
\quad m_{\rm{inv}} > 2 m_{\chi_1} \,.
\end{align}

For the topology with 0 resonance, we have 2D,
$Z\to \chi_2 \chi_2$, with the subsequent decay of $\chi_2 \to \chi_1 \gamma$. 
This topology can be motivated by an extended MIDM model as explained in
\cref{sec:MIDMmodelDetail}. From the event topology, the two photons in 
final state has no resonance feature. However, the photon energy distribution has a 
box shape similar to model 1A. The topology dictates the energy range of both photons,
\begin{align}
E_\gamma^{\rm{max},2D} & =\frac{m_2^2-m_1^2}{4 m_2^2} \left(\sqrt{s} 
+ \sqrt{s-4 m_2^2}   \right) \\
E_\gamma^{\rm{min},2D} & =\frac{m_2^2-m_1^2}{4 m_2^2} \left(\sqrt{s} 
- \sqrt{s-4 m_2^2}   \right) \,.
\end{align}
Therefore, we propose the following cuts besides the pre-selection cuts for model 2D,
\begin{align}
\rm{2D}: \quad    E_\gamma^{\rm{max},2D} > E_\gamma > 
E_\gamma^{\rm{min},2D}, 
\quad m_{\rm{inv}} > 2 m_{\chi_1} \,.
\end{align}

In \cref{fig:ZtoMET2gamma},
we show the $95\%$ C.L. exclusion on exotic Z decay BR for the final 
state $Z \to \slashed{E} \gamma \gamma$. In panel (a) of \cref{fig:ZtoMET2gamma} 
with two resonances and $m_{\chi_1} = 0$ GeV, 
the SM background events are so suppressed that the event number
is typically smaller than 1. As a result, the sensitivity does not scale as $S/\sqrt{B} \sim 
\sqrt{L}$, that the scaling is linear with $\sim L$. This behavior
can be seen from panel (a) in \cref{fig:ZtoMET2gamma} that BR sensitivity of Giga Z is 
around $[10^{-8.4}, 10^{-6.7}]$ and $[10^{-11}, 10^{-9.7}]$ for Tera Z.
The sensitivity difference of these two are about $10^{-2} - 10^{-3}$. 
The best sensitivity appears near the region where $m_{A'} \geq 10~\rm{GeV}$ because of
the pre-selection cut  $E_{\rm{inv}} > 10 ~ \rm{GeV}$. The sensitivity gets better 
when $m_{\phi_d}$ becomes large since the photon becomes more energetic and 
the SM background becomes smaller.
In panel (b) of \cref{fig:ZtoMET2gamma}, we assume $m_{\chi_1} = 0$. 
With only one resonance,  one should expect the sensitivity of figure (b) 
to be weaker than the sensitivity in figure (a) with two resonances. 
We do see this point that sensitivity for figure (a) is better than
figure (b) at the same scalar mass  $m_{\phi_d} = m_{\phi}$ for Tera Z case, but
not for Giga Z.
We have looked into the cut efficiency of signal and background, which explains such
behavior. The cut efficiencies for signal in (a) and (b) are about the same order
$\mathcal{O}(0.1 - 0.8)$. But the figure (b) has slightly larger efficiency than (a), 
because in figure (b) the scalar $\phi$ is easier to get larger energy share by 
competing with massless $\chi_1$ while in figure (a) the scalar $\phi_d$ needs to
compete with massive $A'$. 
For figure (a) and (b), they have the same SM background. The background efficiency 
are $\mathcal{O}(10^{-4})$ and $\mathcal{O}(10^{-2})$ for (a) and (b) respectively, 
which shows that the resonance condition for missing energy does help to reduce the SM background.
In Giga Z case, the background event is already smaller than 1 for figure (b), therefore
it has slightly better sensitivity than figure (a) due to higher signal acceptance. For the case of
Tera Z, the increased luminosity has brought back the needs to suppress the SM background,
therefore figure (a) has better sensitivity than (b).  
In panel (c) of \cref{fig:ZtoMET2gamma}, the limits on exotic Z decay BR is not as good as 
figure (a) and (b), because there is no resonance feature in the topology. 
However, the constraints can still reach $[10^{-8.4}, 10^{-7.4}]$ for Giga Z and 
$[10^{-10.3}, 10^{-9.2}]$ for Tera Z.

For the panels (a), (b) and (c) in \cref{fig:ZtoMET2gamma}, one might expect the
sensitivity on BR decreases because the number of resonance $n_{res}$ decreases.
This is clearly true when comparing $n_{res} = 1, 2$ with $n_{res} =0$. However,
for $n_{res} = 2$ and $n_{res} =1$, the difference in sensitivity is not
very significant, while the sensitivity relies more on the particle mass and the cuts.
For example, the best sensitivity for 2A appears when $m_{A'} \sim 15$ GeV and
$m_{\phi_d} \sim 60$ GeV. The higher  $m_{\phi_d}$ the higher photon energy,
however one should also keep $m_{A'}$ large enough to pass the missing energy
cut $\slashed{E} > 10$ GeV. The best sensitivity for 2C appears when 
$m_{\chi_2} \sim 90 $ GeV and $m_{\phi_d} \sim 80$ GeV if fixing
$m_{\chi_1} =0$ GeV. This high $m_{\phi_d} $ mass can guarantee a harder
photon spectrum than 2A. Therefore, even without the resonance cut on $\slashed{E}$,
 the SM background of 2C is similar to that of 2A, making the sensitivities on BR are similar.

\subsection{$Z\to \slashed{E}+ \ell^+ \ell^-$ }
\label{sec:ZtoMET2l}

In this section, we focus on the exotic Z decay to final state $\slashed{E} + \ell^+ \ell^-$. 
The SM background for this final state is coming from $\ell^+\ell^- \bar{\nu}\nu$, mediated 
by off-shell gauge boson $\gamma^*$, $Z^*$ and $W^*$. Comparing with ISR photon, the 
energy spectrum of leptons are harder. And the spectrum of invariant mass of 
$\ell^+ \ell^-$ is softer than that of $\bar{\nu} \nu$. 
Given the fact that when mediated by $W^*$, $\ell$ and $\nu$ are sharing similar kinetic
distribution, this can not lead to difference in invariant mass. The difference 
is originated from that $\ell^+ \ell^-$ can be
produced from $\gamma^*$ favoring smaller invariant masses, while $\bar{\nu} \nu$ from 
$Z^*$ has much harder spectrum because small $m_{\rm{inv}}$ are suppressed by a factor 
$m_{\rm{inv}}^4/ m^4_Z$.

\begin{figure}[!ht]
	\begin{tabular}{c c}
		\includegraphics[width=0.47\textwidth]{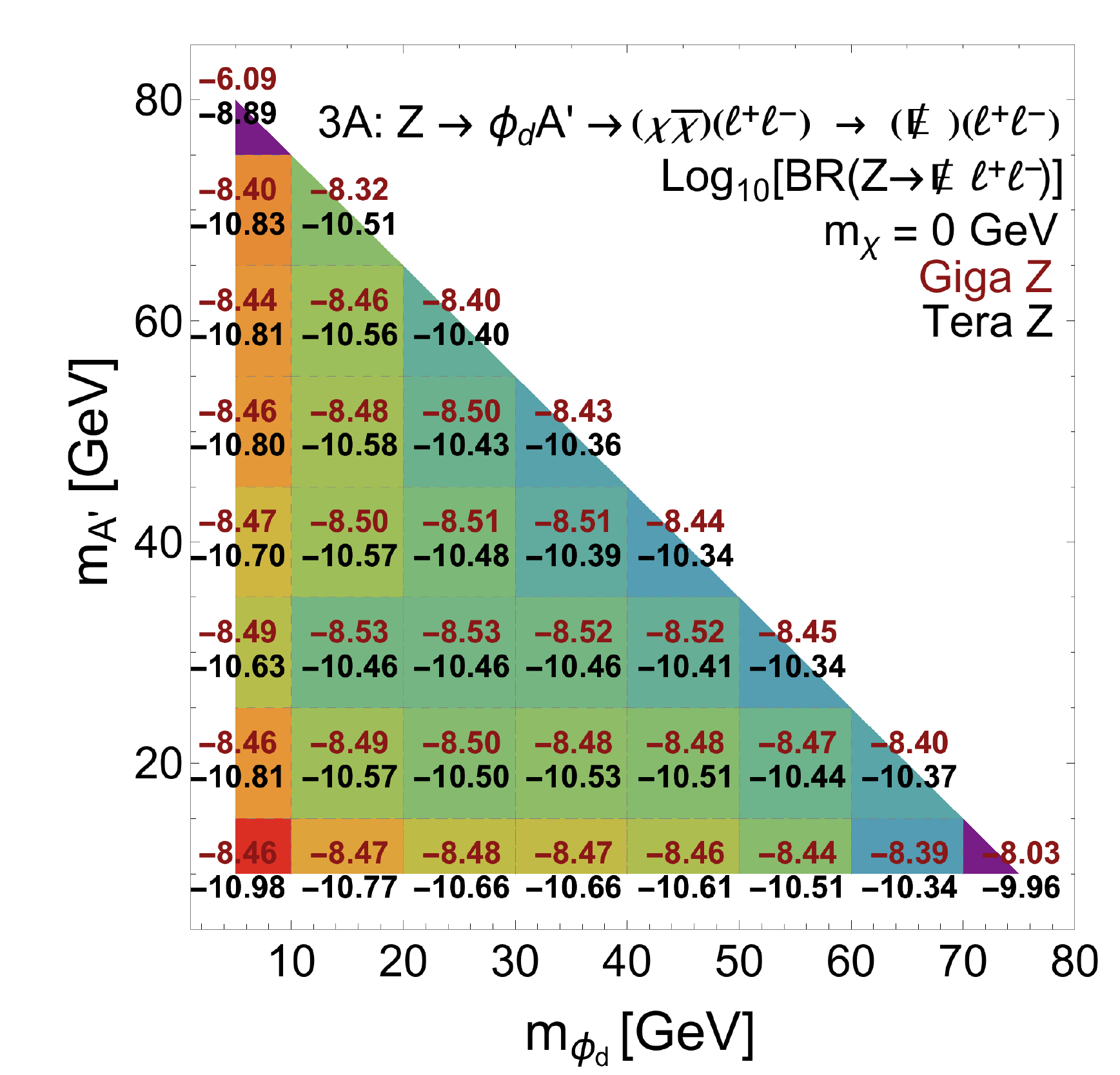} &
		\includegraphics[width=0.47\textwidth]{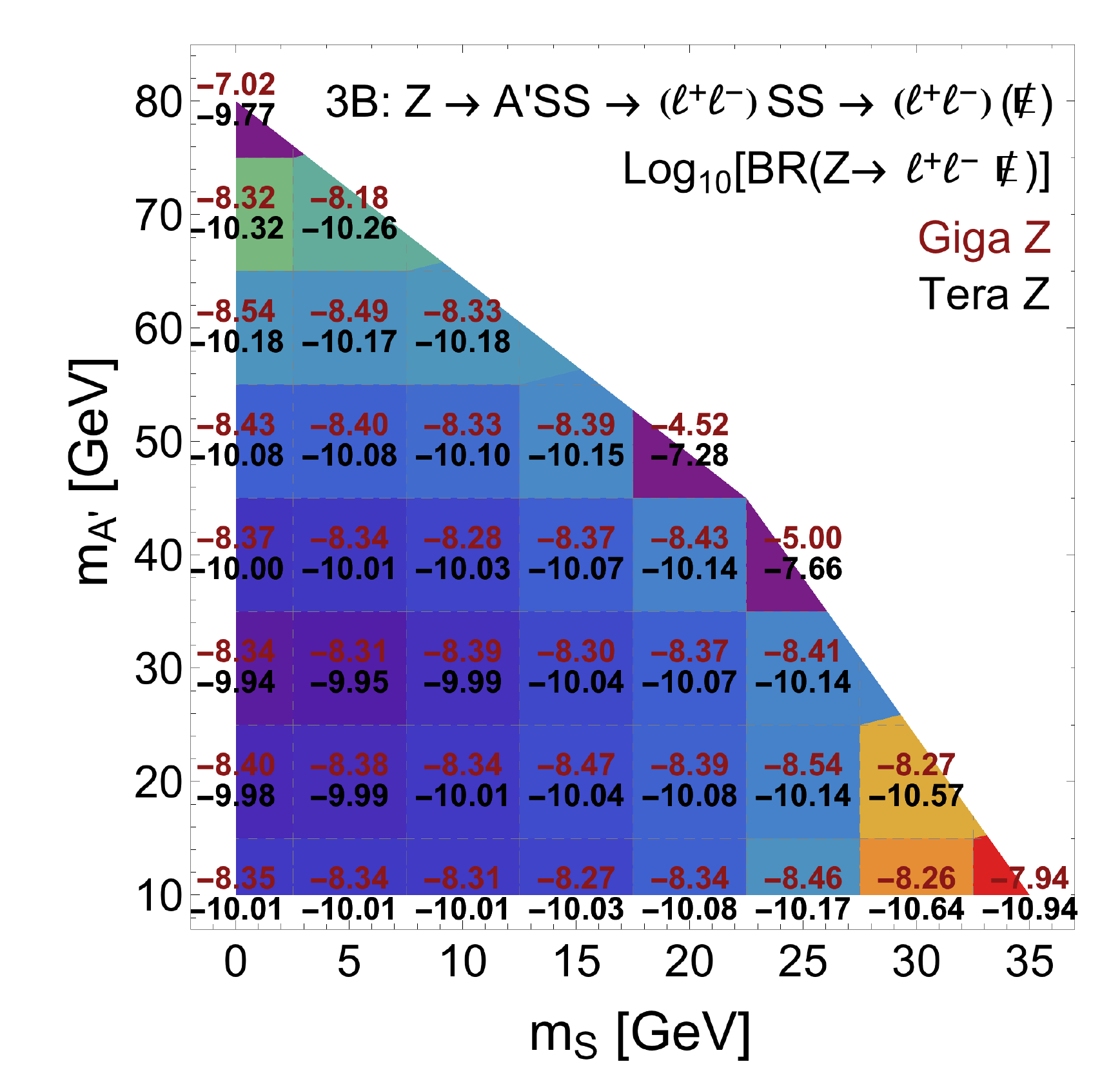} \\
		(a) &  (b) \\
		\includegraphics[width=0.47\textwidth]{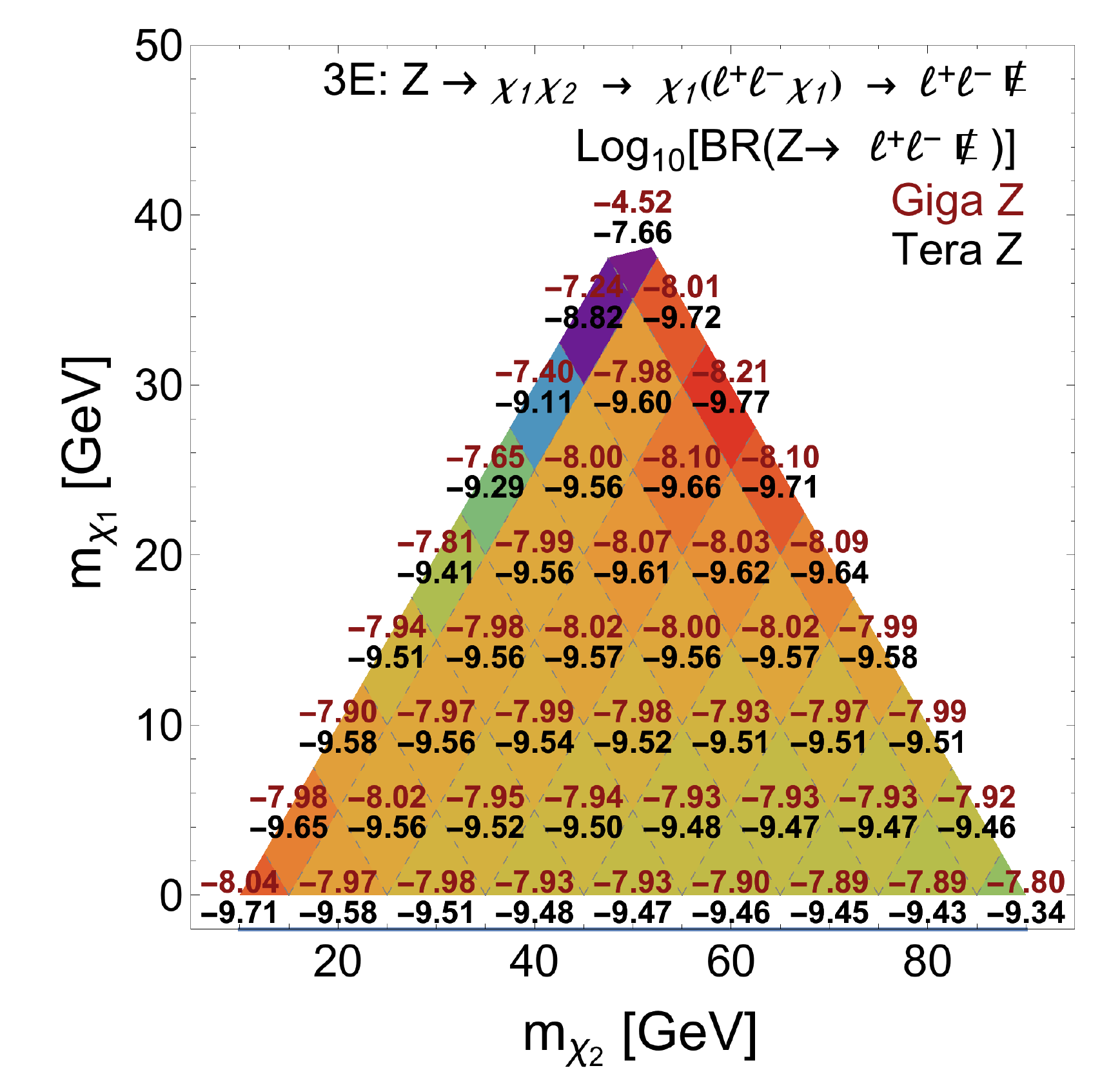} &
		\includegraphics[width=0.47\textwidth]{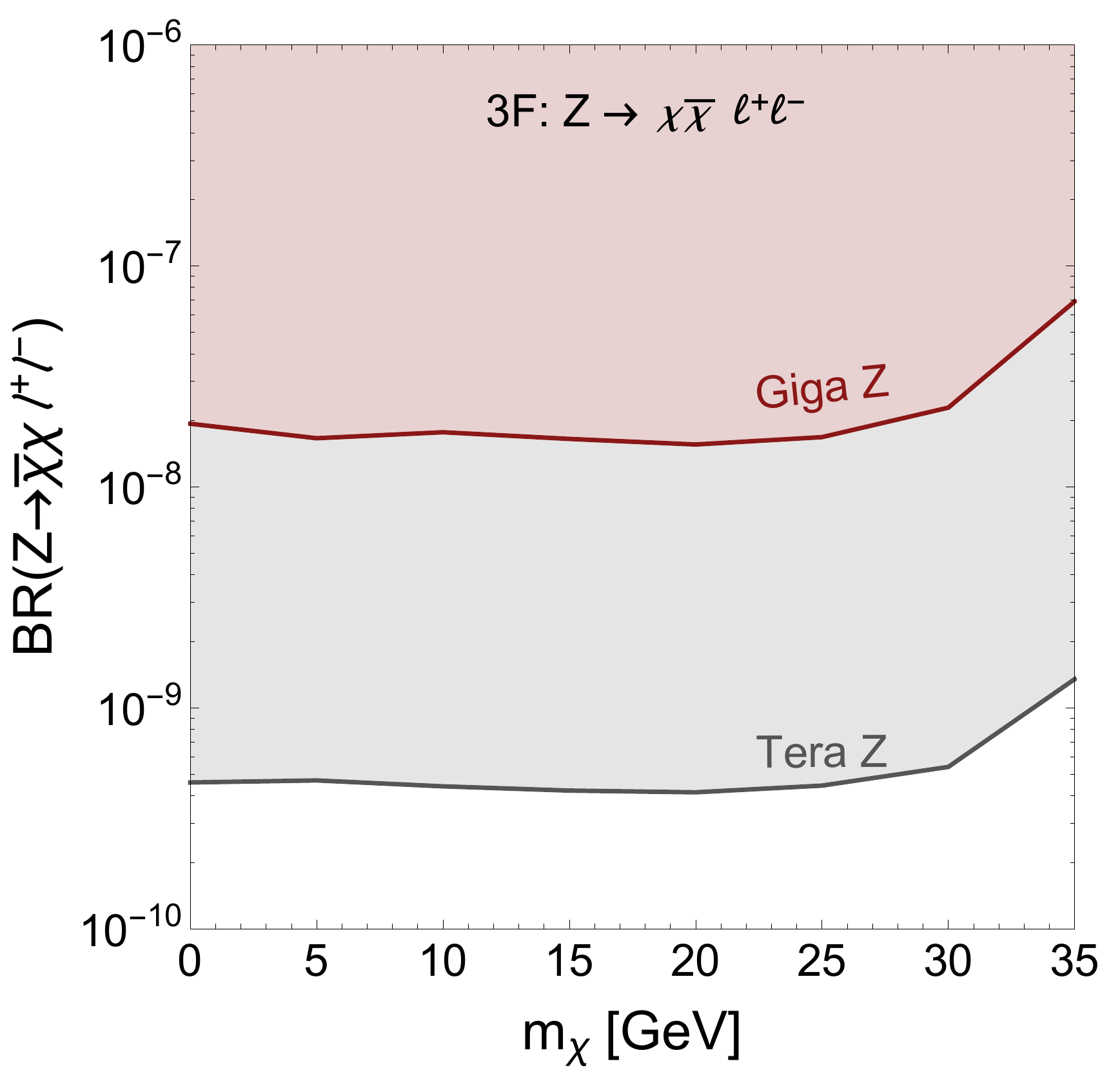} 
		\\
		(c) & (d)
	\end{tabular}
	\caption{ The $95\%$ C.L. exclusion on exotic Z decay BR for the final 
		state $Z \to \slashed{E} \ell^+ \ell^-$. 
		The numbers in each block are the sensitivity reach for the exotic Z 
		decay BR in $\text{log}_{10}$
		for Giga Z and Tera Z respectively, while the color mapping is coded for Tera Z.
		(a): the decay topology 3A, $Z\to \phi_d A' \to (\bar{\chi} \chi)(\ell^+\ell^-)$
		from Vector portal model. The numbers in each block are reaches for exotic Z decay BR in 
		$\text{log}_{10}$ for Giga Z and Tera Z. 
		(b): the decay topology 3B, 3-body decay $Z\to A' S^* S \to (\ell^+ \ell^-) 
		\slashed{E}$ from Vector portal model with scalar DM $S$.  
		(c): the decay topology 3E, a cascade decay $Z\to \chi_2 \chi_1$, with 
		subsequent decay $\chi_2 \to \chi_1 Z^*/\gamma^* \to \chi_1 \ell^+ \ell^- $
		motivated by MIDM operator. 
		(d): the decay topology 3F, a 4-body decay process 
		$Z \to \bar{\chi} \chi (Z^*/\gamma^*) \to \bar{\chi} \chi \ell^+ \ell^-$, 
		which can be motivated from RayDM operator.
	}
	\label{fig:ZtoMET2l}
\end{figure}

We have listed six topologies with the number of resonances from 
2 to 0 in \cref{tab:finalstate}. The event topology with 2 resonances is 3A, 
$Z\to \phi_d A'$, with subsequent decays $A'\to (\ell^+ \ell^-)$ and 
$\phi_d \to (\bar{\chi} \chi)$. The dark Higgs bremsstrahlung process
can be naturally realized by vector portal model with a dark Higgs in 
\cref{sec:VportalSdm}.

The topologies with 1 resonance are 3B, 3C and 3D. The topology 3B is 
a 3-body decay $Z\to A' S^* S \to (\ell^+ \ell^-) 
  \slashed{E}$, which can be motivated from vector portal 
model with scalar DM in \cref{sec:VportalSdm}. The topology 3C is also a 3-body 
process mediated by off-shell Z or photon, 
$Z \to \phi (Z^*/\gamma^*) \to (\slashed{E}) \ell^+ \ell^-$ 
where $\phi$ is assumed to decay outside of detector. It can be motivated by 
axion-like particle model in \cref{AxionPortal}, or Higgs portal model in \cref{sec:ModelSingletS} 
where $\phi$ is a singlet scalar which mixes with SM Higgs and can decay to DM 
pair $\bar{\chi} \chi$. The topology 3D is a 2-body cascade decay,
$Z\to \chi_2 \chi_1 \to A' \chi_1  + \chi_1\to  (\ell^+\ell^-) + \slashed{E}$,
which  can be motivated by Vector portal and Inelastic DM in \cref{sec:inelasticFdm}.

The topologies without a resonance are 3E and 3F. 
The topology 3E is a cascade decay $Z\to \chi_2 \chi_1$, with subsequent decay
$\chi_2 \to \chi_1 Z^*/\gamma^* \to \chi_1 \ell^+ \ell^- $, where the last step 
is a 3-body decay. Such process can be motivated from MIDM operator in
\cref{sec:MIDMmodelDetail}.
The topology 3F is a 4-body decay process 
$Z \to \bar{\chi} \chi (Z^*/\gamma^*) \to \bar{\chi} \chi \ell^+ \ell^-$, which can
be motivated from RayDM operator in \cref{sec:MIDMmodelDetail}.

We will study the constraints from exotic Z decay in topologies 3A, 3B, 3E
and 3F. They are chosen to represent different $n_{res}$ and number of
particles in the cascade decay, from 2-body to 4-body. Besides the same pre-selection
cuts, we propose the following different cuts for different topologies:
\begin{align}
& \rm{3A}: \quad    |m_{\ell^+ \ell^- } - m_{A'}| < 2.5 ~\rm{GeV}, 
\quad |m_{\rm{inv}} - m_{\phi_d}| < 2.5 ~\rm{GeV} \,, \\
& \rm{3B}:  \quad    |  m_{\ell^+ \ell^- } -m_{A'} | < 2.5 ~\rm{GeV}, \quad     
2m_S<m_{\rm{inv}} < m_Z-m_{A'} \,, \\
& \rm{3E}:  \quad    m_{\ell^+ \ell^- }  < m_{\chi_2} - m_{\chi_1}, \quad     
m_{\rm{inv}} >  2 m_{\chi_1} \,, \\
& \rm{3F}:  \quad    m_{\ell^+ \ell^- }  < 20  ~\rm{GeV}, \quad     
m_{\rm{inv}} >  2 m_\chi \,.
\end{align}

In \cref{fig:ZtoMET2l}, we show the constraints on exotic Z decay branching 
ratio $\rm{BR}(Z\to  \slashed{E} \ell^+ \ell^- )$. 
For Giga Z, the topologies with $n_{res} > 0 $ will probe exotic Z decay
BR down to $\sim 10^{-8.5}$, while for $n_{res} =0$ the sensitivity of
BR can reach $\sim 10^{-8}$. With more resonances, 
the SM background events are suppressed a lot that the event number
is typically smaller than 1. As a result, the sensitivity 
reach scales as $L$. For $n_{res} > 0 $ in panels (a) and (b) of \cref{fig:ZtoMET2l}, 
the sensitivity on BR between Giga Z and Tera Z differs by factor 
of $10^2 \to 10^3$ due to small number of SM background. While for
$n_{res} = 0 $ in panels (c) and (d), the sensitivity on BR between Giga Z 
and Tera Z differs by factor of $\sim 10^{1.5}$ which is a very typical
scaling from $S/\sqrt{B} \sim \sqrt{L}$.

\subsection{$Z\to \slashed{E}+ JJ$ }
\label{sec:ZtoMETjj}

In this section, we focus on the exotic Z decay to final state $\slashed{E} + JJ$.
The $J$ includes both the light flavor jets $j$ and bottom quark jets $b$.  
The topologies are similar as $Z \to \slashed{E}+ \ell^+ \ell^-$, and the limits
on exotic Z decay BR are calculated through the same procedure. 
The SM background for this final state is dominantly from $ \bar{\nu}\nu+ J J$, 
mediated by off-shell gauge boson $\gamma^*$, $Z^*$ and $W^*$. 

We choose three topologies 4A, 4B and 4C to study the sensitivity reach of exotic Z decay BR.
The topology 4A is $Z \to \phi_d A' \to (\bar{\chi} \chi) (jj)$, and  
4B is $Z \to \phi_d A' \to (bb)(\bar{\chi} \chi) $. Both topologies can be  
motivated by the vector portal 
model in \cref{sec:VportalDM}. Here we do not use $\phi_d \to jj$ 
because Yukawa coupling is suppressed light quark mass. 
The last topology 4C is $Z \to \chi_2 \chi_1 \to bb \chi_1 + \chi_1 \to bb \slashed{E}$, 
which can be motivated from the MIDM operator in \cref{sec:MIDMmodelDetail}. 
Besides the fiducial cuts, we propose the following cuts for different topologies:
\begin{align}
& \rm{4A}: \quad    |m_{jj } - m_{A'}| < 5 ~\rm{GeV}, 
\quad |m_{\rm{inv}} - m_{\phi_d}| < 5 ~\rm{GeV} \,, \\
& \rm{4B}:  \quad   |m_{bb } - m_{\phi_d}| < 5 ~\rm{GeV}, 
\quad |m_{\rm{inv}} - m_{A'}| < 5 ~\rm{GeV} \,, \\
& \rm{4C}:  \quad    2 m_b <m_{bb}< m_{\chi_2}-m_{\chi_1} \,,  m_{\rm{inv}} >  2 m_{\chi_1} \,.
\end{align}

\begin{figure}[ht]
  \begin{tabular}{c c c}
    \includegraphics[width=0.33\textwidth]{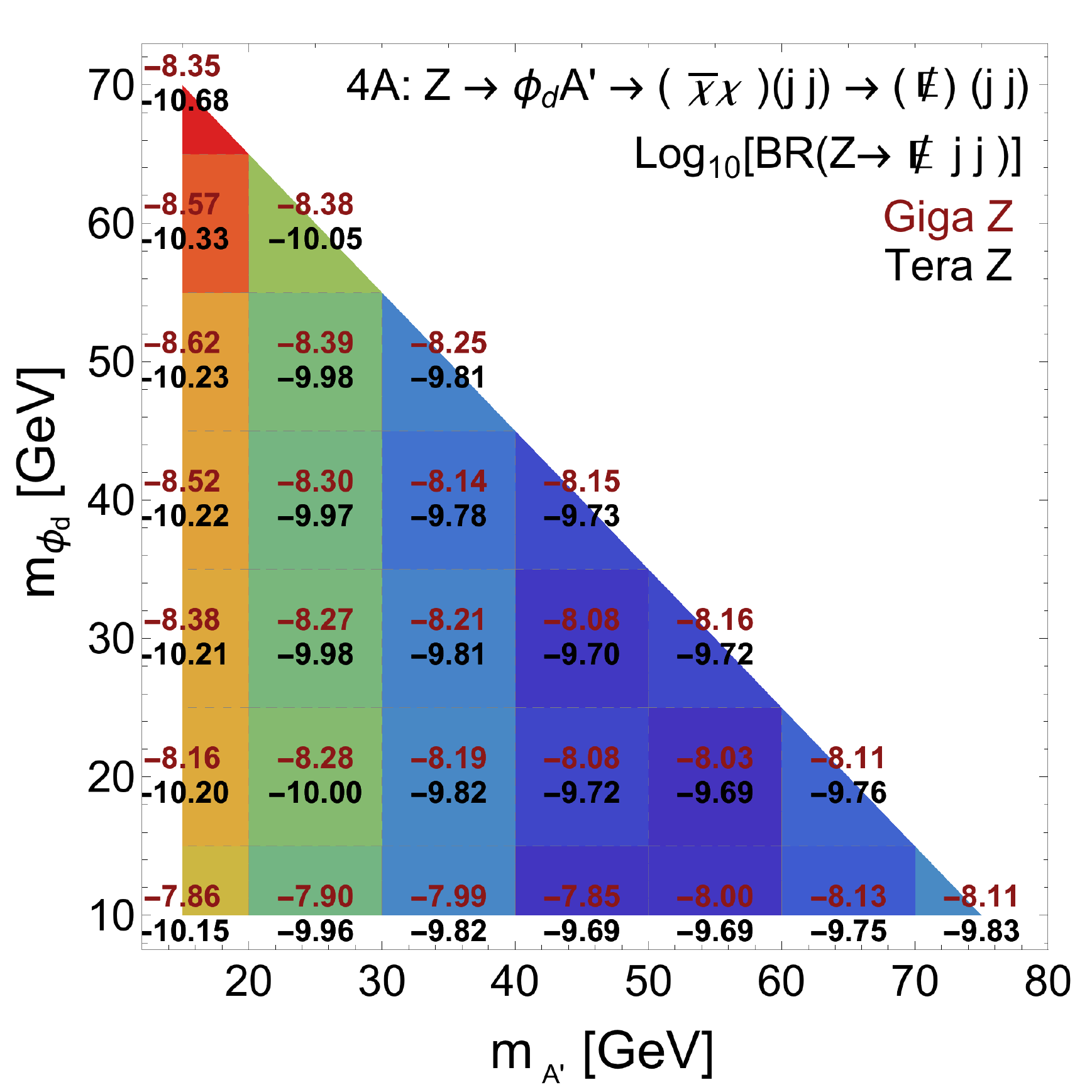} &
    \includegraphics[width=0.33\textwidth]{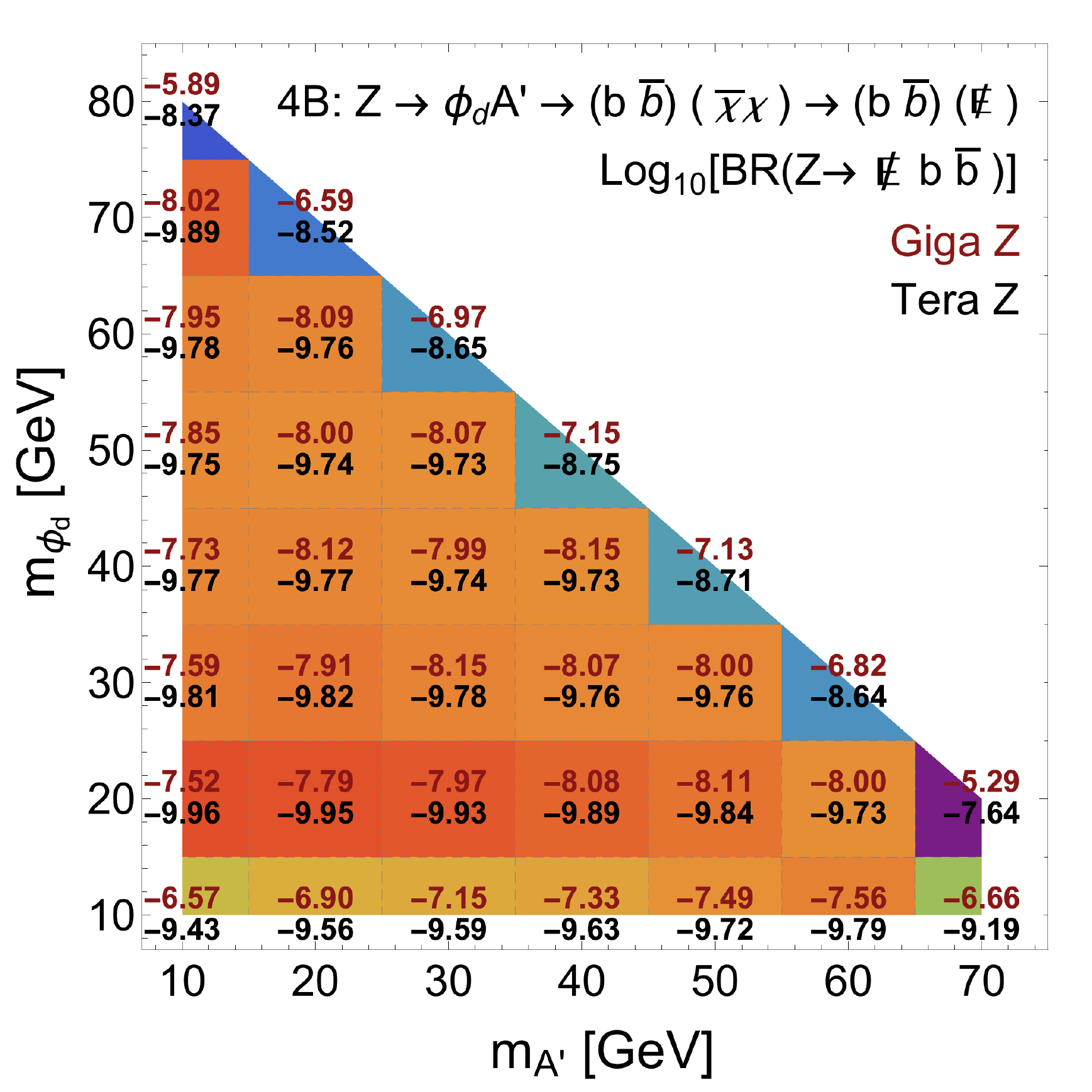} & 
    \includegraphics[width=0.33\textwidth]{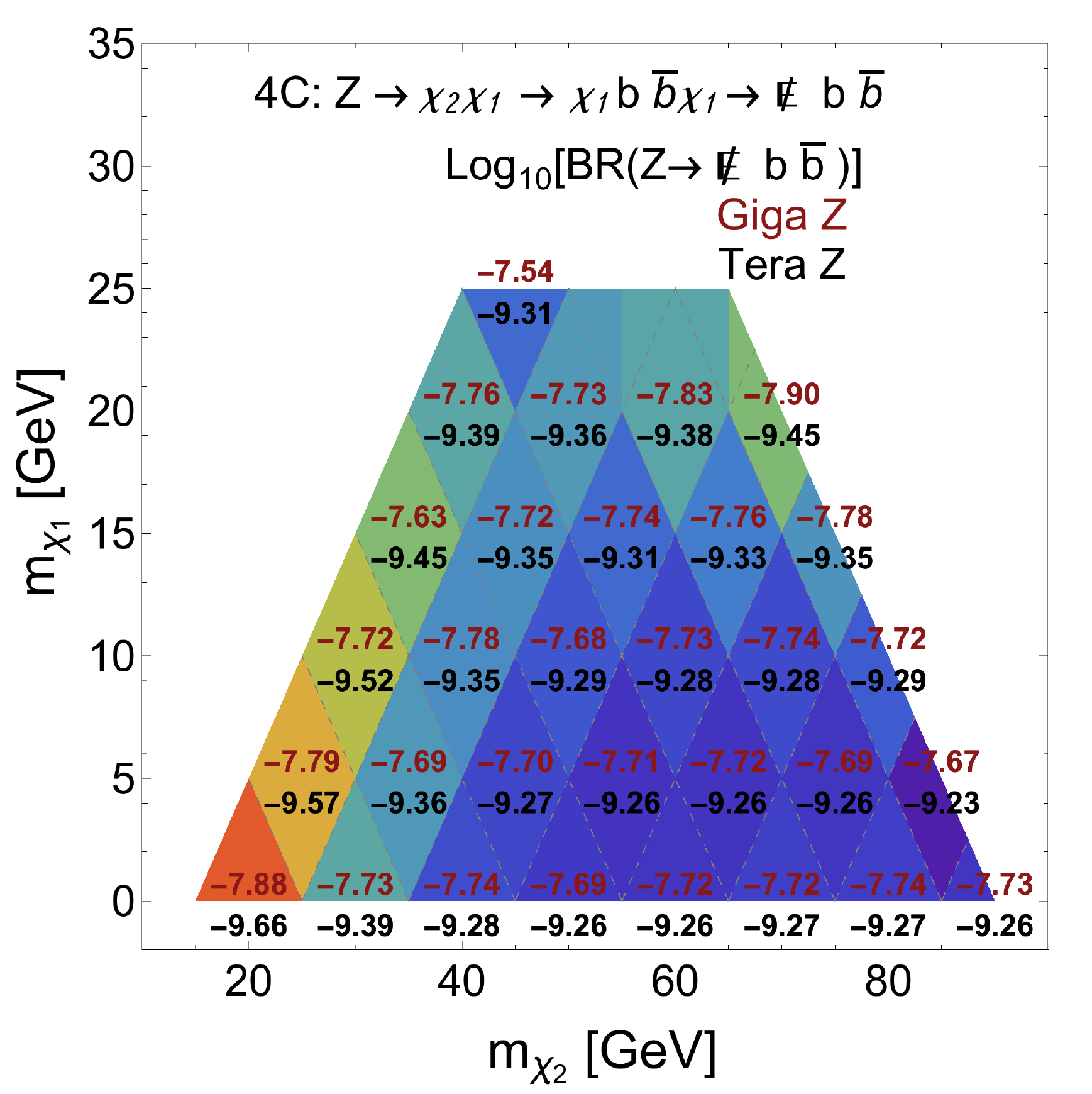} \\
      (a) &  (b)  & (c)
  \end{tabular}
  \caption{ The $95\%$ C.L. exclusion on exotic Z decay BR for the final 
  state $Z \to \slashed{E} JJ$, where $J$ includes both light flavor jet $j$ 
  and bottom quark jet $b$. 
  The numbers in each block are the sensitivity reach for the exotic Z 
  decay BR in $\text{log}_{10}$
  for Giga Z and Tera Z respectively, while the color mapping is coded for Tera Z.
  (a): The decay topology 4A, $Z \to \phi_d A' \to (\bar{\chi} \chi) (jj)$
  from vector portal model. The numbers in each block are reaches for exotic Z decay BR in 
  $\text{log}_{10}$  for Giga Z and Tera Z. 
  (b): The decay topology 4B, $Z \to \phi_d A' \to (bb)(\bar{\chi} \chi) $ from
   vector portal model.  
  (c): The decay topology 4C, $Z \to \chi_2 \chi_1
   \to \chi_1 bb \chi_1 \to \slashed{E} bb$ from MIDM model.}
  \label{fig:ZtoMET2j}
\end{figure}

In \cref{fig:ZtoMET2j}, we show the constraints on exotic Z decay branching 
ratio $\rm{BR}(Z\to  \slashed{E} JJ )$. 
For Giga Z, the exotic Z decay BR can be probed down to $10^{-7}
- 10^{-8}$, while the sensitivity of Tera Z is generally better by 
factor of $\sim 10^{1.5}$ comparing with Giga Z. Comparing the BR sensitivity
of 4A and 4B, we see that the difference between light flavor jet $j$ and heavy 
flavor jet $b$ is not large. One might expect the sensitivity of 
$\slashed{E} (b\bar{b})$ should be better than $\slashed{E} (jj)$, due to smaller
SM background. 
However,  the topologies 4A and 4B are not the same, where in 4A
the jets come from $A'$ while in 4B the b-jets come from $\phi_d$.

\subsection{$Z\to (JJ)(JJ)$}

\begin{figure*}[h]
\centering
 \includegraphics[width=0.32\textwidth]{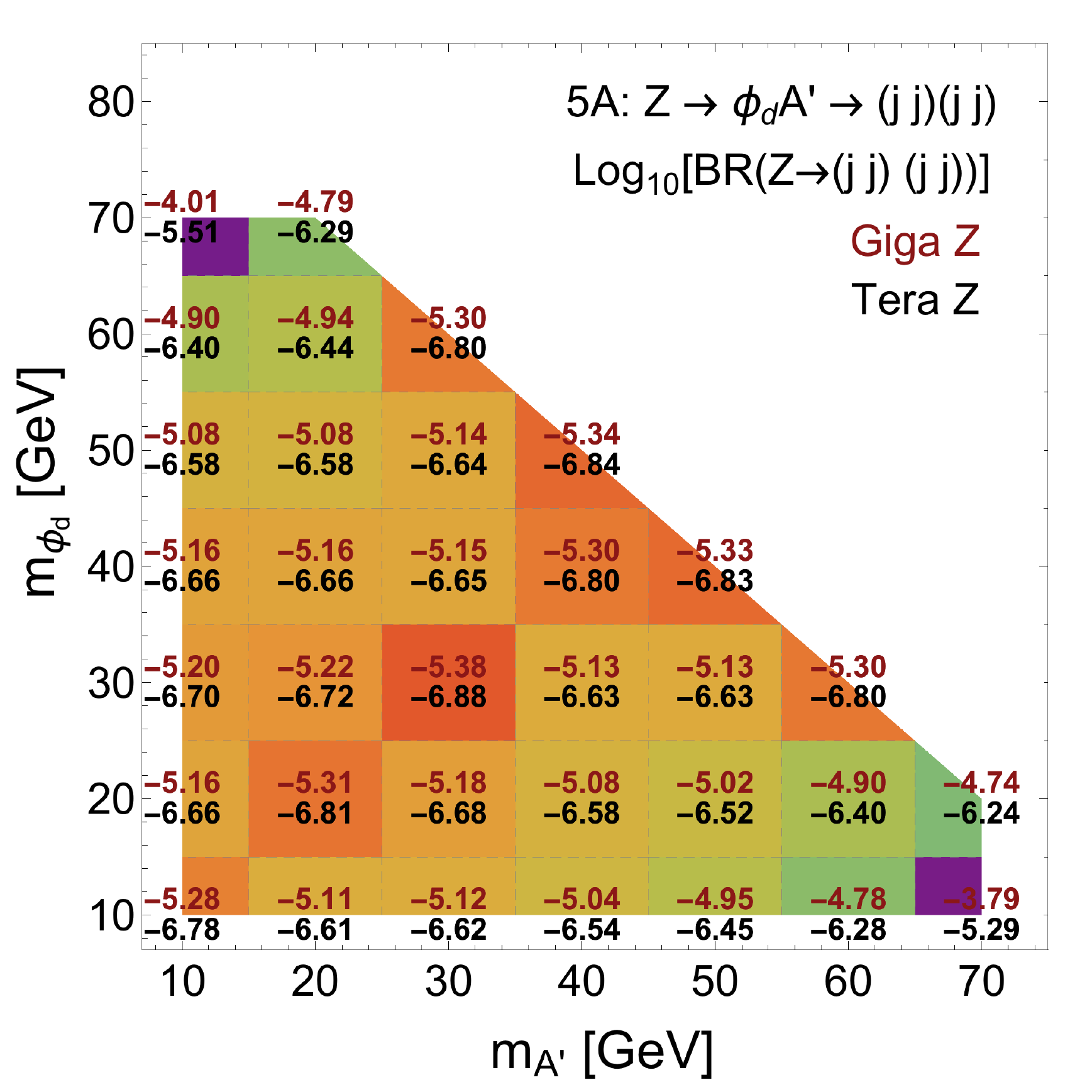} 
 \includegraphics[width=0.32\textwidth]{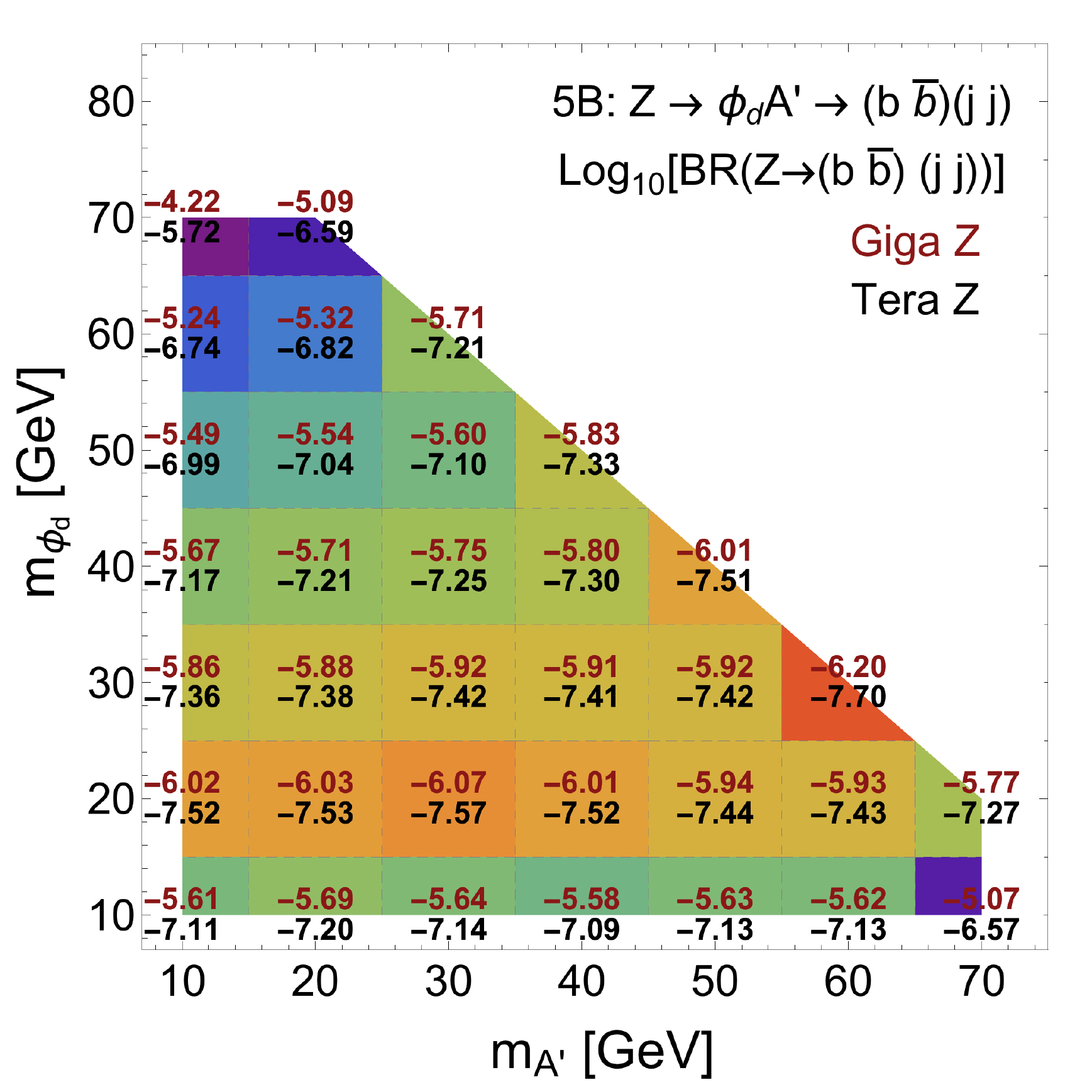} 
 \includegraphics[width=0.32\textwidth]{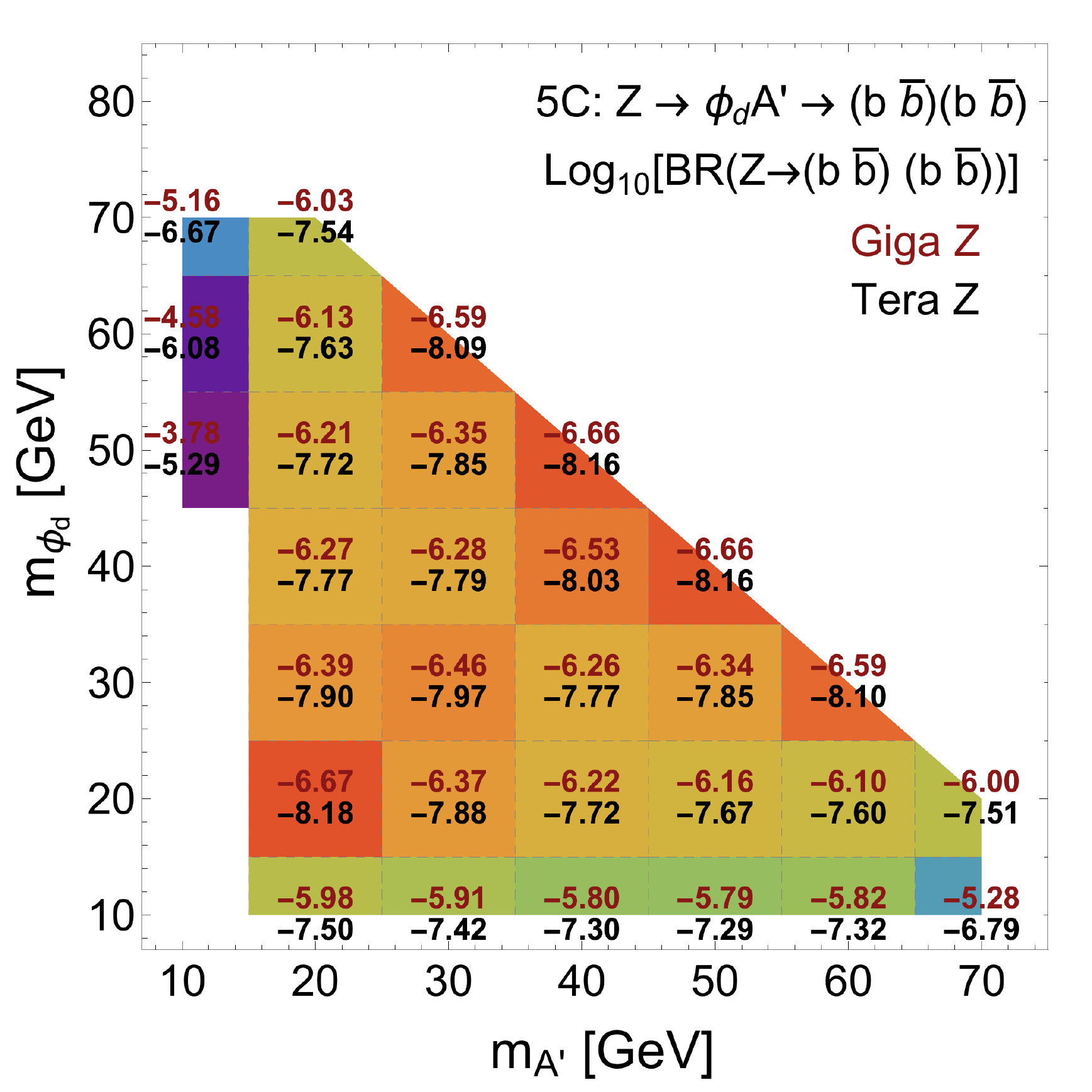} 
  \caption{The $95\%$ C.L. sensitivity for exotic Z decay $Z \to (JJ)(JJ)$, where $J$ 
  	could be light flavor jet or b-jet. 
  	The numbers in each block are the sensitivity reach for the exotic Z 
  	decay BR in $\text{log}_{10}$
  	for Giga Z and Tera Z respectively, while the color mapping is coded for Tera Z.
  The decay process is $Z \to \phi_d A'$ with subsequent decays $\phi_d \to j j$ 
  and $A' \to jj$. We show three combination $(jj)(jj)$, $(jj)(bb)$ and $(bb)(bb)$
  in the figure.}
  \label{fig:Zto4j}
\end{figure*}

In this section, we focus on the exotic Z decay to final state $(JJ) + (JJ)$.
Note that we only discuss the cases where there are two jet resonances in
the final states. The SM background for this final state are mostly from 
electroweak process, mediated by off-shell gauge boson $\gamma^*$, 
$Z^*$ and $W^*$.

In \cref{tab:finalstate}, we have listed the topologies. 5A could
be motivated from Higgs bremsstrahlung in vector and scalar portal model.
We will choose the topology 5A to illustrate the sensitivity
to the BR of $(JJ) + (JJ)$ final state. We divide the final states with three
combinations $(jj)+(jj)$, $(jj)+(bb)$ and $(bb)+(bb)$, where the last two are denoted as 5B and 5C. 
There could be other topologies like $Z \to \phi_A \phi_H \to (JJ)(JJ)$
from 2HDM, but the topology and kinematics are similar, 
therefore their sensitivity should be similar to 5A. 
Beside the pre-selection cuts, we add the following similar cuts 
for the topology 5A, 5B and 5C:
\begin{align}
& \rm{5A}: \quad    |m_{jj } - m_{A'}| < 5 ~\rm{GeV}, 
\quad |m_{jj} - m_{\phi_d}| < 5 ~\rm{GeV}   \\
& \rm{5B}: \quad    |m_{jj } - m_{A'}| < 5 ~\rm{GeV}, 
\quad |m_{bb} - m_{\phi_d}| < 5 ~\rm{GeV}   \\
& \rm{5C}: \quad    |m_{bb} - m_{A'}| < 5 ~\rm{GeV}, 
\quad |m_{bb} - m_{\phi_d}| < 5 ~\rm{GeV}   
\end{align}
The $\chi^2$ method are employed to determine which pair of jets are from
$A'$ decay or $\phi_d$ decay. The mass window that we take is conservative.
For example, at $E_j = 40$ GeV, the jet energy resolution is about $5\%$
leading to $\Delta E_j = 2$ GeV from \cref{eq:jetRES}.

In \cref{fig:Zto4j}, we show the constraints on exotic Z decay branching 
ratio $\rm{BR}(Z\to  (JJ) (JJ))$. For Giga Z, the exotic Z decay BR can be 
probed down to $\sim 10^{-5}$ for $(jj)(jj)$ final state, $\sim 10^{-6}$ for $(jj)(bb)$
and $10^{-6.5}$ for $(bb)(bb)$. The sensitivity of Tera Z is generally better by 
factor of $\sim 10^{1.5}$ comparing with Giga Z, from the integrated
luminosity scaling $S/\sqrt{B} \approx \sqrt{L}$. It is clear that
the sensitivity for heavy flavor jet is slightly better than light flavor jet.
This is because the heavy flavor jet has fewer SM background events,
by a factor of $ N_f^{1/2} \approx 10^{0.5}$, where $N_f$ is the number of
flavor in jets.

\subsection{$Z\to \gamma\gamma\gamma$}
\label{sec:ZTo3gamma}

\begin{figure*}[h!]
\centering
 \includegraphics[width=0.5\textwidth]{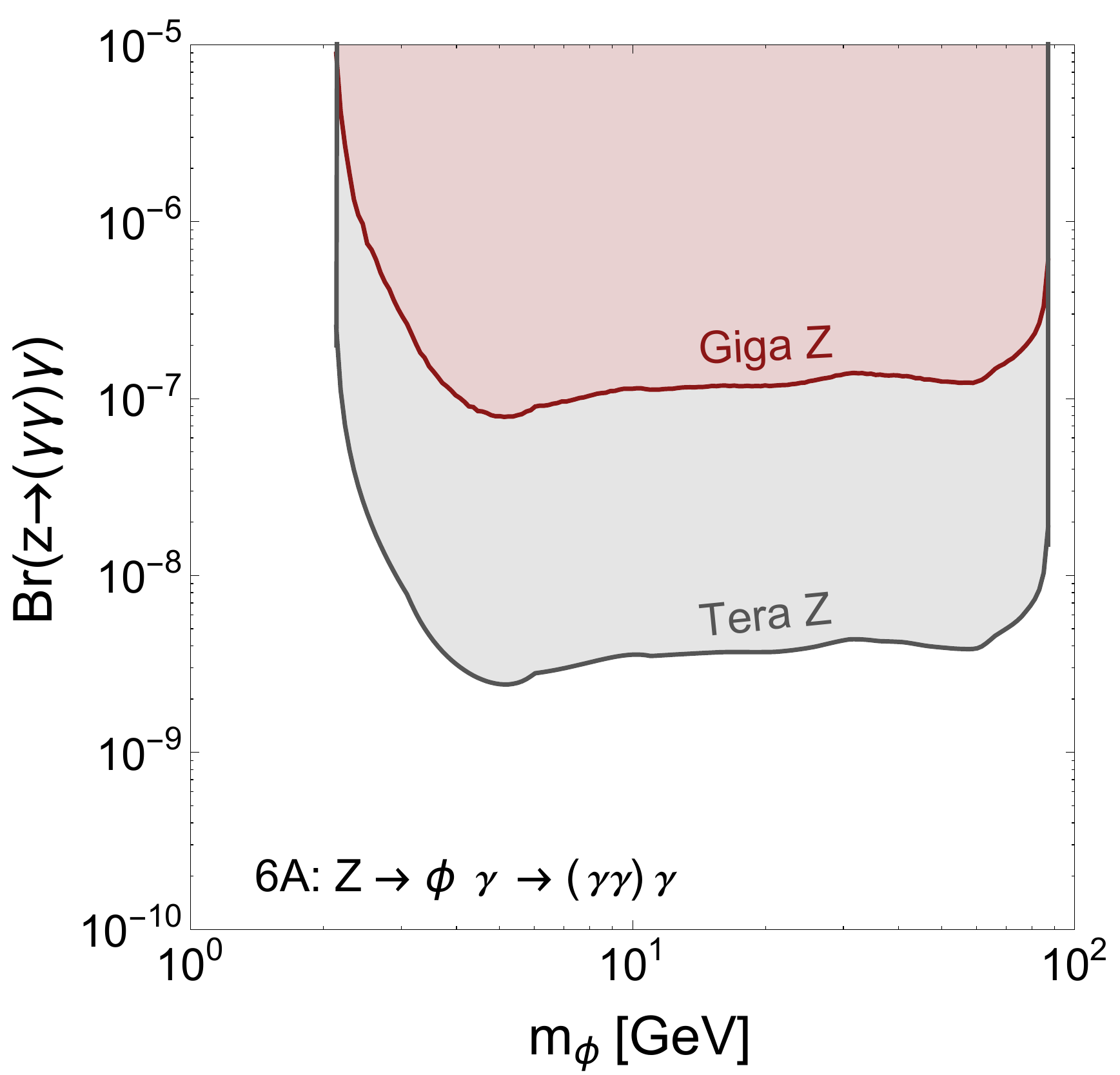} 
  \caption{The $95\%$ C.L. exclusion for exotic Z decay topology 6A, 
  	$Z \to \phi \gamma \to (\gamma \gamma)	\gamma$. The Giga (Tera )Z 
  	exclusion region are shown in dark red (grey) respectively. The $m_\phi < 2$ GeV
  region is not limited due to photon separation failure, but can be constrained by 
  $\gamma\gamma$ search, see \cref{fig:axionconstraint}. }
  \label{fig:Zto3a}
\end{figure*}

In this section, we discuss the exotic Z decay to final state $(\gamma \gamma)
\gamma$. The SM background for this final state $\gamma \gamma \gamma$ 
are dominated by QED process $e^+ e^- \to \gamma\gamma$ with an extra 
$\gamma$ from initial state radiation, therefore the photon energy 
generally tends to be \textit{soft}.
The signal topology 6A in \cref{tab:finalstate}, $Z \to \phi \gamma \to (\gamma \gamma)
\gamma$, could be motivated from axion-like particle, or from Higgs
portal scalar, which can decay to $\gamma\gamma$ from top loop. 
We take axion-like particle as an example in \cref{fig:Zto3a}, and the result should
also apply to Higgs portal scalar. Besides the pre-selection cuts, we propose 
the following cuts for topology 6A
\begin{align}
\rm{6A}: \quad    |m_{\gamma \gamma } - m_{\phi}| < 1~ \rm{GeV}, \quad
|E_\gamma^{\text{3rd}} -  E_\gamma^{\text{6A}}  | < 1~\rm{GeV} ,
\end{align}
where $E_\gamma^{\text{6A}} = (s-m_\phi^2)/(2\sqrt{s})$. 
We use the $\chi^2$ method to determine the pair of photons from
$\phi$ decay and single out the $\text{3rd}$ photon.
The energy of the $\text{3rd}$ photon $E_\gamma^{\text{3rd}}$ is very close 
to $E_\gamma^{\text{6A}}$, therefore we add an energy window cut.
In \cref{fig:Zto3a}, we see the sensitivity on BR for exotic Z decay 
for topology 6A can reach $\sim 10^{-7}$ for Giga Z and $3 \times 10^{-9}$
for Tera Z. For $m_\phi < 2$ GeV, it is hard to separate the two photons from
$\phi$ decay and the signal efficiency goes to zero. Instead of three photons
in the final state, one could look for two photons because the photons from 
$m_\phi$ can not be distinguished and therefore cover this mass range as in 
\cref{fig:axionconstraint}.

\subsection{The sensitivity reach of the HL-LHC}
\label{sec:LHCcompare}

The HL-LHC ($3 ~\text{ab}^{-1}$) also produce a lot of  Zs, which can be sensitive to some of 
the exotic Z decay modes. In this section we would like to 
study the sensitivities for the HL-LHC, and 
compare its reaches with the ones from 
Z-factories. 
A full fledged study with 
realistic detector simulations is beyond the scope of this paper. Instead, we perform simplified 
simulations aiming at gain an order of magnitude estimation. As we will see, the capabilities of 
the HL-LHC and Z-factories are very different. Our approach is sufficient to highlight
the relative strengths of the two experiments. 
For each topology we only pick up one benchmark parameter (zero for DM mass and 
40 GeV for other new physics mediate particles) to set the HL-LHC sensitivity,
and did not scan the parameter spaces of models, because the cut efficiency is not strongly depending 
on the mass. We have chosen the benchmark mass parameters to give the most energetic Z-decay 
products. In addition,  we do not consider fake photons from QCD, which will significantly 
reduce the HL-LHC sensitivity. In this sense, our projection for the HL-LHC should be considered as optimistic.
In order to suppress the huge QCD background and avoid pre-scaling, we search Z production in association 
with a high $p_T$ jet or high $p_T$ photon. For all the visible particle, we require $|\eta|< 2.5$.

\noindent $Z\to \gamma + \slashed{E}$:

For exotic decay $Z\to \gamma + \slashed{E}$, we generate $jZ$ event with the Magdraph 5 at 13 TeV LHC, 
and require the jet to be $p_T^{j_1} > 60$ GeV to make $Z $ have enough $p_T$ produce 
the energetic photon and large enough $\slashed{E}$ suppress the SM background. Specifically, 
we require $\slashed{E}_T > 50$ GeV and $p_T^\gamma > 20$ GeV together with $p_T^{j_1} > 60$ 
GeV as the basic cuts.  After the parton level event generation, it is passed to Pythia v6.4 
\cite{Sjostrand:2006za} for showering and hadronization, and to Delphes v3.2 \cite{deFavereau:2013fsa} 
for detector simulation. In the detector, missing energy could come from the jet reconstruction
due to jet energy resolution and uncertainty. Therefore, we include the SM background 
$j \gamma$ and irreducible SM background $j \gamma \nu \bar{\nu}$. We list the cross-sections
after basic cuts for signal $j Z \to j + \gamma + \slashed{E}$ and each SM backgrounds in
\cref{tab:LHC-a} in the column labeled with ``$\sigma_{\text{basic}}$". 

\begin{table*}[htbp]
  \centering
 \resizebox{\textwidth}{!}{%
  \begin{tabular}{|c|c|c||c|c|c||c|c|c|}
  \hline
  \multicolumn{3}{|c||}{$Z\to \gamma \slashed E$} & \multicolumn{3}{|c||}{$Z\to \gamma\gamma  \slashed E $}& \multicolumn{3}{|c|}{$Z\to l^+l^-  \slashed E $}\\  \hline
  &  $\sigma_{\text{basic}}$(pb) & $\epsilon$ & & $\sigma_{\text{basic}}$(pb)& $\epsilon$  & &    $\sigma_{\text{basic}}$(pb)& $\epsilon$ \\ \hline
 bkg($j \gamma$) &  14.6   &   0.15 & bkg($j \gamma\gamma$) &    0.037   &      0.083 $ \left( \times (0.05-0.2) \right)$        & bkg ($j \ell^+\ell^-$) & 0.68& 0.1  $\left( \times (0.03-0.8) \right)$               \\ \hline
 bkg($j \gamma  \nu\bar{\nu}$) &  0.23   &  0.16   &  bkg($j \gamma\gamma \nu\bar{\nu}$)  & 0.001   &     0.097 $\left( \times (0.084-0.17) \right)$  & bkg($j \ell^+\ell^-\nu \bar \nu$) & 0.37  & 0.28 $\left( \times (0.13-0.2) \right)$    \\ \hline
 1A& 459$ \times \rm BR$ & 0.54 &  2A & 124$ \times \rm BR$ &  0.2 & 3A  &  $101.6\times \rm BR$  &  0.63 \\ \hline
  1B & 108$ \times \rm BR$   &    0.55   &  2C  & 52.8$ \times \rm BR$ &  0.21  & 3B   &   $92.6\times \rm BR $   &     0.62\\ \hline
   1C& 471$ \times \rm BR$& 0.52  &   2D  &89.7$ \times \rm BR$ &  0.43       & 3D    &  $ 60.8\times \rm BR $     &    0.69 \\ \hline
    & & & & & & 3F & 85$\times \rm BR $   & 0.613  \\ \hline
     \multicolumn{3}{|c||}{$Z\to jj \slashed E$} & \multicolumn{3}{|c||}{$Z\to j j j j $} & & &\\  \hline
     bkg($j(j)\gamma$) & 32.23 & 0.11     & bkg ($\gamma j j( j)$)& 159.3  & 0.069 & &  &\\ \hline
     bkg ($b(b)\gamma$) & 0.67  &   0.156     &  bkg ($\gamma b b( j)$)& 5.1  & 0.071 & & &\\ \hline
      bkg($j (j)\gamma \nu\bar\nu$) & 0.185   & 0.22   & bkg ($\gamma b b( b)$)& 0.0023  & 0.076 & &  & \\ \hline
     bkg($b(b) \gamma \nu\bar\nu$) &  0.0023    & 0.256 & &   &   & & &\\ \hline
     5A &  0.27 $ \times \rm BR$  & 0.491 & 6A & 0.6$\times $BR  &0.43 & & &\\ \hline
     5B &  0.26 $ \times \rm BR$  & $0.50$  & 6B&   0.13$\times $BR & 0.39& & & \\ \hline
     5C &  0.19 $ \times \rm BR$  & $0.48 $ & 6C &   0.03$\times $BR & 0.26 & & &\\ \hline
       \end{tabular}}
  \caption{The exotic Z decay final states are listed for both SM backgrounds and signals. 
  The ``$\sigma_{\text{basic}}$" column gives the cross-section after basic cuts, and the $\epsilon$
  gives the cut efficiency for the further optimized cuts. The above cut efficiencies do not including the b-tagging efficiency. In the final sensitivity calculation, we use the b-tagging efficiency 0.7 and mis-tagged efficiency 0.015 \cite{Chatrchyan:2012jua} to re-weight the events according to the signal. 
  For SM background $j \gamma\gamma$, $j \gamma\gamma \nu\bar{\nu}$, $j \ell^+\ell^-$ and $j \ell^+\ell^-\nu \bar{\nu}$, there is an additional invariant mass window cut for $\gamma\gamma $ or $\ell^+ \ell^-$, which should multiply the efficiency given in the parentheses $\left(\times () \right)$. This additional efficiency is given as a range, because the mass window changes with the mediator mass in the signal topology. 
  Such change is indicated by the light brown shaded region for HL-LHC in \cref{fig:summary}.}
  \label{tab:LHC-a}
\end{table*}

\begin{figure}
	\centering
	\includegraphics[width=0.32 \columnwidth]{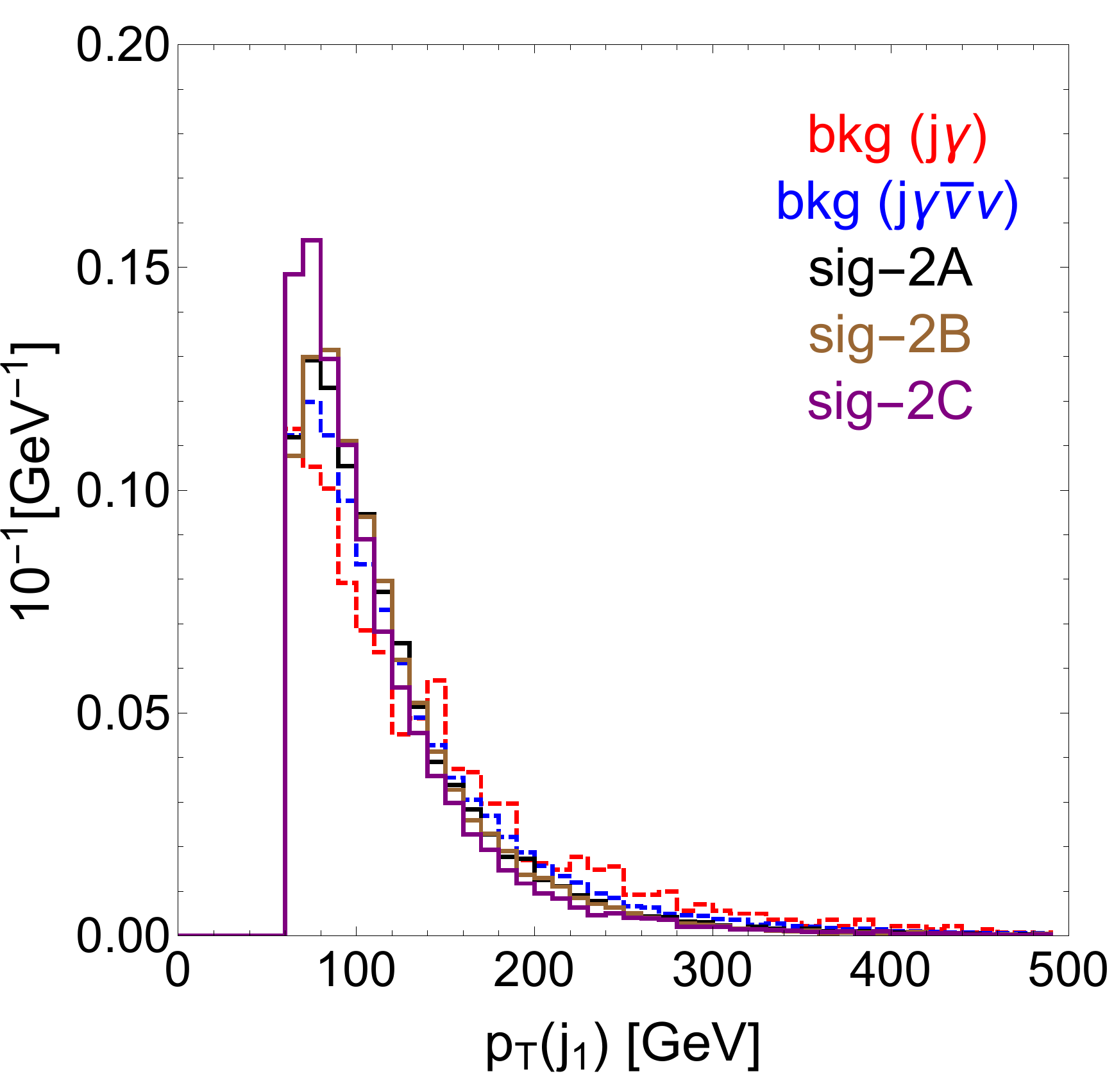} 
	\includegraphics[width=0.32 \columnwidth]{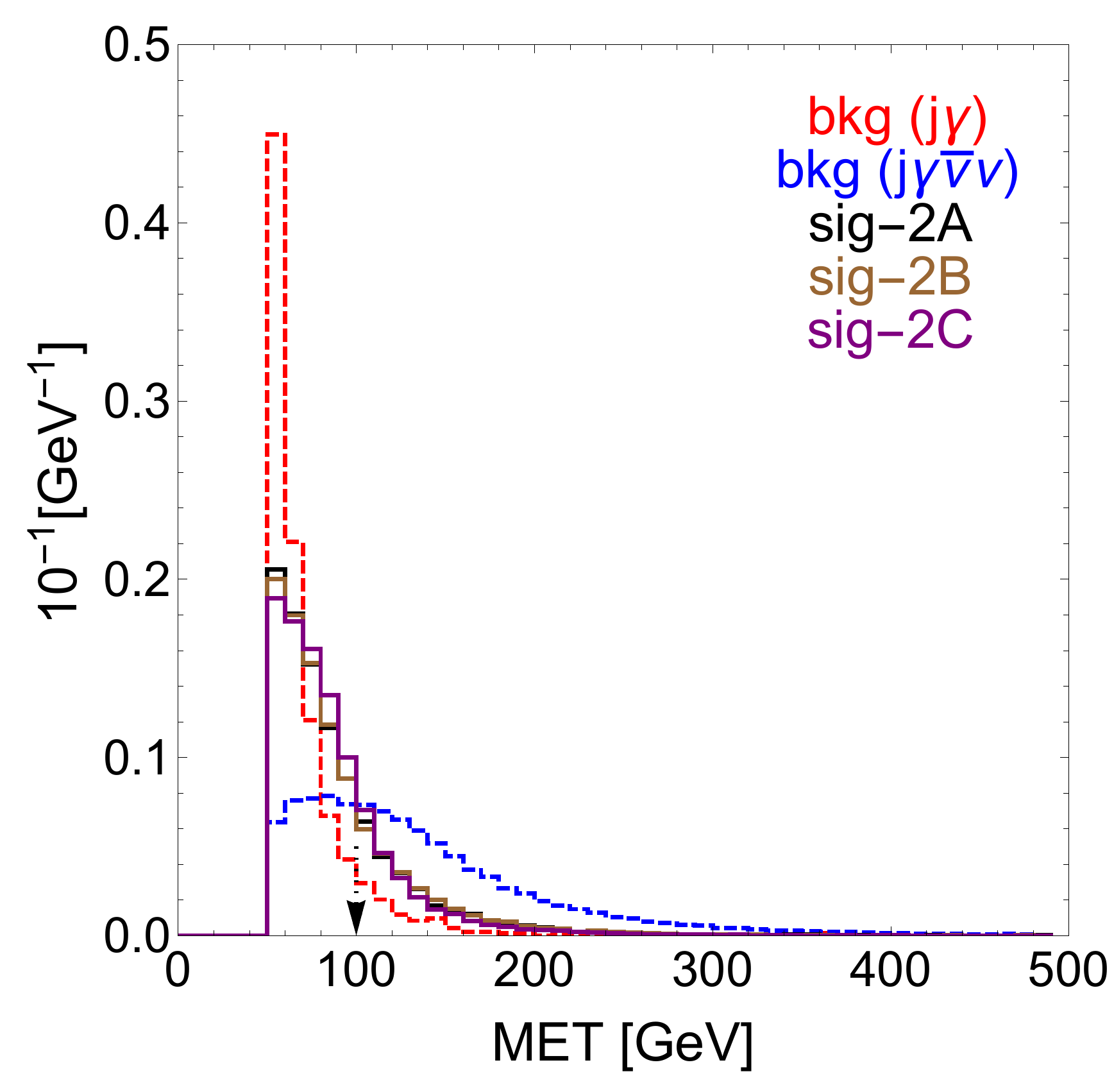} 
	\includegraphics[width=0.32 \columnwidth]{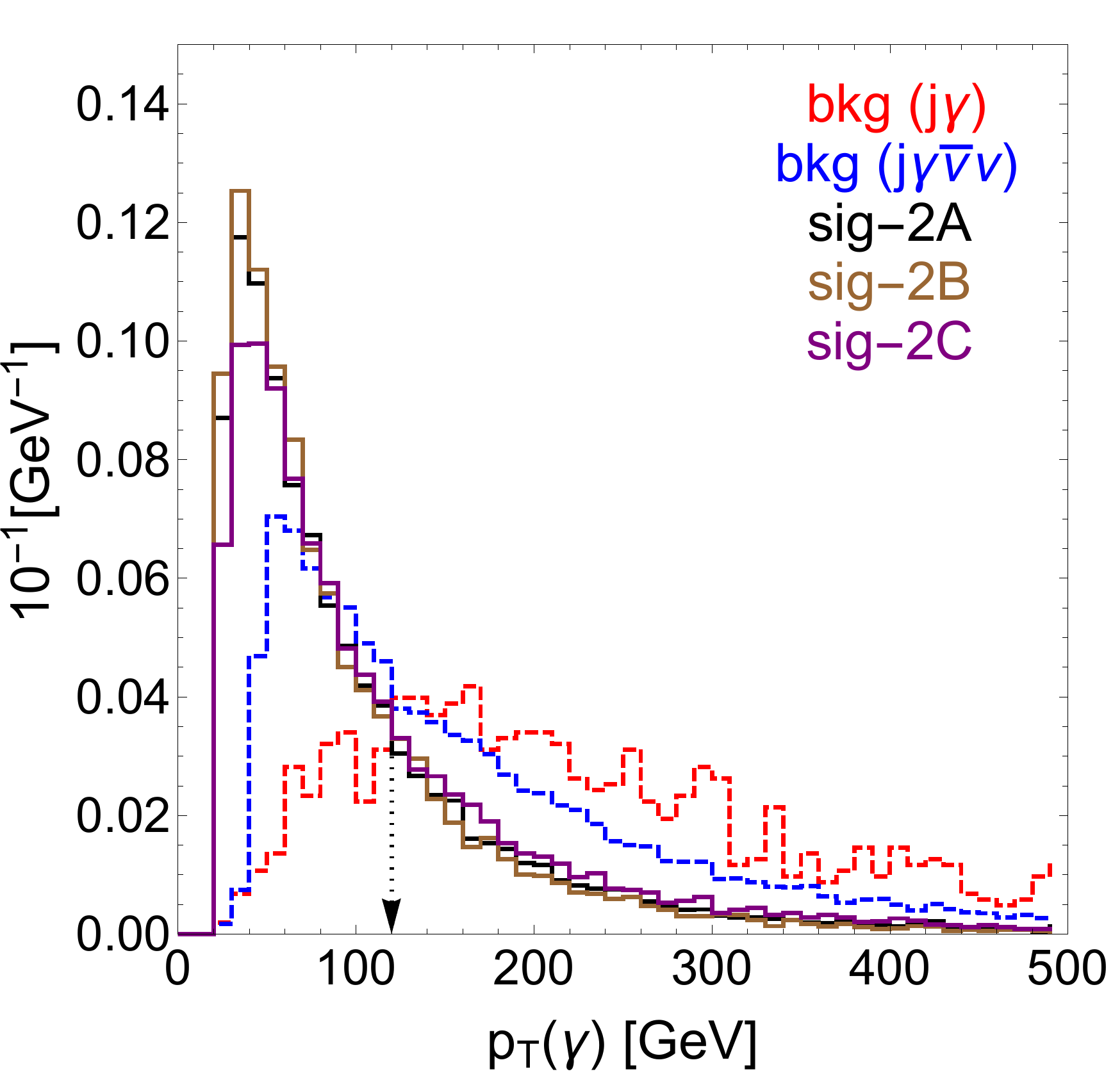} \\
	(a) ~~~~~~~~~~~~~~~~~~~~~~~~~~~~~~~~~~~~~~~~~~~~~~(b)~~~~~~~~~~~~~~~~~~~~~~~~~~~~~~~~~~~~~~~(c)
	\caption{ The normalized event distributions for kinematic variable $p_T^{j_1}$, $p_T^{\gamma}$ and $\slashed{E}_T$ for signal $jZ\to j+\gamma+ \slashed{E}$ and the corresponding SM background. The  distributions have been normalized to 1.
	}
	\label{fig:kinematic-LHC}
\end{figure}

\begin{figure*}[h!]
	\centering
	\includegraphics[width=0.98\textwidth]{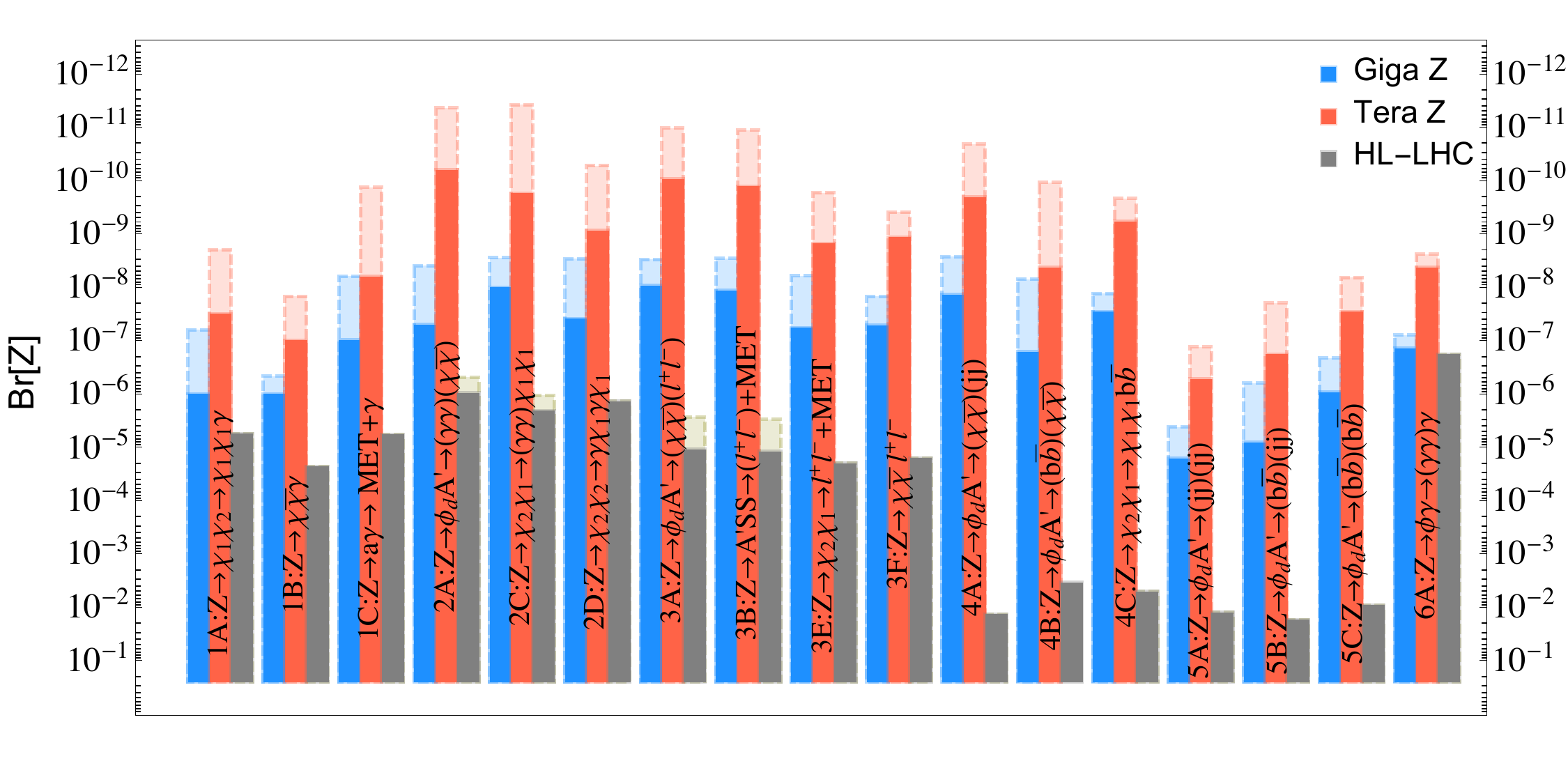}
	\caption{The sensitivity reach for BR for various exotic Z decay topologies at 
	the future Z-factory (Giga Z and Tera Z) and the HL-LHC at 13 TeV with $\mathcal{L} = 3  ~\rm{ab^{-1}}$.
    The BR sensitivity generally depends on model parameter, for example mediator mass and 
    dark matter mass.  The dark color
    region with solid line as boundary indicates the worst reach for the topology, while the lighter region with dashed line indicates the best reach. For HL-LHC, we add the light shaded region for the topology 2A, 2C
    , 3A and 3B to indicate the effect of an invariant mass window cut for diphoton and dilepton. For the topology 6A, the HL-LHC limit is obtained by rescaling the ATLAS study at 8 TeV LHC \cite{Aad:2015bua} with $\mathcal{L}=20 ~\rm{fb^{-1}}$. }
	\label{fig:summary}
\end{figure*}

To further optimize the signal, we make the differential distribution for kinetic variables $p_T^{j_1}$,
$\slashed{E}_T$ and $p_T^\gamma$ in \cref{fig:kinematic-LHC}. We compare the distribution of
SM background $j\gamma$ and $j\gamma \nu \bar{\nu}$ with signal $j Z$ with exotic Z decay
topology 1A, 1B and 1C. Based on \cref{fig:kinematic-LHC}, we further impose the following cuts,
\begin{align}
\slashed{E}_T< 100 ~{\rm GeV}, \quad p_T^\gamma < 150 ~\rm{GeV} .
\end{align}
We did not use additional cuts on $p_T^{j_1}$ because the distributions of SM background and
signal are quite similar. After applying the above cuts, we list the corresponding cut efficiency
in \cref{tab:LHC-a} in the column labeled ``$\epsilon$". For the HL-LHC ($3 ~\text{ab}^{-1}$), we can
reach the sensitivity for exotic Z decay BR of $5.6\times 10^{-6}$, $2.3\times 10^{-5}$ 
and $5.76 \times 10^{-6}$ for signal topology 1A, 1B and 1C. The sensitivities for the HL-LHC for each
topology are given in the summary plot \cref{fig:summary}.

\noindent $Z\to \gamma \gamma + \slashed{E}$:

The SM background we consider are $j \gamma \gamma$ with $\slashed{E}$ from mis-reconstruction and irreducible 
$j \gamma \gamma \nu\bar{\nu}$. The basic cuts are $p_T^j>60$ GeV,  $\slashed{E}_T>50$ GeV, 
and two photons with $p_T^\gamma > 20$ GeV. The cross-sections after cuts for signal and SM
background are again listed in \cref{tab:LHC-a}.
We use the following  cuts to further optimize our signal,
\begin{equation}
p_T^{j_1} > 80 ~{\rm GeV}, \quad 50 ~{\rm GeV} <\slashed{E}_T < 100 ~{\rm GeV},  
\quad 40 ~{\rm GeV} < p_T^{\gamma_1}< 100 ~{\rm GeV}.
\end{equation}
The cut efficiencies for signal and SM background are listed in \cref{tab:LHC-a}. For topology 2A and 2C, 
we can make an additional $5$ GeV window cut on the invariant mass of diphoton to suppress 
the SM background, while the signal is remain unaffected.  The corresponding efficiency has been 
listed in the parentheses in the $\epsilon$ column in \cref{tab:LHC-a}. It is a range for SM background 
due to the change of mediator mass. For the HL-LHC ($3 ~\text{ab}^{-1}$), the future 
sensitivity reach for exotic Z decay topologies 2A, 2C and 2D are $(5 - 10) \times 10^{-7}$, 
$(1 - 2) \times10^{-6}$, and $1.4 \times 10^{-6}$ respectively, and have been plotted in 
\cref{fig:summary}. 
The sensitivity range for the topology 2A and 2C has been indicated by light brown shaded
region.

\noindent $Z\to \ell^+ \ell^- + \slashed{E}$:

For decay topology $Z\to \ell^+ \ell^- +\slashed{E}$, we consider SM background $j \ell^+ \ell^- $ and 
irreducible $j \ell^- \ell^+ \nu\bar\nu$ with the
same reason. The basic cuts are one jet with $p_T^j> 60$ GeV, missing energy $\slashed E_T > 50$ 
GeV, and two leptons with $p_T^\ell> 20$ GeV. After checking the kinematic variable distribution,
we propose further cuts to optimize our signal,
\begin{align}
p_T^j> 90 ~{\rm GeV},    p_T^{\ell_1} < 80 ~{\rm GeV} . 
\end{align}
For topology 3A and 3B, we have added the same additional $5$ GeV window cut on the invariant 
mass of dilepton. The corresponding efficiency has been listed in the parentheses in the $\epsilon$ 
column in \cref{tab:LHC-a}. For the HL-LHC ($3 ~\text{ab}^{-1}$), the future sensitivity reach for exotic 
Z decay topologies 3A, 3B, 3D and 3F are $(3 - 11) \times 10^{-6}$, $(3 \sim 12) \times 10^{-6}$, 
$2.0\times 10^{-5}$ and $1.6\times 10^{-5}$ respectively, and have been plotted in \cref{fig:summary}. 
The sensitivity range for the topology 3A and 3B has been indicated by light brown shaded
region in \cref{fig:summary}.

\noindent $Z\to j j + \slashed{E}$:

For decay topology $Z\to j j +\slashed{E}$, we generate signal events $\gamma Z$ to suppress
QCD background and consider SM background $\gamma j $ and irreducible $\gamma j j \nu\bar\nu$. 
The basic cuts are two jets with $p_T^j> 30$ GeV, missing energy $\slashed E_T > 50$ 
GeV, and one photon with $p_T^\gamma > 60$ GeV. After checking the kinematic variable distribution,
we propose further cuts to optimize our signal,
\begin{align}
p_T^{j_1}< 100 ~{\rm GeV},   \slashed E_T> 60 ~{\rm GeV} ,   p_T^\gamma > 90 ~{\rm GeV}.
\end{align}
For the HL-LHC ($3 ~\text{ab}^{-1}$), the future sensitivity reach for exotic Z decay topologies 
4A, 4B, 4C are $0.0136$, $3.45 \times10^{-3}$ and $5.07 \times 10^{-3}$, respectively, and have 
been plotted in \cref{fig:summary}. 

\noindent $Z\to(JJ)(JJ)$:

For decay topology $Z\to (JJ)(JJ)$ which is fully hadronic, 
we generate signal events $\gamma Z$ to suppress QCD background and consider 
SM background $\gamma J $ matched with $\gamma J J$ by Pythia and irreducible 
$\gamma  J \nu\bar\nu$ matched with $\gamma  J J \nu\bar\nu$, where $J$ can be
light flavor jets $j$ or b-tagged jet $b$.  
We require at least four jets with $p_T^J> 60$ GeV, and 
one photon with $p_T^\gamma > 60$ GeV. We propose further cuts to optimize our signal,
\begin{align}
p_T^{J_1} > 120 ~{\rm GeV} ,    m_{JJJJ}< 250 ~{\rm GeV},
\end{align}
and the cut efficiencies for signal and SM background are given in \cref{tab:LHC-a}.
Note we have generated the SM backgrounds with light flavor jet and b-jet separately.
Both of them can contribute to background of the corresponding signal topologies
5A, 5B and 5C with b-tagging efficiency re-weighting.  
For the HL-LHC ($3 ~\text{ab}^{-1}$), the sensitivity reach for exotic Z decay topologies 
5A, 5B, 5C are $0.0126$, $0.0172$ and $0.00915$, respectively, and have 
been plotted in \cref{fig:summary}. It is not surprising that sensitivity for fully hadronic decay of Z at the
HL-LHC can not compete with future $e^+e^-$ collider, because of the huge QCD background.

Using jet substructure technique can probably achieve better sensitivities in the 
exotic hadronic Z decay topologies. CMS at 13 TeV LHC has searched for light vector
resonance which decay into quark pair in association with a high $p_T$ jet to make
the light vector gauge boson highly boosted \cite{CMS-PAS-EXO-17-001}, which decay 
products are merged into a single jet. The characteristic feature of the signal is a single 
massive jet with two-prong substructure produced in association with a jet from 
initial state radiation. The SM process $jZ \to j (jj)$ has been nicely reconstructed. 
In the exotic decay topology $Z\to(JJ)(JJ)$, one would look for four-prong 
substructure in the fat jet to suppress the SM $jZ$ background. For final state
including b-jets, b-tagging techniques in the jet substructure could further help
in reducing the SM QCD background, which already help observing with a local
significance of 5.1 standard deviations for the first time in the single jet topology
in $Z\to bb$ process \cite{CMS-PAS-HIG-17-010}.

\noindent $Z\to\gamma\gamma\gamma$:

The last exotic Z decay search is $Z\to \gamma\gamma\gamma$, which has been 
performed by  ATLAS at 8 TeV LHC \cite{Aad:2015bua} with $\mathcal{L}=20 ~\rm{fb^{-1}}$. 
The corresponding constraint is $BR(Z\to \gamma\gamma \gamma) < 2.2\times 10^{-6}$.
It is hard for us to reliably study this topology due the difficulty in simulating the fake photons
from QCD backgrounds. Instead, we do a simple rescaling according to the HL-LHC
integrated luminosity $3 ~\text{ab}^{-1}$, which gives the limit 
$BR(Z\to \gamma\gamma \gamma) < 1.8 \times 10^{-7}$.

\noindent \textit{Summary}:

In \cref{fig:summary}, we see the sensitivity of exotic Z decay branching ratios at the HL-LHC 
generally can not compete with future Z-factory because of the large QCD background. The Z exotic decay products are typically rather soft. Requiring another hard radiation can help with triggering and making the Z-decay products more energetic. At the same time, it will reduce the signal rate significantly.  
For exotic Z decay with missing energy in final state, another important background can come from mis-measurement of the QCD jets. Since the missing energy from Z-decay tends to be small, this background can be significant.  
For photons in final states,
there can be fake photons from QCD which we have not considered. For hadronic exotic
Z decay, the situation at the HL-LHC is even worse.

\section{Conclusion and discussion}
\label{sec:conclusion}

We have presented a comprehensive study on exotic Z decay at future Z-factories,
with emphasis on its prospects to exploring dark sector models. There are many dark
sector models can give rise to exotic Z decay modes, many of which contain missing
energy in the final states. A Z-factory provides a clean environment for decay modes 
which can be overwhelmed by large background at hadron colliders.  Another advantage 
of searching for such  exotic Z decay at future $e^+ e^-$ colliders is the ability of 
reconstructing the full missing 4-momentum, while we can only reconstruct missing 
transverse momentum at hadron colliders. We have demonstrated the capability of exotic 
Z decay at future Z-factory to provide the leading constraint in comparison with existing 
collider limits, future HL-LHC projections, and current DM searches.

We  classify final states of the exotic decays with
the number of resonances, and possible topologies it could have. 
We make projections on  the sensitivity on the branching ratio of exotic Z decay 
at future Z-factory. For final states with missing energy, it can provide limits on BR down to $10^{-6}
- 10^{-8.5}$ for Giga Z and $10^{-7.5} - 10^{-11}$ for Tera Z. 
The sensitivities on BR for different final states are roughly ordered from
high to low as $\slashed{E} \ell^+\ell^- \sim \slashed{E} \gamma \gamma$,
$\slashed{E} JJ$ and $\slashed{E} \gamma$, due to the size of the SM backgrounds for each mode. 
In the same final states, it is quite clear the SM backgrounds for signal with more resonances can be better suppressed. 
In addition to the final states with missing energy, we also selectively studied
the fully visible final states $(JJ)(JJ)$ and $(\gamma \gamma) \gamma$,
where the first one contains two resonances and the second one contains
one resonance. It is interesting to look for purely hadronic final states at future Z-factories,
because it has much less QCD background in comparison with hadron collider.
We found it can provide limits on BR down to $10^{-5} - 10^{-6.5}$
for Giga Z and $10^{-6.5} - 10^{-8}$ for Tera Z. The sensitivity to the final states with $b$ jet is better than 
those with light flavor jets due to smaller SM backgrounds. 

We have also made estimates of the reach of the HL-LHC on the exotic Z decay modes. The decay products tend to be 
soft and difficult for the LHC searches. There is also large QCD backgrounds.  
We considered the cases with additional energetic initial state radiation, which can help with suppressing these
backgrounds. However, this also reduced the signal rate. Therefore, for the channels we considered here,
it is very hard for the HL-LHC to compete with future Z-factory. 
The one exception is the $(\gamma \gamma) \gamma$ channel. The sensitivity on BR
can reach $10^{-7}$ for Giga Z and a few $10^{-9}$ for Tera Z. 
The corresponding HL-LHC sensitivity is rescaled from an exiting
study at 8 TeV by rescaling, and it can be comparable to that of the Z-factory.

We have studied four representative models in \cref{sec:models}, 
namely Higgs portal with scalar DM, vector portal DM model, MIDM and RayDM, 
and axion-like particle model. 
In Higgs portal model with DM, the decay topology $Z \to \tilde{s} Z^* \to 
(\bar{\chi} \chi) + \ell^+ \ell^-$
has been
studied.  Future Z-factories can provide the leading constraint on mixing
angle $\sin \alpha$ between SM Higgs and dark singlet scalar mediator. The
constraint from $Z \to \tilde{s} \gamma$ via loop effect has also been
considered, but is weaker due to loop suppression and larger SM background.
In vector portal DM model, the decay topologies $\tilde{Z} \to \tilde{A}'
S S^* \to (\ell^+\ell^-)\slashed{E}$ and $\tilde{Z} \to \tilde{A}' \tilde{\phi}
 \to \ell^+ \ell^- (\slashed{E})$ are studied. The first one simply arises
when DM is a scalar and charged under $U(1)_D$, and the second one
is a dark Higgs bremsstrahlung process. We found that the limits from the
exotic Z decay provides a competitive and complementary constraints with
DM direct detection, while the other collider limits are much weaker.   
In MIDM and RayDM model, the decay topologies $Z \to \chi_2 \chi_1 \to 
(\chi_1 \gamma) \chi_1$ from MIDM operator and $Z \to \chi_1 \chi_1
\gamma $ from RayDM operator has been considered. 
Both operators can be originated from heavy fermions and scalars in the
loop, which couples to DM. The  constraint on MIDM operator is much stronger 
than the constraint on RayDM.
It is also much better than gamma-line search in indirect 
detection and future hadron collider projections. 
In axion-like particle model, the decay topologies $Z \to a \gamma \to 
(\gamma \gamma) \gamma$ and $Z \to a \gamma \to (\slashed{E}) 
\gamma$ have been considered, where in the first one the axion-like
particle decay promptly into two photons and in the second one it
decays outside the detector. We find future Z-factory can provide
the leading constraint on $\Lambda_{\text{aBB}}$ comparing with
limits from LEP and LHC.

All in all, the exotic Z decay searches can provide unique tests on 
dark sector models at future Z-factory, especially when missing energy 
and/or hadronic objects appears in the final states. 
We explicitly analyze four representative dark
sector models and find the exotic Z decay searches can provide the leading 
and complementary limits to the current and future collider searches and DM
searches. It can also cover parameter spaces of DM models with the relic 
abundance requirement, which provides a complementary cross-check on
DM problem.

\section*{Acknowledgments}
\label{sec:acknowledgments}

We would like to thank Oliver Fischer, Manqi Ruan and Felix Yu 
for useful discussions and  comments. 
JL is supported by the Oehme Fellowship.
LTW is supported by the DOE grant No. DE-SC-0013642.
XPW is supported  by the U.S. Department of
Energy under Contract No. DE-AC02-06CH11357.
WX is supported by the U.S. Department of Energy under grant contract numbers 
DE-SC-00012567 and DE-SC-00015476, and 
the grant 669668-NEO-NAT-ERC-AdG-2014. 
The work of JL and XPW was partially
supported by the Cluster of Excellence Precision Physics, Fundamental 
Interactions and Structure of Matter (PRISMA-EXC 1098), the German 
Research Foundation (DFG) under Grants
No. \mbox{KO 4820/1--1}, and No. FOR 2239, and from the European
Research Council (ERC) under the European Union’s Horizon 2020
research and innovation program (Grant No. 637506, ``$\nu$Directions'').

\section{Appendix}
\label{sec:appendix}

\subsection{The annihilation cross-section for scalar DM with vector portal}
\label{sec:SigmaOFscalarVportal}

We calculate the annihilation cross-sections of scalar DM into SM fermions. 
The scalar DM is charged under $U(1)_D$ as in \cref{eq:lAS}, and the kinetic
mixing induced interactions with SM sector are given in \cref{eqn:currents},
which includes both s-channel $\tilde{A}'$ and $\tilde{Z}$ mediation. The
annihilation cross-sections for one generation are given,

\begin{align}
\sigma v_{SS\to u\bar u}&=\frac{g_D^2e^2\epsilon^2\sqrt{s-4m_u^2}\left(s-4m_S^2\right)}{576\pi s^{3/2} \cos\theta_w^4(m_{\tilde K}^2-m_Z^2)^2\left((s-m_{\tilde K}^2)^2+m_{\tilde K}^2\Gamma_{\tilde K}^2\right)\left((s^2-m_{\tilde Z}^2)^2+m_{\tilde Z}^2\Gamma_{\tilde Z}^2 \right)} \\ \nonumber
&\left[(17s+7m_u^2)\left(s^2(m_{\tilde K}^2-m_{\tilde Z}^2)^2+m_{\tilde K}^2m_{\tilde Z}^2(m_{\tilde Z}\Gamma_{\tilde K}-m_{\tilde K}\Gamma_{\tilde Z})^2\right) \right. \\ \nonumber
&-40\cos\theta_w^2m_{\tilde Z}^2(s+2m_u^2)\left(s(m_{\tilde K}^2-m_{\tilde Z}^2)^2+m_{\tilde K}m_{\tilde Z}(m_{\tilde Z}\Gamma_{\tilde K}-m_{\tilde K}\Gamma_{\tilde Z})(m_{\tilde K}\Gamma_{\tilde K}-m_{\tilde Z}\Gamma_{\tilde Z})\right) \\ \nonumber
&\left. +32\cos\theta_w^4m_{\tilde Z}^4(s+2m_u^2)\left(m_{\tilde K}^4 + m_{\tilde K}^2(\Gamma_{\tilde K}^2 -2 m_{\tilde Z}^2)-2m_{\tilde K}m_{\tilde Z}\Gamma_{\tilde K}\Gamma_{\tilde Z}+m_{\tilde Z}^2(m_{\tilde Z}^2+\Gamma_{\tilde Z}^2)\right) \right] ,
\end{align}

\begin{align}
\sigma v_{SS\to d\bar d}&=\frac{g_D^2e^2\epsilon^2\sqrt{s-4m_d^2}(s-4m_S^2)}{576\pi \cos\theta_w^4s^{3/2}(m_{\tilde K}^2-m_Z^2)^2\left((s-m_{\tilde K}^2)^2+m_{\tilde K}^2\Gamma_{\tilde K}^2\right)\left((s-m_{\tilde Z}^2)^2+m_{\tilde Z}^2\Gamma_{\tilde Z}^2 \right)} \\ \nonumber
&\left[(5s-17m_d^2)\left(s^2(m_{\tilde K}^2-m_{\tilde Z}^2)^2+m_{\tilde K}^2m_{\tilde Z}^2(m_{\tilde Z} \Gamma_{\tilde K}-m_{\tilde K}\Gamma_{\tilde Z})^2\right)\right. \\ \nonumber
&-4\cos\theta_w^2m_{\tilde Z}^2\left(s+2m_d^2\right)\left(s(m_{\tilde K}^2-m_{\tilde Z}^2)^2+m_{\tilde K}m_{\tilde Z}(m_{\tilde Z}\Gamma_{\tilde K}-m_{\tilde K}\Gamma_{\tilde Z})(m_{\tilde K}\Gamma_{\tilde K}-m_{\tilde Z}\Gamma_{\tilde Z})\right) \\ \nonumber
&\left.+8\cos\theta_w^4m_{\tilde Z}^4(s+2m_d^2)\left(m_{\tilde K}^4+m_{\tilde K}^2(-2m_{\tilde Z}^2+\Gamma_{\tilde K}^2)-2m_{\tilde K}m_{\tilde Z}\Gamma_{\tilde K}\Gamma_{\tilde Z}+m_{\tilde Z}^2(m_{\tilde Z}^2+\Gamma_{\tilde Z}^2)\right)\right] ,
\end{align}

\begin{align}
\sigma v_{SS\to \nu\bar\nu}&=   \frac{g_D^2e^2\epsilon^2(s-4m_S^2)\left(s^2(m_{\tilde K}^2-m_{\tilde Z}^2)^2+m_{\tilde K}^2m_{\tilde Z}^2(m_{\tilde Z}\Gamma_{\tilde K}-m_{\tilde K}\Gamma_{\tilde Z})^2\right)}{ 192 \cos\theta_w^4\pi(m_{\tilde K}^2-m_{\tilde Z}^2)^2\left((s-m_{\tilde K}^2)^2+m_{\tilde K}^2\Gamma_{\tilde K}^2\right)\left((s-m_{\tilde Z}^2)^2+m_{\tilde Z}^2\Gamma_{\tilde Z}^2\right)}
,
\end{align}

\begin{align}
\sigma v_{SS\to l\bar l}&=\frac{g_D^2e^2\epsilon^2\sqrt{s-4m_l^2}(s-4m_S^2)}{192\pi\cos\theta_w^4s^{3/2}(m_{\tilde K}^2 -m_{\tilde Z}^2)^2\left((s-m_{\tilde K}^2)^2+m_{\tilde K}^2\Gamma_{\tilde K}^2\right)\left((s-m_{\tilde Z}^2)^2+m_{\tilde Z}^2\Gamma_{\tilde Z}^2 \right)}     \\ \nonumber
&\left[ (5s+7m_l^2)\left(s^2 \left(m_{\tilde K}^2-m_{\tilde Z}^2\right)^2+m_{\tilde K}^2 m_{\tilde Z}^2 (m_{\tilde Z} \Gamma_{\tilde K}-m_{\tilde K} \Gamma_{\tilde Z})^2\right)\right. \\ \nonumber
&-12\cos\theta_w^2m_{\tilde Z}^2(s+2m_l^2)\left(s(m_{\tilde K}^2-m_{\tilde Z}^2)^2+m_{\tilde K}m_{\tilde Z}(m_{\tilde Z}\Gamma_{\tilde K}-m_{\tilde K}\Gamma_{\tilde Z})(m_{\tilde K}\Gamma_{\tilde K}-m_{\tilde Z}\Gamma_{\tilde Z})\right) \\ \nonumber
&\left.+8\cos\theta_w^4m_{\tilde Z}^4(s+2m_l^2)\left(m_{\tilde K}^4+m_{\tilde K}^2(-2m_{\tilde Z}^2+\Gamma_{\tilde K}^2)-2m_{\tilde K}m_{\tilde Z}\Gamma_{\tilde K}\Gamma_{\tilde Z}+m_{\tilde Z}^2(m_{\tilde Z}^2+\Gamma_{\tilde Z}^2)\right)\right]  ,
\end{align}
where we see the cross-sections are p-wave suppressed.

\newpage
\bibliography{ref}

\bibliographystyle{JHEP}

\end{document}